\documentclass[a4paper,11pt]{article}
\usepackage{jheppub} 
\usepackage{lineno}
\usepackage{booktabs}
\usepackage{amsmath}
\usepackage{graphicx, caption, subcaption}
\usepackage{array}
\usepackage{float}
\usepackage{multirow}
\usepackage{makecell}

\arxivnumber{1234.56789} 

\title{\boldmath Probing triple Higgs production via $4\tau 2b$ decay channel at a 100 TeV hadron collider}

\author{Zhenyu Dong,}
\author{Xiaohu Sun$^*$,}
\author{Botao Guo,}
\author{Licheng Zhang,}
\author{Zhiyuan Li,}
\author{Jin Wang,}
\author{Zhe Li,}
\author{Yong Ban and}
\author{Yajun Mao}
\affiliation{%
Department of Physics and State Key Laboratory of Nuclear Physics and Technology, Peking University, Beijing 100871, China
}%

\emailAdd{Xiaohu.Sun@pku.edu.cn}

\abstract{A comprehensive study of triple Higgs boson production in the $4\tau 2b$ decay final state is performed for the first time at a future 100 TeV hadron collider. The analysis incorporates modified Higgs self-couplings via trilinear Higgs self-coupling $c_3$ and quartic Higgs self-coupling $d_4$, enabling for a model-independent investigation of potential new physics effects. Higgs bosons are reconstructed using both resolved and boosted techniques. To optimize sensitivity across different kinematic regions, we introduce a novel event categorization strategy based on the triple Higgs invariant mass spectrum and the multiplicity of boosted Higgs bosons. In addition to a traditional cut-based analysis, a Boosted Decision Tree (BDT) approach is employed to exploit multivariate correlations among kinematic observables, leading to a significant improvement in sensitivity. Our result demonstrates that the $4\tau 2b$ channel provides a viable pathway for probing the Higgs quartic coupling, complementing the existing multi-Higgs production studies, and could reach 5 $\sigma$ in significance for $c_3 \lesssim -1$ and $d_4 \gtrsim 10$ in the scanned range.}

\begin{document}
\maketitle
\flushbottom

\section{Introduction}
\label{sec:intro}

The discovery of the Higgs boson at the Large Hadron Collider (LHC) marked a monumental achievement in particle physics \cite{CMS:2012qbp,ATLAS:2012yve}. This discovery completed the Standard Model (SM) and opened new frontiers in understanding the origin of mass and the nature of electroweak symmetry breaking. Since the discovery of the Higgs boson, precise measurements of its properties have been a cornerstone of particle physics \cite{CMS:2022dwd,ATLAS:2022vkf}. In particular, Higgs self-interactions through Higgs pair and triple Higgs production, have become crucial probes for investigating the fundamental structure of the Higgs potential and understanding the mass origin of the universe.

The measurement of Higgs pair production serves as the first step towards understanding the Higgs self-interactions. This process is directly sensitive to the trilinear Higgs self-coupling $c_3$, a key parameter that determines the shape of the Higgs potential. At the LHC, extensive searches for Higgs pair production have been conducted through various decay channels, such as $bbbb$, $bb\gamma\gamma$, $bb\tau\tau$, $bbWW$, etc, providing important constraints on $c_3$ \cite{ATLAS:2022jtk,ATLAS:2022xzm,ATLAS:2023gzn,ATLAS:2024lsk,ATLAS:2018fpd,ATLAS:2018hqk,ATLAS:2023elc,ATLAS:2022fpx,CMS:2024awa,CMS:2017rpp,CMS:2024rgy,CMS:2022cpr,CMS:2022gjd,CMS:2022hgz,CMS:2022kdx,CMS:2020tkr,CMS:2022omp,CMS:2024fkb}. 
The constraints on the Higgs trilinear self-coupling are often expressed in terms of the parameter $\kappa_\lambda$ ($\kappa_\lambda = 1$ for the SM). ATLAS and CMS collaborations provide significant constraints on $\kappa_\lambda$, derived from combined analyses of various Higgs pair decay channels, leveraging the full Run 2 dataset at $\sqrt{s} = 13$ TeV. ATLAS reports limits of $\kappa_\lambda \in [-0.4, 6.3]$ at 95\% confidence level (CL), and CMS places limits of $\kappa_\lambda \in [-1.3, 8.1]$ at 95\% CL \cite{ATLAS:2022jtk,CMS:2024awa}. The relationship between $\kappa_\lambda$ and the coupling parameter $c_3$ used in this study is given by $c_3 + 1 = \kappa_\lambda$. 
These efforts have demonstrated the feasibility of probing Higgs self-interactions at hadron colliders, though the observation of the SM Higgs pair production remains challenging due to its small cross section. The low production rate in the SM also means that any excess in the di-Higgs signal could point to new physics scenarios. Motivated by this possibility, searches for resonant Higgs pair production and other beyond the Standard Model (BSM) effects have been extensively carried out \cite{ATLAS:2022hwc,ATLAS:2020jgy,ATLAS:2020azv,ATLAS:2021ifb,ATLAS:2023vdy,ATLAS:2018ili,ATLAS:2023tkl,ATLAS:2024auw,CMS:2024pjq,CMS:2016jvt,CMS:2017hea,CMS:2017aza,CMS:2018tla,CMS:2018qmt,CMS:2018vjd}, relevant to various BSM scenarios \cite{Grojean:2004xa,Cao:2013si,Gouzevitch:2013qca,Gupta:2013zza,Han:2013sga,Nishiwaki:2013cma,Goertz:2014qta,Hespel:2014sla,Cao:2014kya,Carena:2015moc,Grober:2015cwa,Wu:2015nba,He:2015spf,Carvalho:2015ttv,Zhang:2015mnh,Huang:2015tdv,Nakamura:2017irk,DiLuzio:2017tfn,Huang:2017nnw,Buchalla:2018yce,Borowka:2018pxx,Chang:2019vez,Blanke:2019hpe,Li:2019tfd,Capozi:2019xsi,Alves:2019igs,Kozaczuk:2019pet,Barducci:2019xkq,Huang:2019bcs,Cheung:2020xij,Cao:2015oaa,Cao:2016zob,Li:2019uyy,Lu:2015qqa,Ren:2017jbg,Guo:2022biq}.

Theoretical predictions suggest that the cross section for triple Higgs production is significantly smaller than di-Higgs production, making it a challenging target for current colliders \cite{Plehn:2005nk}. However, the advent of a 100 TeV hadron collider offers new possibilities for accessing to this challenging but crucial process. The proposed Future Circular hadron-hadron Collider (FCC-hh) at CERN, aiming for an unprecedented center-of-mass energy up to 100 TeV and high luminosity, is expected to produce over 50,000 triple Higgs events with an integrated luminosity of $30 \, \text{ab}^{-1}$ \cite{Papaefstathiou:2015paa}, which is a promising opportunity to probe the Higgs potential structure \cite{Contino:2016spe,Mangano:2016jyj,FCC:2018byv,FCC:2018evy,Benedikt:2651300}. The Super proton-proton Collider (SppC) in China, with a planned center-of-mass energy of up to 125 TeV, represents another potential avenue for exploring high-energy Higgs processes \cite{Tang:2022fzs, CEPC-SPPCStudyGroup:2015csa,CEPC-SPPCStudyGroup:2015esa,CEPCStudyGroup:2023quu,Tang:2022qku}.


The triple Higgs production, though challenging to observe, provides a unique window into the Higgs quartic coupling $d_4$. ATLAS performed the first direct search for triple Higgs boson production in the six b-quark final state, using $126 \, \text{fb}^{-1}$ of the LHC Run2 data at $\sqrt{s} = 13$ TeV, setting an upper limit of $59 \, \text{fb}$, at 95\% confidence level, on the SM triple Higgs production cross-section \cite{ATLAS:2024xcs}.
Looking ahead to future colliders, phenomenological studies at 100 TeV have explored several promising channels \cite{Papaefstathiou:2015iba,Papaefstathiou:2019ofh,Fuks:2015hna,Fuks:2017zkg,Chen:2015gva,Dicus:2016rpf,Agrawal:2017cbs,Kilian:2017nio,Belyaev:2018fky}. The $4b2\gamma$ channel, characterized by its clean signature, benefits from efficient b-tagging and excellent photon energy resolution, offering the potential to achieve a $2\sigma$ significance in the SM \cite{Chen:2015gva,Fuks:2015hna,Papaefstathiou:2015paa}. The $2b2l4j+\text{MET}$ (Missing Transverse Energy) channel has also been studied, although it remains challenging to observe in the SM scenario, it could serve as a sensitive probe for BSM physics scenarios \cite{Kilian:2017nio}. Among the multi-b/$\tau$-jet final states -- specifically, the $6b$ ($\sim 19.21\%$ branching ratio), $4b2\tau$ ($\sim 6.31\%$) and $4\tau2b$ ($\sim 0.69\%$) channels -- the $6b$ and $4b 2\tau$ channels have been investigated, achieving a $2\sigma$ significance within the SM \cite{Fuks:2017zkg,Fuks:2015hna,Papaefstathiou:2019ofh}. 

In this analysis, we present a first study of triple Higgs production in the $4\tau2b$ final state with hadronically-decaying $\tau$ leptons. A coupling-dependent partitioning of the triple Higgs invariant mass space is introduced. Additionally, a categorization scheme based on the number of boosted Higgs bosons is employed to optimize the signal acceptance across various final states topologies. Both resolved and boosted reconstruction strategies are incorporated to maximize the sensitivity across different kinematic regions.

The structure of this paper is as follows: In Section~\ref{sec:mc}, we describe the theoretical framework, Monte Carlo (MC) sample generation and the simulation setup. Section~\ref{sec:analysis} details the analysis strategy, including both resolved and boosted reconstruction techniques. A cut-based optimization is performed in both cases.
A Boosted Decision Tree (BDT) training is further performed in the resolved regions. 
In Section~\ref{sec:results}, we present the results of both cut-based and BDT training analyses, including event yields and significance for both SM and BSM scenarios, as well as sensitivity contours for triple Higgs production alone and in combination with VHH production in the $(c_3,d_4)$ plane. Finally, Section~\ref{sec:conclusion} summarizes our findings and discusses prospects for future research. 

\section{MC Simulation}
\label{sec:mc}

\subsection{Theoretical framework}
In this study, new physics effects in multi-Higgs interactions are investigated by modifying the SM Higgs potential in a model-independent manner, described in Eq.~\ref{eq:HHH}. Two parameters, the trilinear $(c_3)$ and quartic $(d_4)$ Higgs self-couplings, are of our great interests. These two couplings provide a parametric framework to investigate deviations from the SM predictions. For the SM case, $c_3 = 0$ and $d_4 = 0$. Here, $h$ represents the Higgs boson field, $m_{H}$ is its mass, $v_{0}$ is the vacuum expectation value, and $\lambda_{\text{SM}} = \frac{m_H^2}{2v_0^2} $ denotes the SM self-interaction strengths.

\begin{equation}
\label{eq:HHH}
\begin{aligned}
    V(h) = \frac{1}{2} m_H^2 h^2 + \lambda_{\text{SM}}(1 + c_3)v_0 h^3 + \frac{1}{4} \lambda_{\text{SM}}(1 + d_4)h^4,
\end{aligned}
\end{equation}

\begin{figure}[tp]
\centering
\begin{subfigure}{.3\textwidth}
\centering
\includegraphics[height=2cm]{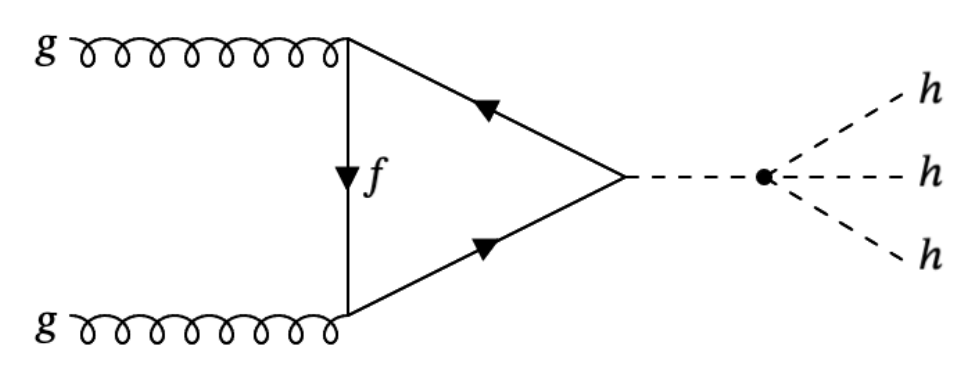}
\caption{}
\end{subfigure}%
\hfill
\begin{subfigure}{.3\textwidth}
\centering
\includegraphics[height=2cm]{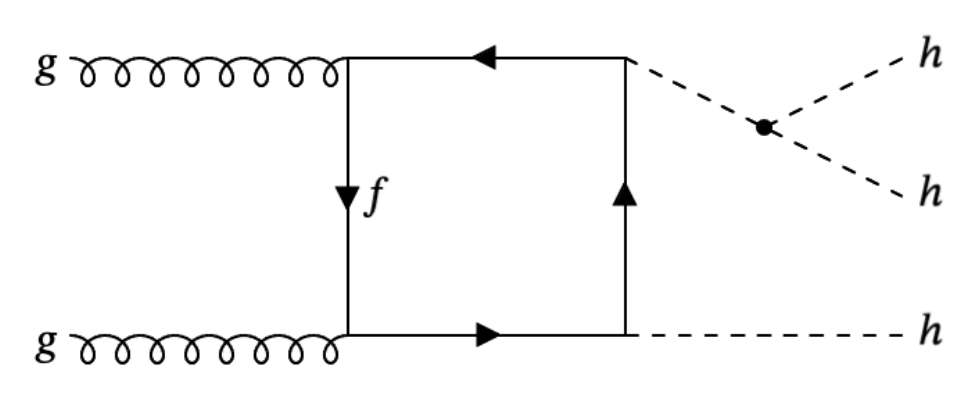}
\caption{}
\end{subfigure}%
\hfill
\begin{subfigure}{.3\textwidth}
\centering
\includegraphics[height=2cm]{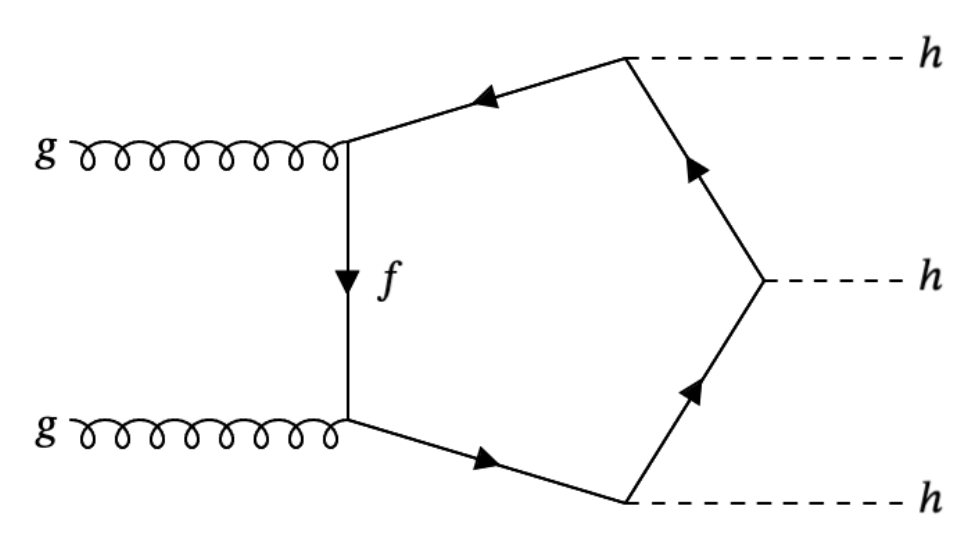}
\caption{}
\end{subfigure}%
\hfill
\begin{subfigure}{.3\textwidth}
\centering
\includegraphics[height=2cm]{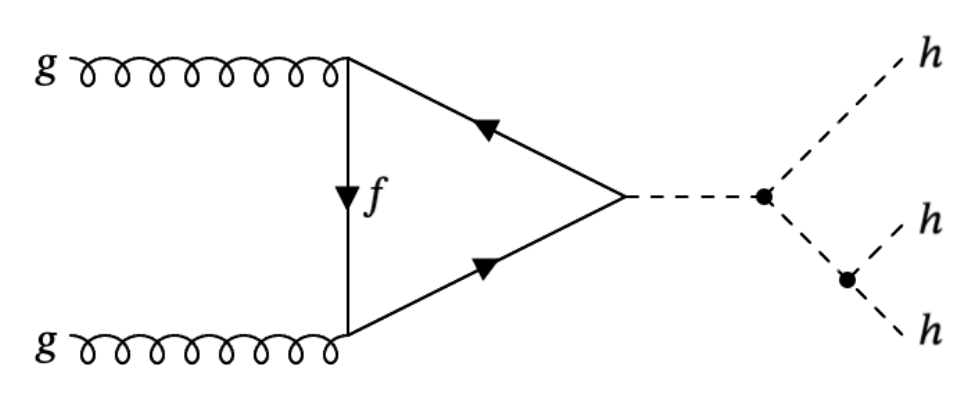}
\caption{}
\end{subfigure}%
\hspace{6em}%
\begin{subfigure}{.3\textwidth}
\centering
\includegraphics[height=2cm]{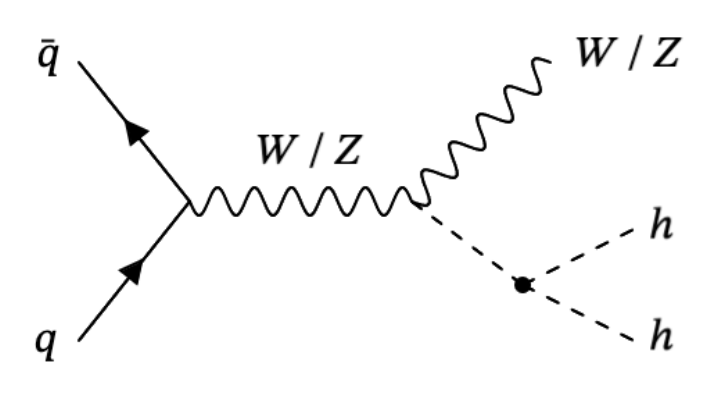}
\caption{}
\end{subfigure}
\caption{Feynman diagrams contributing to gluon fusion triple Higgs production HHH and VHH process. Only representative leading diagrams are shown here. Diagrams (a-d) represent for the HHH production, including the dominant pentagon process (a), the processes containing a trilinear coupling vertex (b,c) and a quartic coupling vertex (d). Diagram (e) of the VHH process also contains a trilinear coupling vertex.}
\label{fig:HHH_diagram}
\end{figure}

Both gluon fusion triple Higgs production (HHH) and vector boson associated di-Higgs production (VHH) channels are considered in this analysis. Fig.~\ref{fig:HHH_diagram}(a-d) shows the representative Feynman diagrams for gluon fusion triple Higgs production, where diagrams containing three-Higgs and four-Higgs vertices provide direct sensitivity to the trilinear and quartic Higgs self-couplings, $c_3$ and $d_4$. Additionally, one representative diagram of the VHH process as presented in Fig.~\ref{fig:HHH_diagram}(e) also contains the trilinear Higgs self-coupling. This process can also decay to the $4\tau 2b$ final state through subsequent Higgs and Z decays. Due to the limited resolution in hadronic final states, this VHH process is not distinguishable from HHH. However, the inclusion of this specific process enhances the sensitivity to the trilinear coupling parameter $c_3$. This approach is generally applicable to other hadronic decay channels, such as $6b$ and $4b2\tau$, as well.

\begin{figure}[tp]
    \centering
    \includegraphics[width=.6\textwidth]{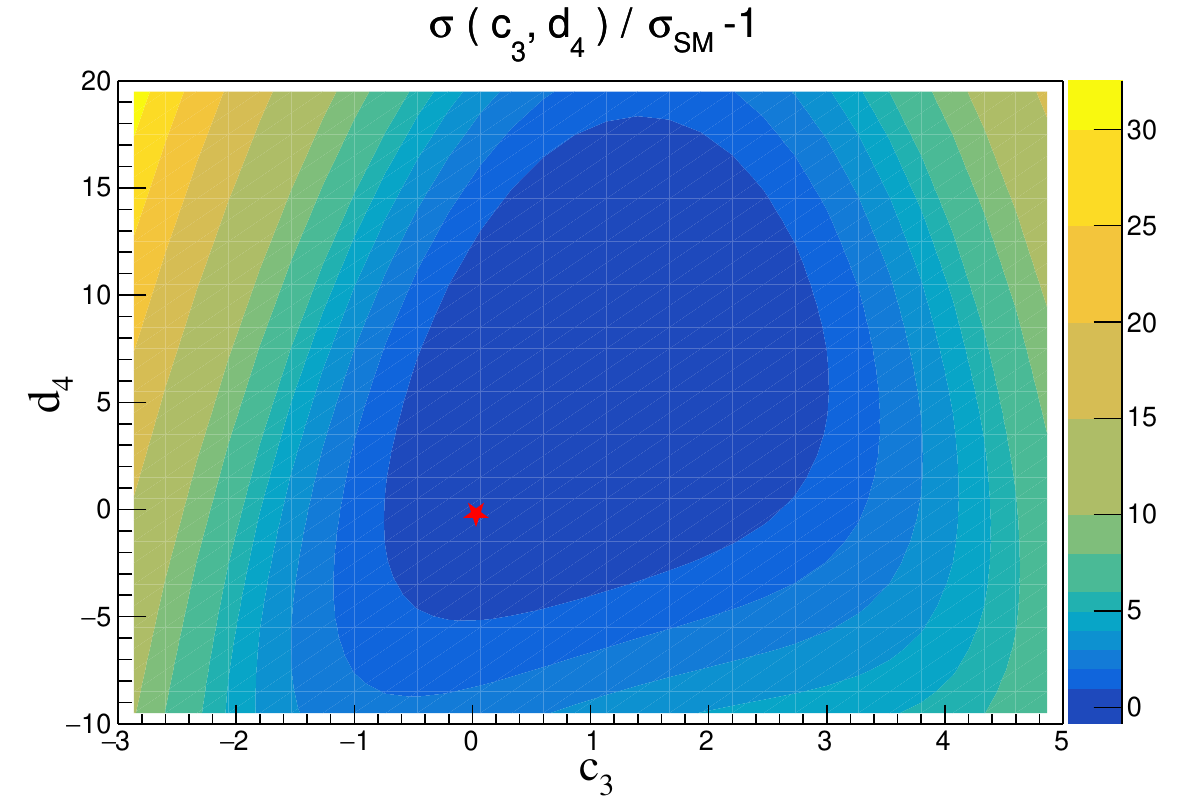}
    \caption{Triple Higgs production cross section as a function of self-couplings parameters $(c_3,d_4)$ at a 100 TeV collider, in the fractional deviation from the SM value. The red star indicates the SM point.}
    \label{fig:HHH_production}
\end{figure}

The cross section of gluon fusion triple Higgs production at next-to-leading-order in the $(c_3, d_4)$ plane at a 100 TeV proton-proton collider is shown in the fractional deviation from the SM value in Fig.~\ref{fig:HHH_production}. The figure reveals that the triple Higgs production cross section exhibits different sensitivities to the two coupling parameters: stronger dependence on $c_3$ compared to $d_4$. Theoretical calculations have shown that diagrams with $c_3$ yield much larger contributions to the total cross section compared to the diagram with $d_4$ \cite{Agrawal:2017cbs}. The dependence on $c_3$ and $d_4$ of the variation of the triple Higgs cross section at a future 100 TeV proton-proton collider, in the fractional deviation from the SM value, is fitted as:

\begin{equation}
\label{eq:HHH_variation}
\begin{split}
     \frac{\sigma_{\text{HHH}(c_3,d_4)}}{\sigma_{\text{HHH(SM)}}}-1 = 0.0297\times c_3^4 - 0.2017\times c_3^3 
     + 0.0395\times c_3^2d_4 + 0.7236\times c_3^2 \\
     + 0.0154\times d_4^2 - 0.1409\times c_3d_4 - 0.6658\times c_3 - 0.1119\times d_4,
\end{split}
\end{equation}

The parametrization shown in Eq.~\ref{eq:HHH_variation} provides a convenient way to evaluate the triple Higgs production cross section for any SM-like model where the main modifications are in the Higgs self-couplings. Our parametrization shows good agreement with previous studies \cite{Papaefstathiou:2019ofh}.

In this study, we focus on  the \(4\tau 2b\) final state with four hadronic tau leptons. Compared to b-enriched final states such as $6b$ and $4b 2\tau$ channel, the \(4\tau 2b\) channel benefits from relatively fewer background processes, though it faces the challenge of a lower rate of events. The primary SM backgrounds arise from top quark pair (\(t\bar{t}\)) and \(W\) boson pair production, which yield final states that contain tau leptons. Additionally, backgrounds include processes such as \(X_{b\bar{b}}\) \(Y_{\tau\tau}\) \(Y_{\tau\tau}\), where $X_{b\bar{b}}$ can be a Higgs boson or a \(Z\) boson which decays to $b\bar{b}$ and \(Y_{\tau\tau}\) can be a virtual photon $\gamma$ or a Z boson which decays to $\tau\tau$. Here we exclude VHH which is taken as a part of signal. 

Tab.~\ref{tab:sample} summarizes samples used in this analysis and the corresponding cross sections (taking into account the branching ratios to the final state of 4 hadronic tau and 2 b-quarks 
\cite{PhysRevD.110.030001}) at a 100 TeV proton-proton collider. For signal processes, a list of representative benchmark points are used: triple Higgs production scenarios with different combinations of $(c_3, d_4)$ and VHH process with various $c_3$ values (only ZHH is considered in the $4\tau 2b$ channel). The dominant SM backgrounds are categorized into two groups: the t/W-related processes ($t\bar{t}\tau\tau + \text{jets}$, $t\bar{t}H$, $t\bar{t}\tau\tau\nu\nu + \text{jets}$, and $t\bar{t}t\bar{t}$) and $X_{b\bar{b}}Y_{\tau\tau}Y_{\tau\tau}$ samples ($Z_{bb}\tau\tau\tau\tau$, and $HZZ$).


\begin{table}
\centering
\begin{tabular}{lll}
\toprule
\text{Class} & \text{Process} & $\sigma \times \text{BR.}~(\text{ab})$ \\
\midrule
\multirow{7}{*}{HHH signal} 
 & $HHH \to (b\bar{b})(\tau_h\tau_h)(\tau_h\tau_h)$, $c_3=0.0$, $d_4=0.0$ (SM) & 3.0 \\
 & $HHH \to (b\bar{b})(\tau_h\tau_h)(\tau_h\tau_h)$, $c_3=1.0$, $d_4=0.0$ & 2.6 \\
 & $HHH \to (b\bar{b})(\tau_h\tau_h)(\tau_h\tau_h)$, $c_3=1.0$, $d_4=9.0$ & 0.67 \\
 & $HHH \to (b\bar{b})(\tau_h\tau_h)(\tau_h\tau_h)$, $c_3=2.0$, $d_4=19.0$ & 7.6 \\
 & $HHH \to (b\bar{b})(\tau_h\tau_h)(\tau_h\tau_h)$, $c_3=-1.0$, $d_4=0.0$ & 7.9 \\
 & $HHH \to (b\bar{b})(\tau_h\tau_h)(\tau_h\tau_h)$, $c_3=-1.0$, $d_4=-6.0$ & 8.3 \\
 & $HHH \to (b\bar{b})(\tau_h\tau_h)(\tau_h\tau_h)$, $c_3=-2.0$, $d_4=-11.0$ & 16.7 \\
\midrule
\multirow{5}{*}{VHH signal}
 & $HHZ \to (b\bar{b})(\tau_h\tau_h)(\tau_h\tau_h)$, $c_3=0.0$ (SM) & 2.7 \\
 & $HHZ \to (b\bar{b})(\tau_h\tau_h)(\tau_h\tau_h)$, $c_3=1.0$ & 4.2 \\
 & $HHZ \to (b\bar{b})(\tau_h\tau_h)(\tau_h\tau_h)$, $c_3=2.0$ & 6.1 \\
 & $HHZ \to (b\bar{b})(\tau_h\tau_h)(\tau_h\tau_h)$, $c_3=-1.0$ & 1.8 \\
 & $HHZ \to (b\bar{b})(\tau_h\tau_h)(\tau_h\tau_h)$, $c_3=-2.0$ & 1.3 \\
\midrule
\multirow{4}{*}{t/W samples}
 & $t\bar{t}\tau\tau$ + jets (LO)& $7.609 \times 10^4$ \\
 & $t\bar{t}H$ (LO)& $1.598 \times 10^4$ \\
 & $t\bar{t}\tau\tau\nu\nu$ + jets (LO)& $5.381 \times 10^2$ \\
 & $t\bar{t}t\bar{t}$ (NLO)& $3.869 \times 10^2$ \\
\midrule
\multirow{2}{*}{$X_{b\bar{b}}Y_{\tau\tau}Y_{\tau\tau}$ samples}
 & $Z\tau\tau\tau\tau(Z \to b\bar{b})$ (NLO)& $1.140 \times 10^2$ \\
 & $HZZ$ (NLO)& $0.518 \times 10^2$ \\
\bottomrule
\end{tabular}
\caption{Cross sections for gluon fusion triple Higgs production, VHH production (only ZHH is considered in the $4\tau 2b$ final state) and SM background processes 
in the $4\tau2b$ final state at a 100 TeV proton-proton collider. Signal processes with different Higgs self-coupling parameters are considered.}
\label{tab:sample}
\end{table}

\subsection{Event generation and detector simulation}

The triple Higgs signal samples are generated using the loop-induced module of the MadGraph5\_aMC@NLO package \cite{Alwall:2014bza,Alwall:2014hca,Artoisenet:2012st,Hirschi:2015iia,Artoisenet:2010cn}.
Background samples are generated at leading-order (LO) and next-to-leading-order (NLO) using MadGraph5\_aMC@NLO. The parton-level events are then interfaced to PYTHIA 8 \cite{Bierlich:2022pfr} for parton showering and hadronization. To simulate detector effects, the generated events are processed through DELPHES 3.5.0 \cite{deFavereau:2013fsa,Selvaggi:2014mya,Mertens:2015kba}. An official configuration card prepared by FCC Collaboration is applied, which implements the baseline detector design for the FCC-hh. 

\section{Physics analysis}
\label{sec:analysis}

\subsection{Analysis strategy}
In this analysis, Higgs reconstruction is performed using both resolved and boosted approaches, with different categorizations, as listed in Tab.~\ref{table:cate}. In the resolved group, Higgs is reconstructed from two small-radius b or tau tagged jets. In the boosted groups, Higgs may appear as a single large-radius fat jet in the detector. We categorize events based on the number of boosted Higgs: zero (as the resolved group), one, two, or three boosted Higgs. 

In the resolved group, we further classify events based on the invariant mass of the triple Higgs system ($m_{\text{HHH}}$), forming low-mass and high-mass categories to enhance sensitivity in different kinematic regions.
In the one boosted Higgs group, only one boosted fatjet is expected to be well reconstructed. There are two primary categories defined: events with one boosted $H_{b\bar{b}}$ and events with one leading boosted $H_{\tau\tau}$. The second leading boosted $H_{\tau\tau}$ can not be considered as a single category due its relatively lower energy. 
In the two boosted Higgs group, exactly two boosted fatjets are expected to be well reconstructed. The category is defined with the same strategy above. 
Eventually, in the three boosted Higgs group, three Higgs bosons are fully reconstructed as boosted fat jets, and only a single category is defined.

\begin{figure}[tp]
    \centering
    \qquad
    \includegraphics[width=.6\textwidth]{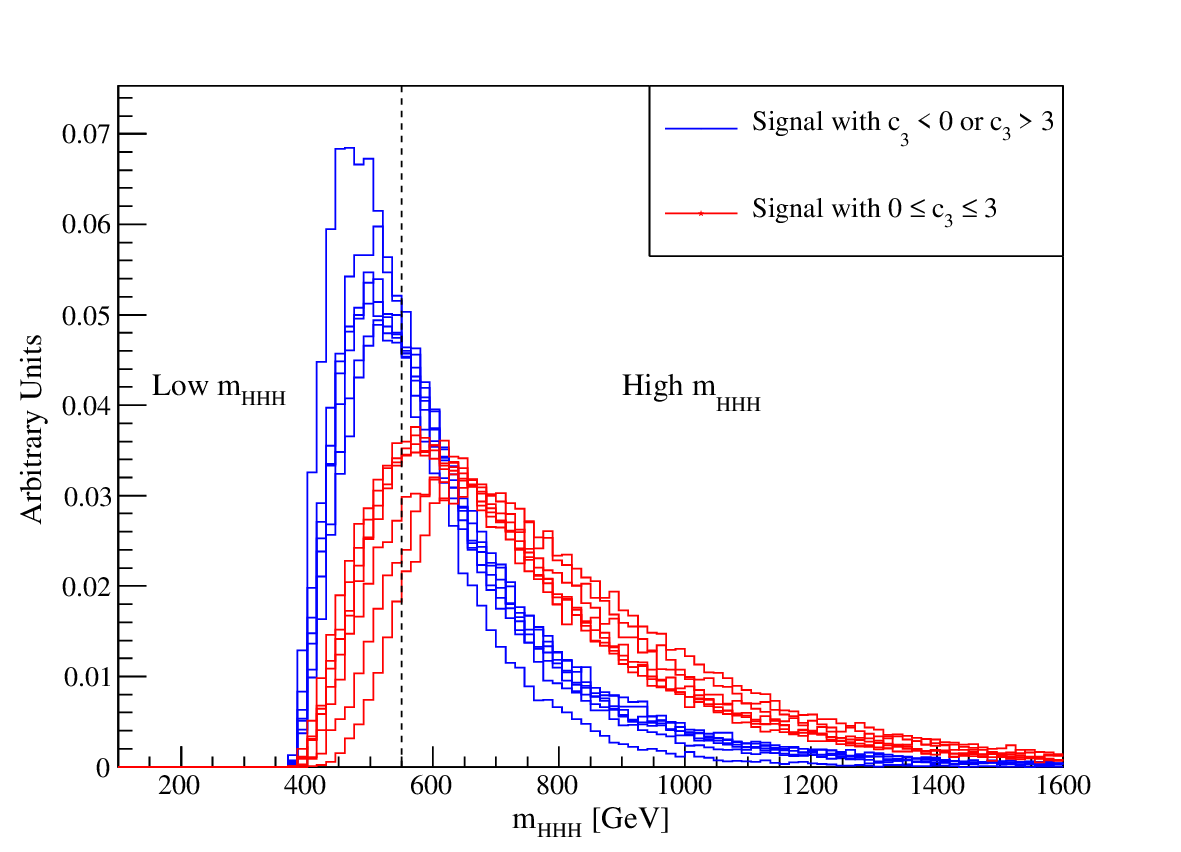}
    \qquad
    \caption{Invariant mass distribution of triple Higgs at the generator level. Signal samples with $c_3 < 0$ or $c_3 > 3$ are represented in blue, samples with $0 \leq c_3 \leq 3$ in red. The vertical dashed line separates at 550 GeV the low and high $m_{\text{HHH}}$ regions.}
    \label{fig:truth_HHH}
\end{figure}

The kinematic distributions of triple Higgs production at the generator level are shown in Fig.~\ref{fig:truth_HHH}. It is seen that the distribution of invariant mass of triple Higgs varies with the coupling parameters. This feature distinguishes between two signal scenarios: one represented in blue, with $c_3 < 0$ or $c_3 > 3$, and the other in red, with $0 \leq c_3 \leq 3$. The vertical dashed line at 550 GeV separates the low and high $m_{\text{HHH}}$ regions that carry different sensitivities to different coupling intervals. This insight allows us to design the analysis by categorizing the invariant mass of triple Higgs into low and high regions in the resolved scenario. However, due to the indistinct $m_{\text{HHH}}$ distribution in boosted events, no mass-based categorization is applied in this case. This strategic approach enhances sensitivity by leveraging coupling dependencies and optimizing specific mass regions.

Moreover, the relatively high statistics in the resolved categories allow us to apply BDT training to exploit multivariate correlations among kinematic variables. The details of the BDT training procedure are provided in Sec.~\ref{sec:bdt}.

\begin{table}[tp]
\centering
\begin{tabular}{ll}
\toprule
\text{Group} & \text{Category} \\
\midrule
\multirow{2}{*}{Resolved} & $m_{\text{HHH}} \leq 550 $ GeV \\ \cline{2-2}
                          & $m_{\text{HHH}} > 550 $ GeV \\
\midrule
\multirow{2}{*}{1 Boosted Higgs} & 1 boosted $H_{bb}$ \\ \cline{2-2}
                                 & 1 boosted $H_{\tau\tau}$ (leading $\tau\tau$ pair) \\
\midrule
\multirow{2}{*}{2 Boosted Higgs} & 1 boosted $H_{bb}$ + 1 boosted $H_{\tau\tau}$ (leading $\tau\tau$ pair) \\ \cline{2-2}
                                 & 2 boosted $H_{\tau\tau}$ \\
\midrule
\multirow{1}{*}{3 Boosted Higgs} & 2 boosted $H_{\tau\tau}$ + 1 boosted $H_{bb}$ \\
\bottomrule
\end{tabular}
\caption{Descriptions of each category based on resolved or boosted Higgs reconstruction.}
\label{table:cate}
\end{table}

\subsection{Physics object definition}
Jets were reconstructed using the $ \text{anti-}k_T $ algorithm from the FastJet package \cite{Cacciari:2011ma,Cacciari:2005hq}, setting the radius parameter to $R=0.4$ and $0.8$ for small-radius jets and large-radius jets, respectively. In this analysis, only jets with $p_T> 20$ GeV and $|\eta|<2.5$ are considered. Additionally, a b-tagging efficiency of 70 \% and a tau-tagging efficiency of 80 \% are applied \cite{Dolan:2012rv}. 

To concentrate on the hadronic decay mode of tau leptons and suppress the leptonic background, events containing reconstructed electrons with $p_T > 12$ GeV and $|\eta| < 2.5$ or muons with $p_T > 8$ GeV and $|\eta| < 2.4$ are vetoed to reject tau leptonic decays \cite{CMS:2020uim,CMS-DP-2020-040}.

For the baseline selections in the resolved scenario, events were preselected with at least four taus and two b-jets, each having $p_T> 20$ GeV and $|\eta|<2.5$. In the boosted scenario, an additional requirement was imposed: the reconstructed fat jet must have $p_T> 300$~GeV.

Furthermore, a pairing algorithm for the four tau system is employed in the resolved scenario, inspired by the di-Higgs to four b jets analysis \cite{CMS:2022cpr}. The four hadronic taus were paired into two tau pairs. Among all possible combinations, the one that minimizes $d=\frac{|m_{H_0}-km_{H_1}|}{\sqrt{1+k^2}}$ is chosen, where $m_{H_0}$ and $m_{H_1}$ are the invariant mass of the two tau pairs, the constant k represents the ratio of the expected peak values of the reconstructed Higgs boson masses for events where the tau pairs are correctly matched. The chosen tau pairs are then ordered based on $p_T$, with the higher $p_T$ pair designated as $m_{\tau\tau1}$ and the lower-$p_T$ pair as $m_{\tau\tau2}$ for their mass, as an example.

\begin{figure}[tp]
    \centering
    \includegraphics[width=.32\textwidth]{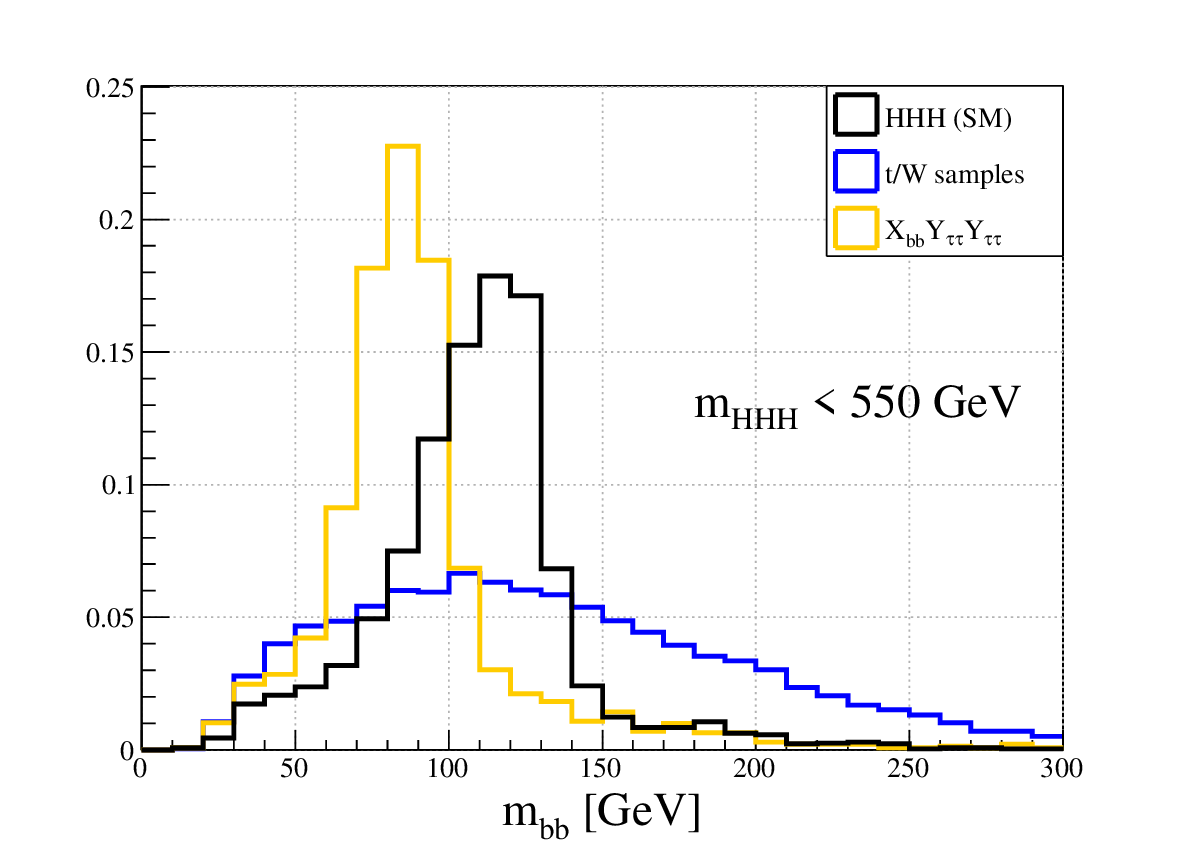}
    \includegraphics[width=.32\textwidth]{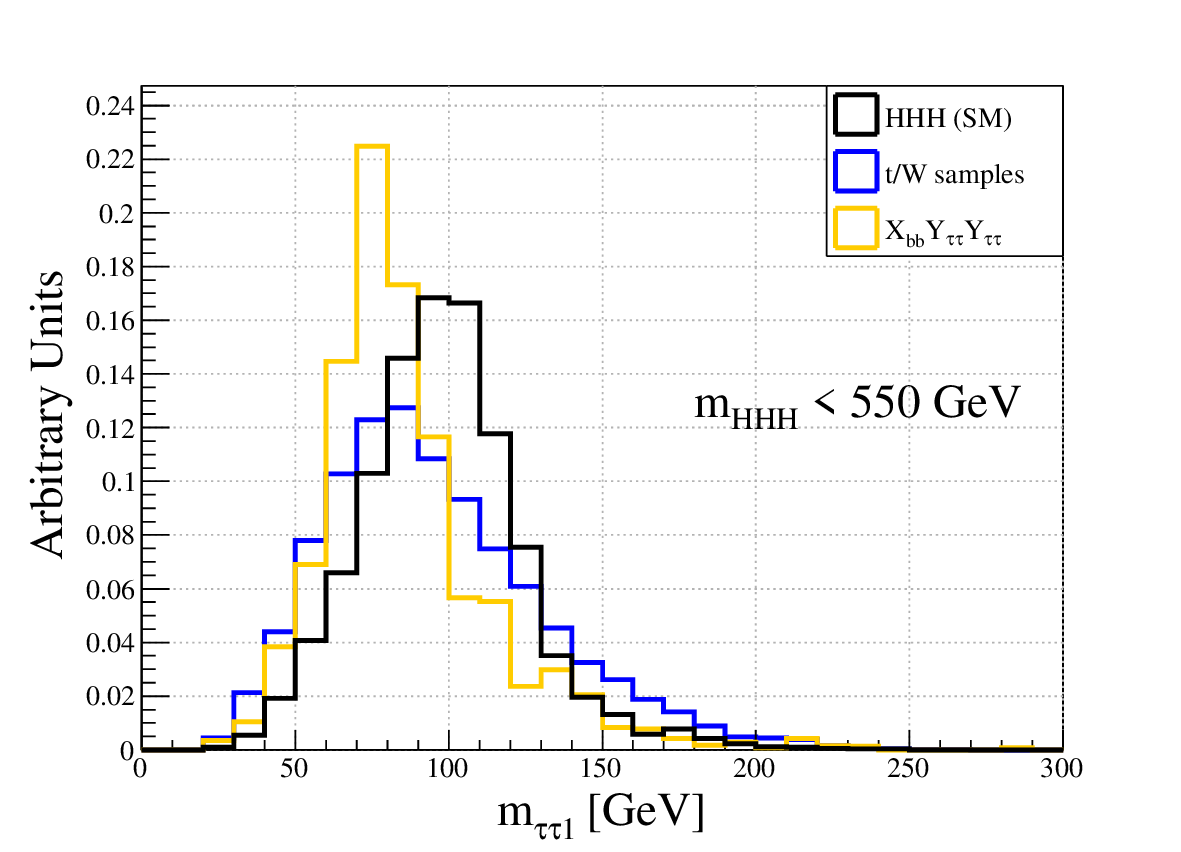}
    \includegraphics[width=.32\textwidth]{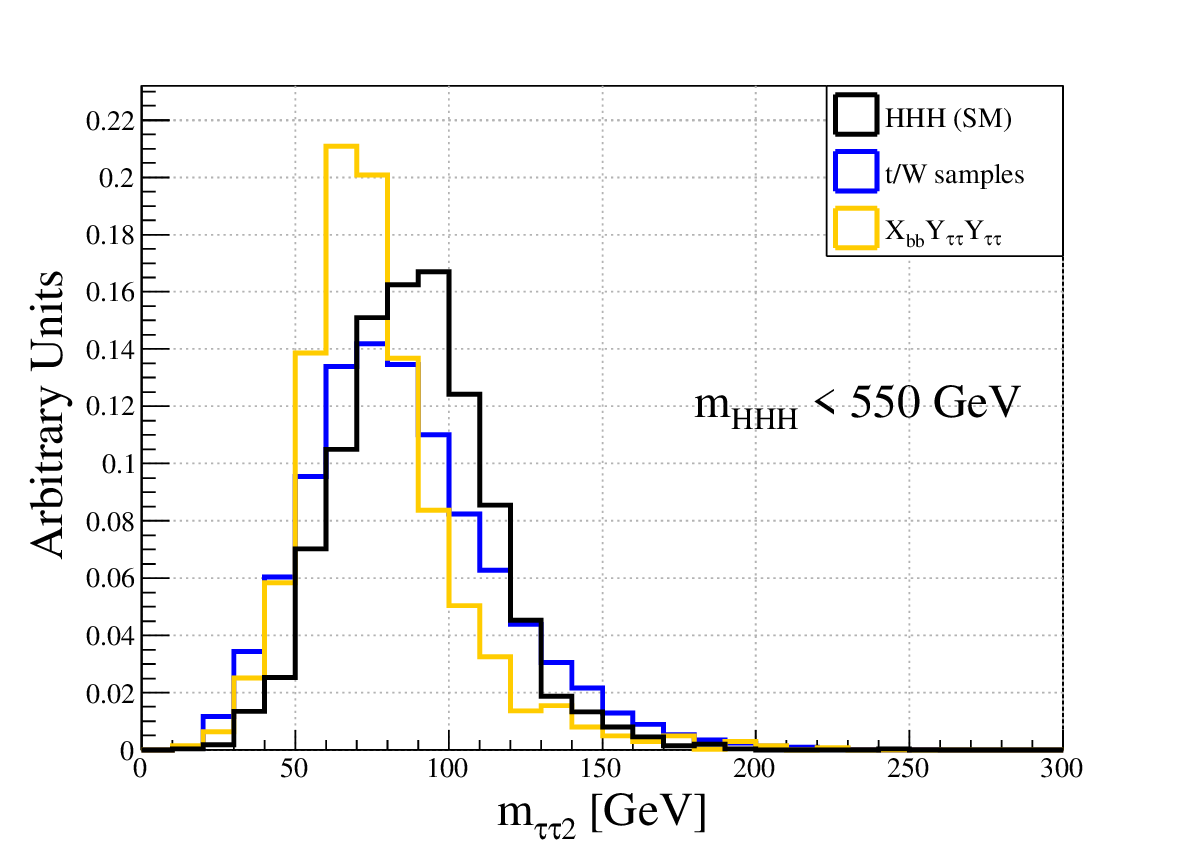}
    \qquad
    \includegraphics[width=.32\textwidth]{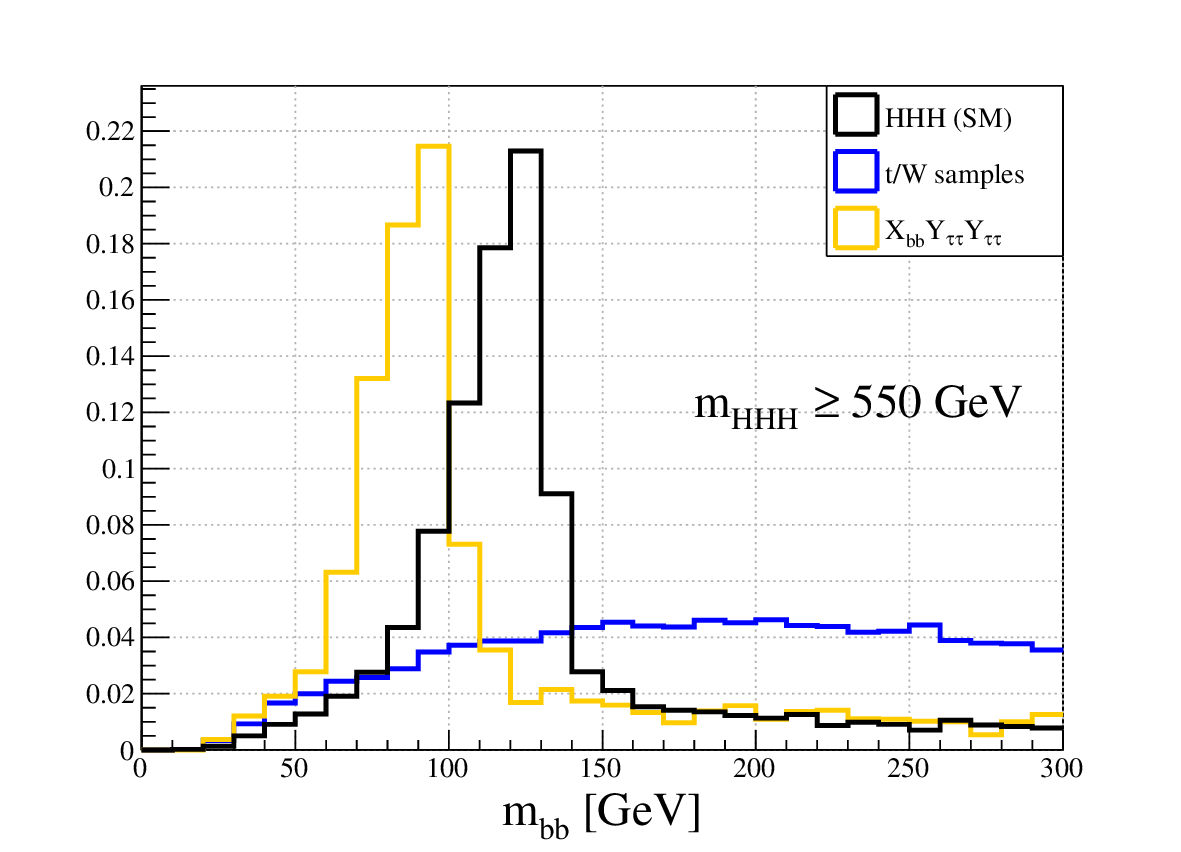}
    \includegraphics[width=.32\textwidth]{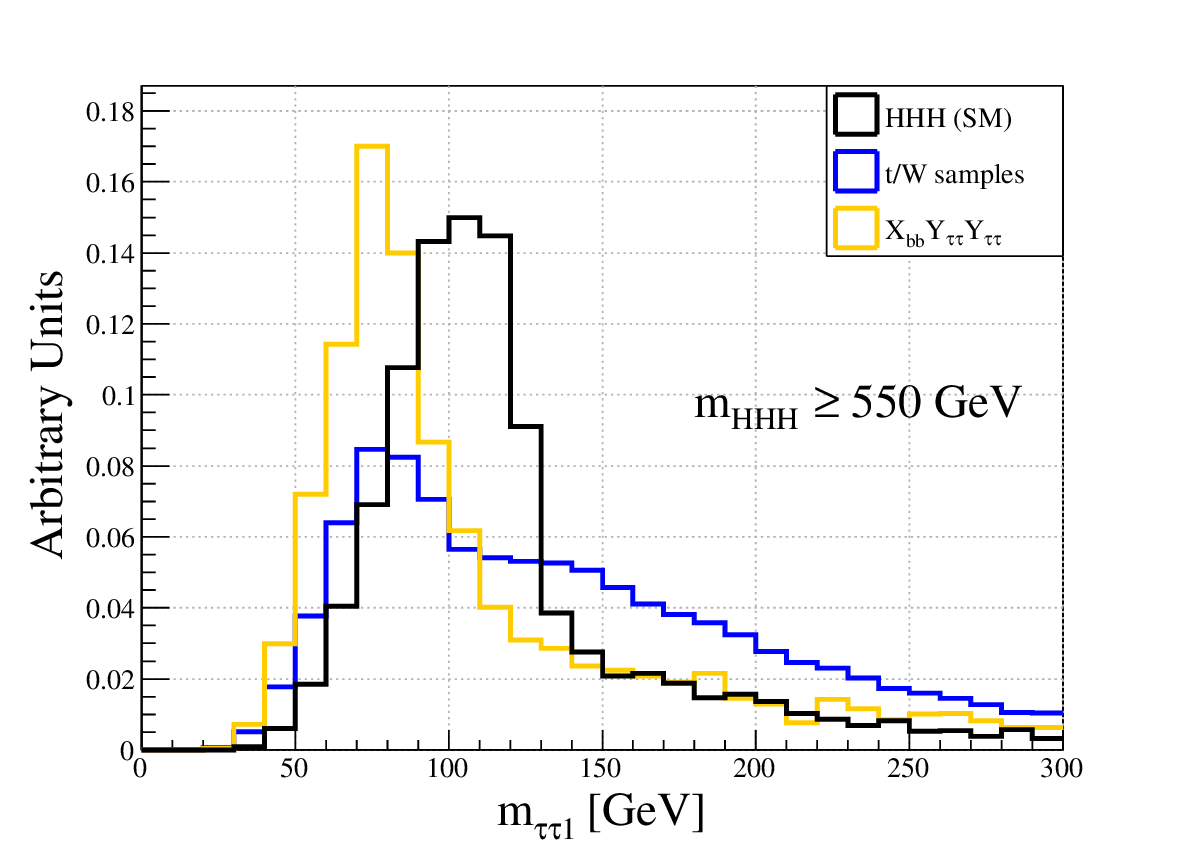}
    \includegraphics[width=.32\textwidth]{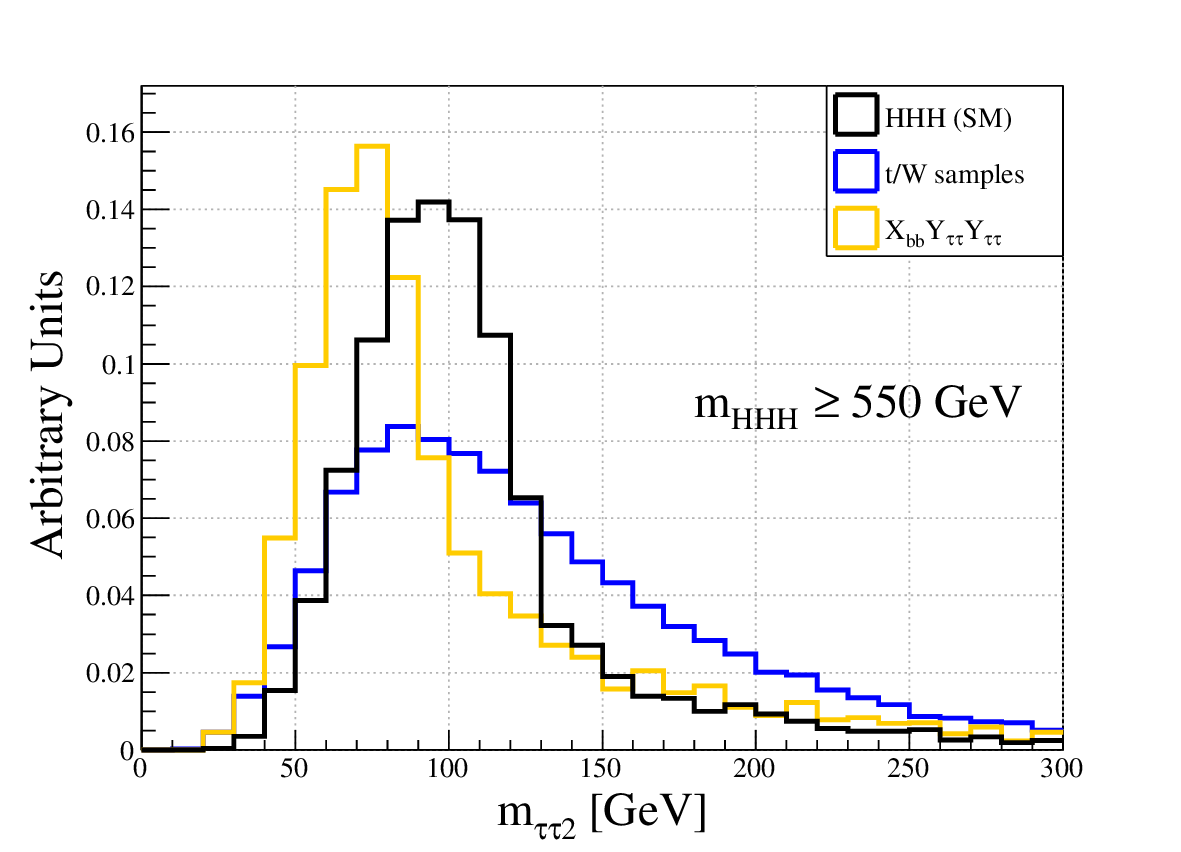}
    \caption{The $m_{b\bar{b}}$ and $m_{\tau\tau}$ distributions for the resolved HHH scenario. The upper (bottom) row corresponds to the low (high) $m_{\text{HHH}}$ category. Black lines represent the signal, while other colors represent background processes, including $t/W$ samples (Orange) and $X_{b\bar{b}}Y_{\tau\tau}Y_{\tau\tau}$ samples (Blue).}
    \label{fig:higgs_mass}
\end{figure}

\begin{figure}[tp]
    \centering
    \includegraphics[width=.32\textwidth]{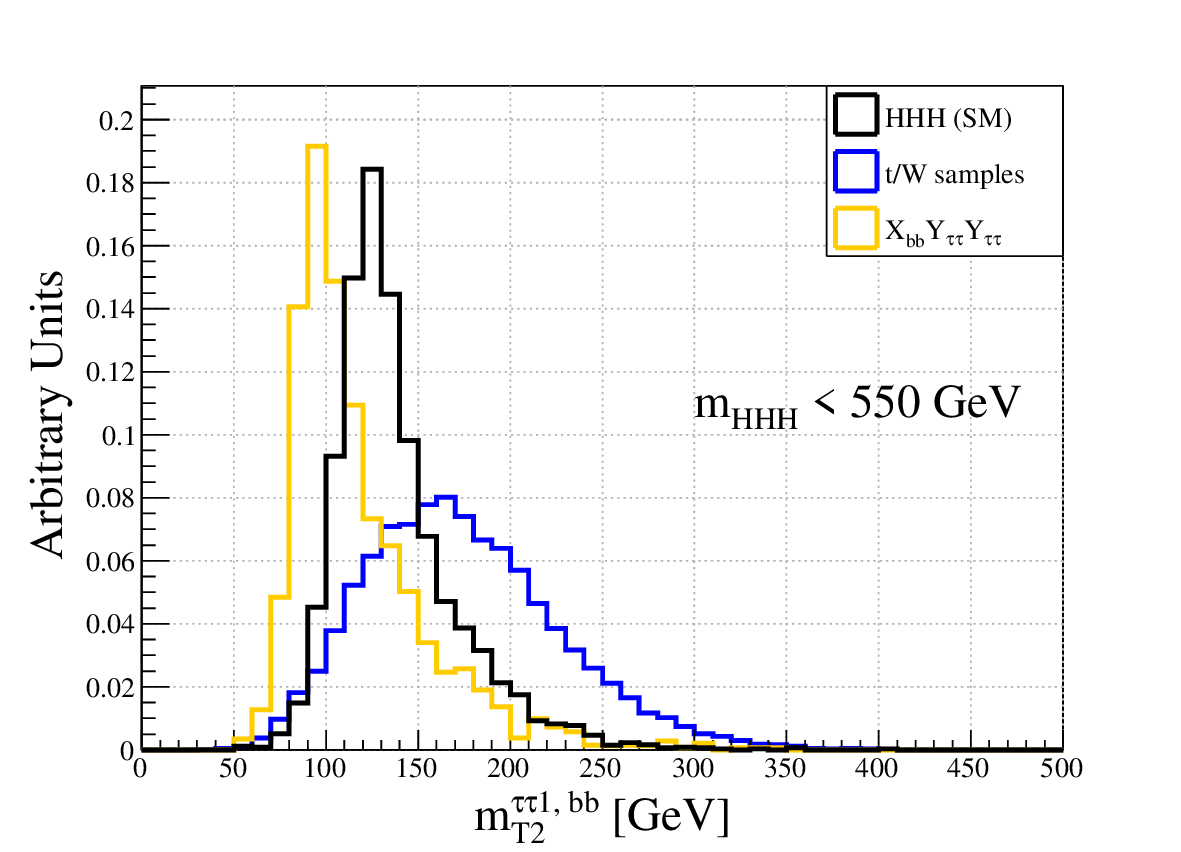}
    \includegraphics[width=.32\textwidth]{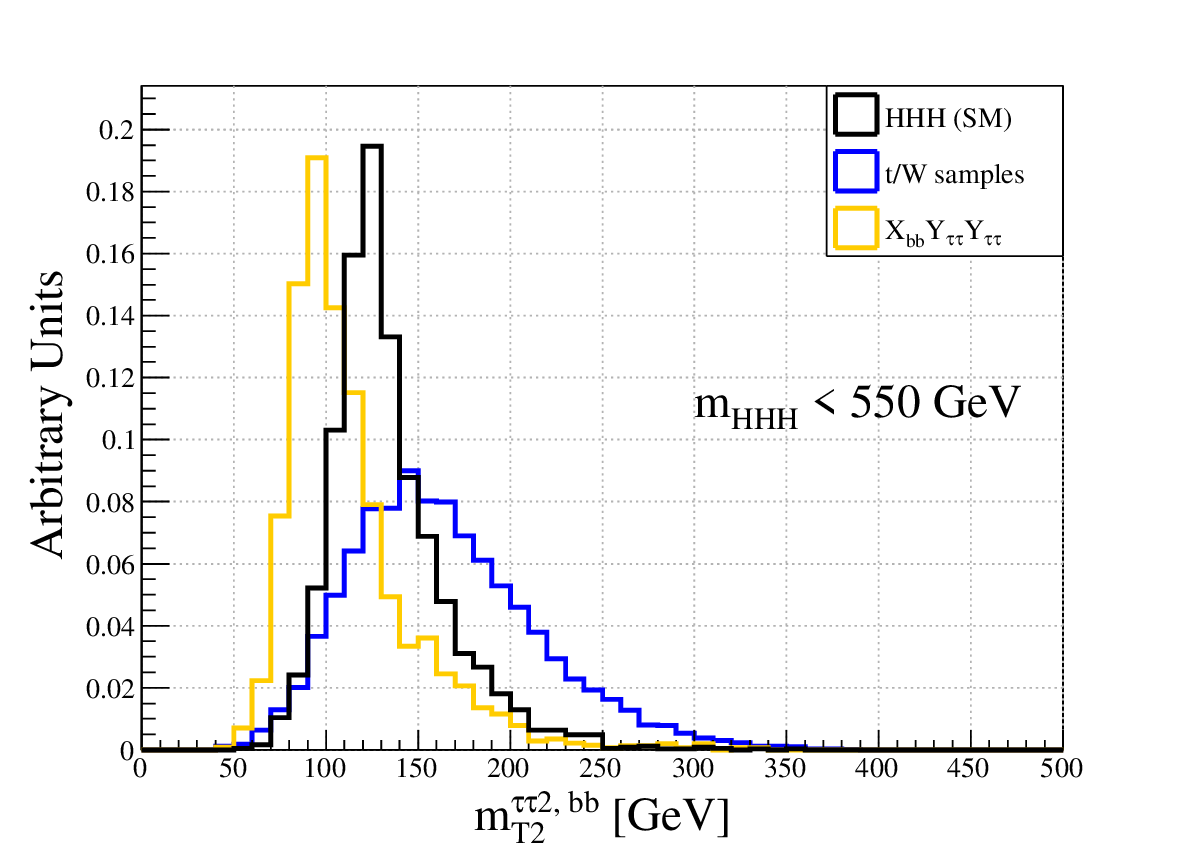}
    \includegraphics[width=.32\textwidth]{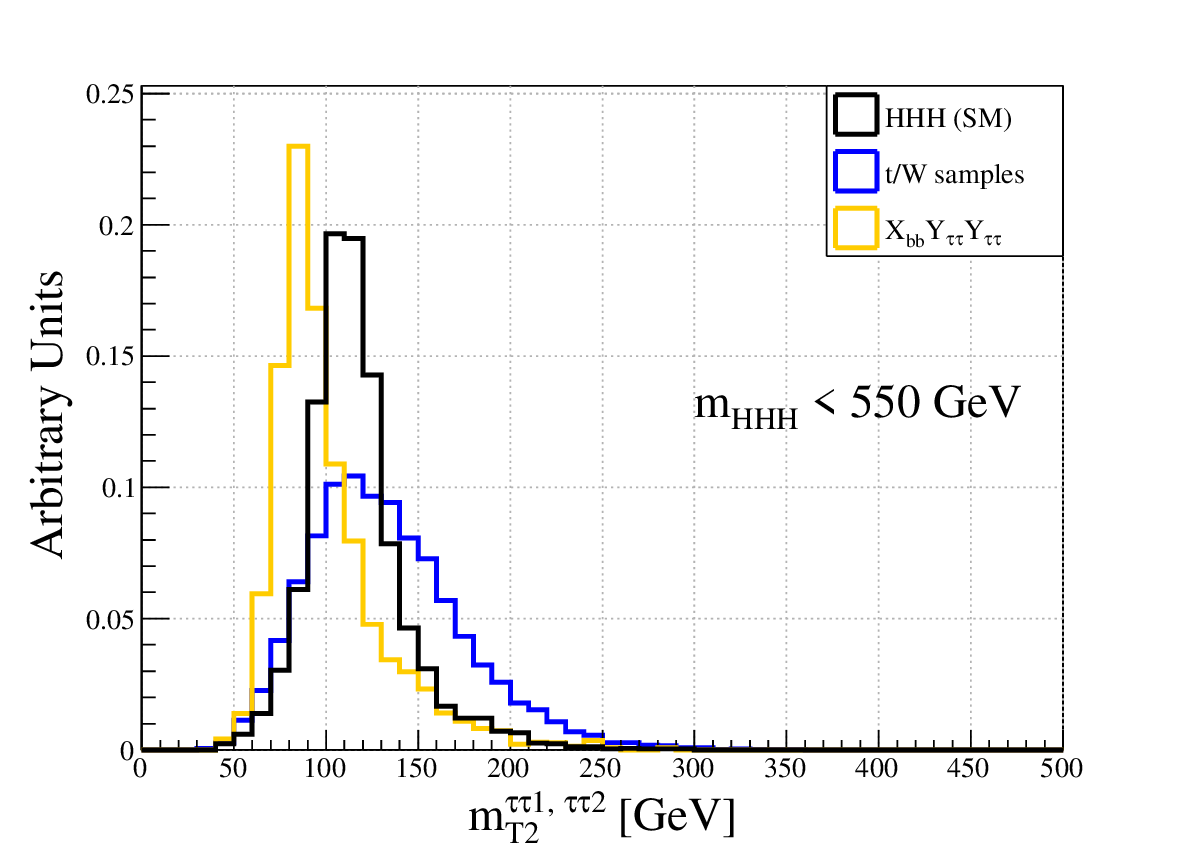}
    \qquad
    \includegraphics[width=.32\textwidth]{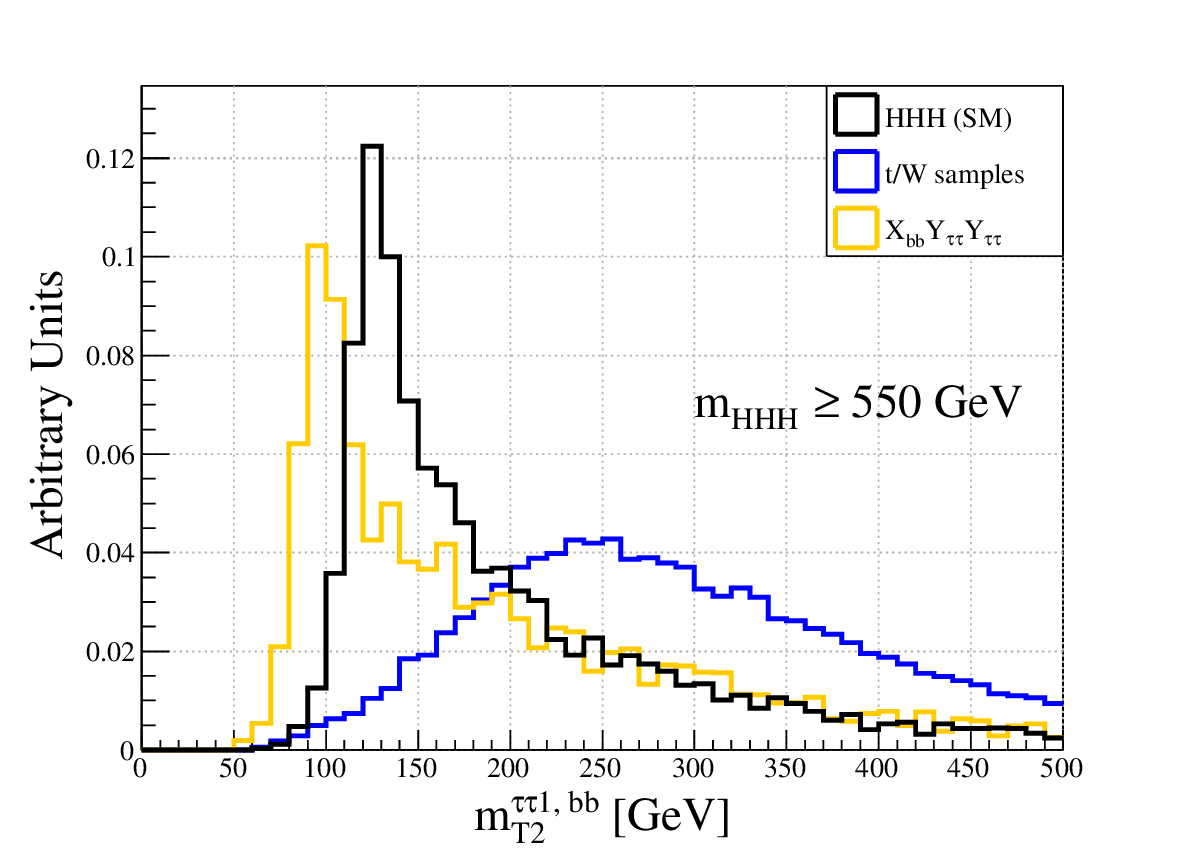}
    \includegraphics[width=.32\textwidth]{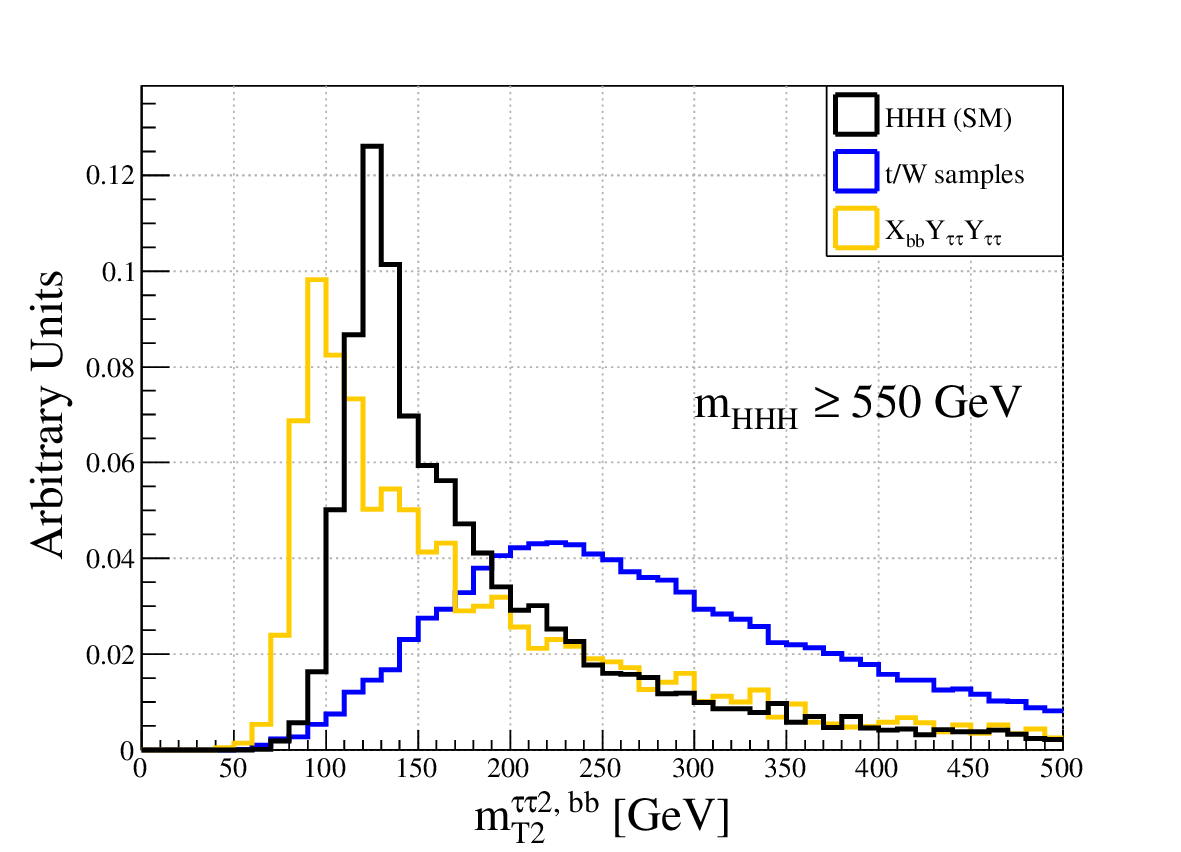}
    \includegraphics[width=.32\textwidth]{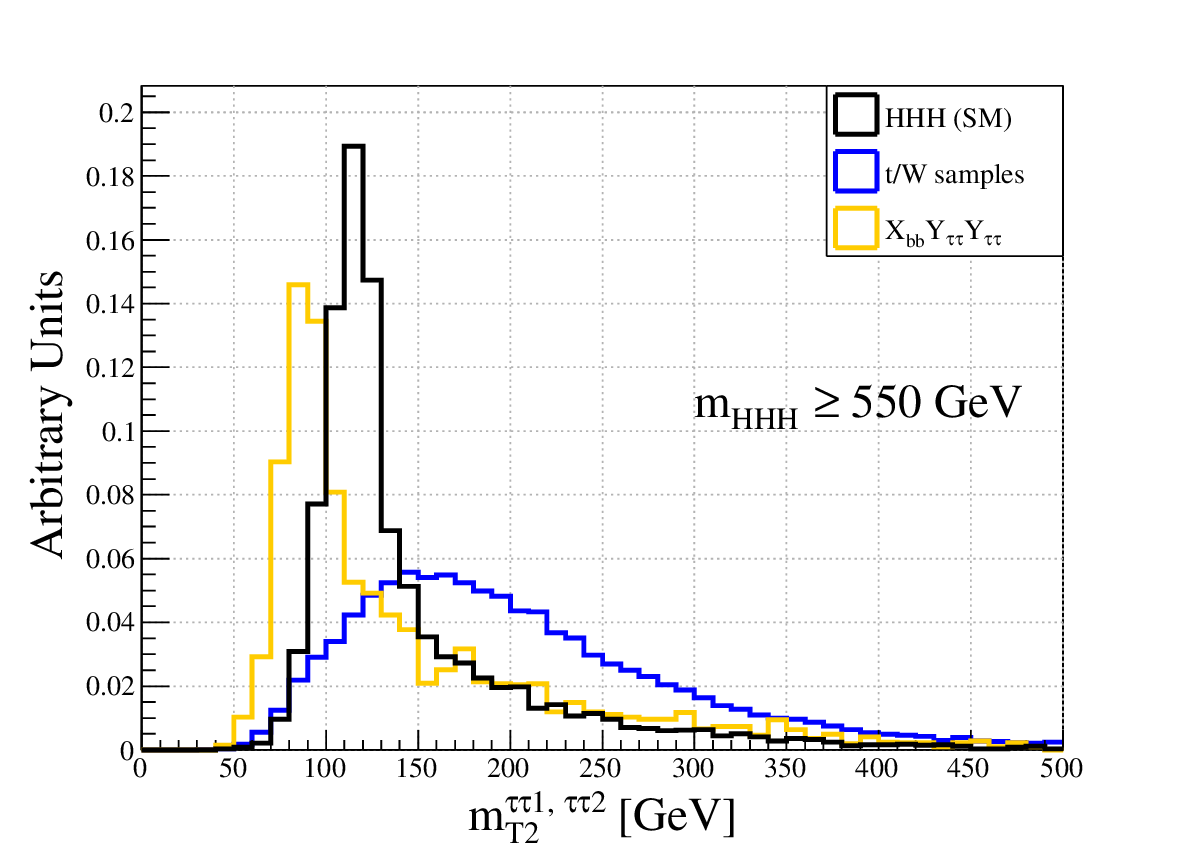}
    \caption{The $ m_{T2}^{h_i, h_j} $ distributions for the resolved HHH scenario. The upper (bottom)  row corresponds to the low (high) $m_{\text{HHH}}$ category. Black lines represent the signal, while other colors represent background processes, including $t/W$ samples (Orange) and $X_{b\bar{b}}Y_{\tau\tau}Y_{\tau\tau}$ samples (Blue).}
    \label{fig:higgs_mT2}
\end{figure}

\subsection{Kinematic distributions}

After analyzing various kinematic distributions of the signal and background, we found two effective discriminators: the mass of the reconstructed Higgs $m_\text{H}$ and a high level transverse mass $m_{T2}$, sometimes also called the "Stransverse Mass". The $m_{T2}$ variable is particularly useful in events where two or more particles have escaped detection \cite{Lester:1999tx,Barr:2003rg}. It calculates a lower bound on the square of the transverse mass $m_{T}$ by distributing the MET in two-body decays. For this calculation, the mass of the invisible particle, assumed to be 0 (the mass of the neutrino), is used in our analysis.

Fig.~\ref{fig:higgs_mass} and Fig.~\ref{fig:higgs_mT2} illustrate the invariant mass distributions of b-jet pairs ($m_{bb}$) and tau pairs ($m_{\tau\tau1}$ and $m_{\tau\tau2}$), along with $m_{T2}$ distributions in the resolved group, categorized by $m_{\text{HHH}}$.
The discriminating power of these kinematic distributions becomes more pronounced in the high $m_{\text{HHH}}$ category, where the decay products are better reconstructed due to the higher transverse momenta of the final-state particles.
The $m_{bb}$ distribution shows a clear peak around 125 GeV for signal events, while the $m_{\tau\tau}$ distribution exhibits a broader structure due to the presence of neutrinos in tau decays. 
For the $m_{T2}$ distributions, signal events display a clear endpoint around the Higgs mass, but shows different features of symmetry, which originated from two types of Higgs pair decay topology: the relatively symmetric case of two Higgs decay to two tau jets and the asymmetric case of two Higgs decay to two tau jets and two b jets respectively. The $m_{T2}$ distributions of $t/W$-related backgrounds extend to higher values and have a broader distribution, which can be understood from two physical aspects: the higher mass scale of the $t\bar{t}$ system naturally leads to higher $m_{T2}$ values, while the presence of missing energy from W boson decays results in a broader $m_{T2}$ distribution.

\begin{figure}[tp]
    \centering
    \includegraphics[width=.32\textwidth]{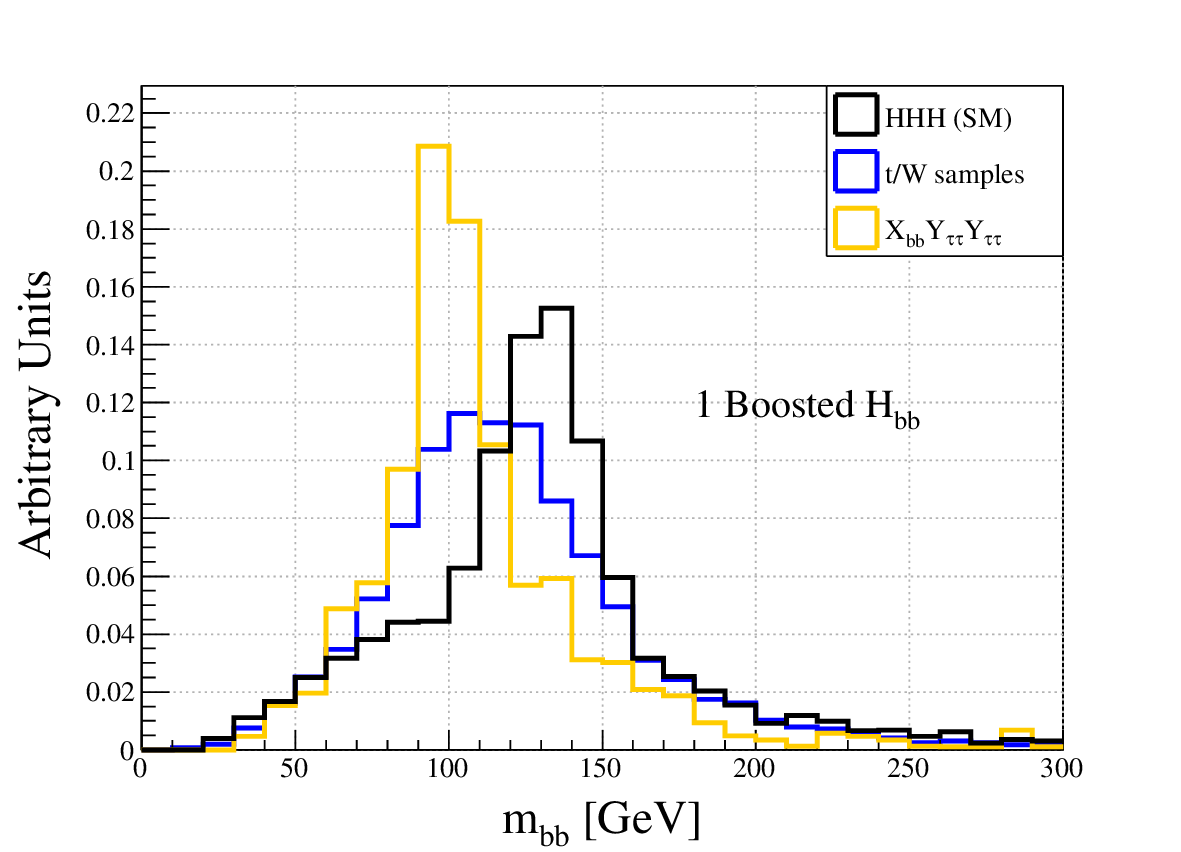}
    \includegraphics[width=.32\textwidth]{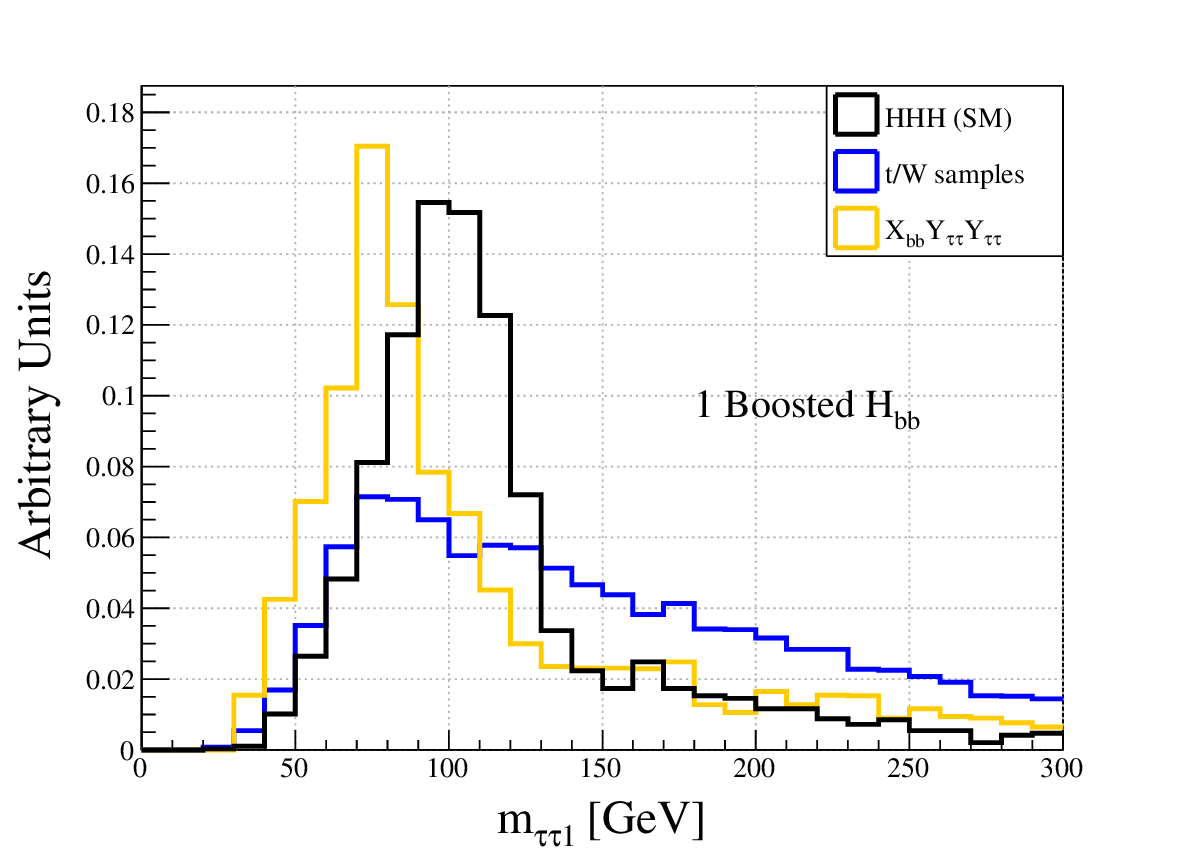}
    \includegraphics[width=.32\textwidth]{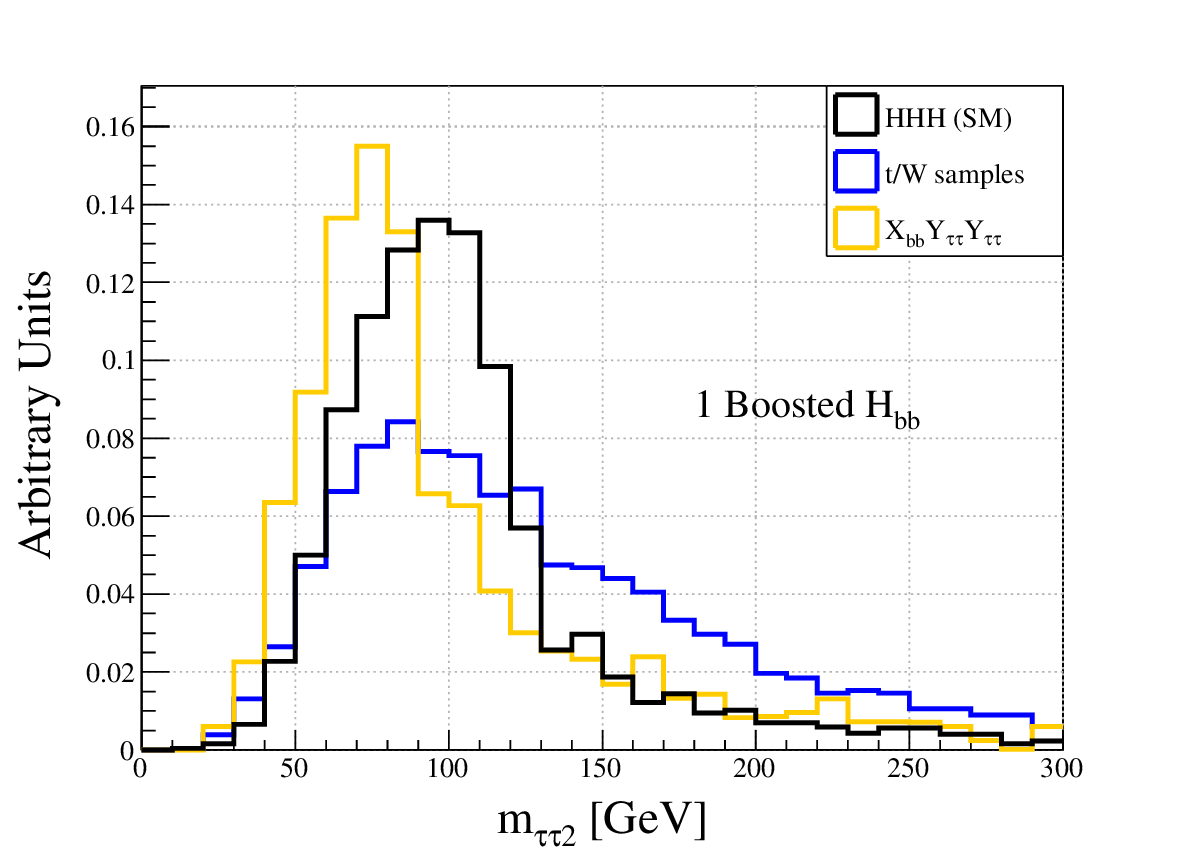}
    \qquad
    \includegraphics[width=.32\textwidth]{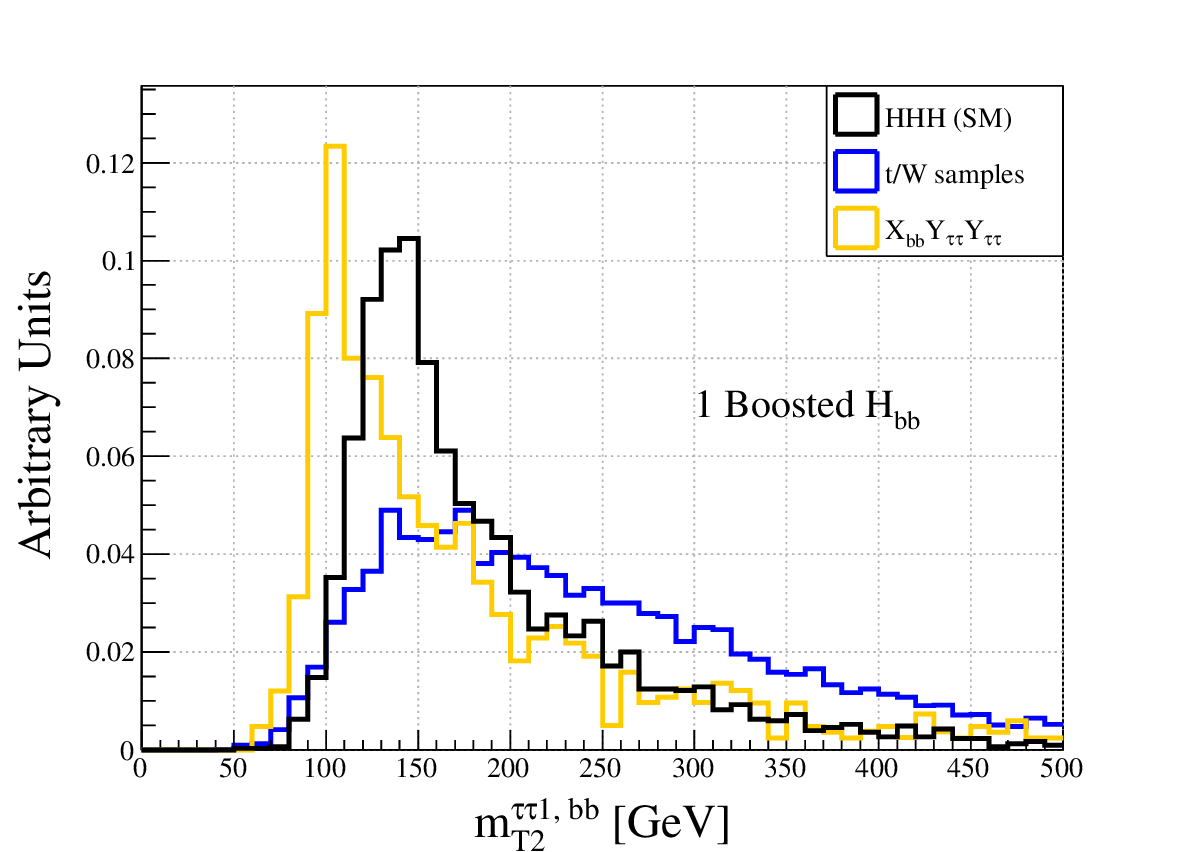}
    \includegraphics[width=.32\textwidth]{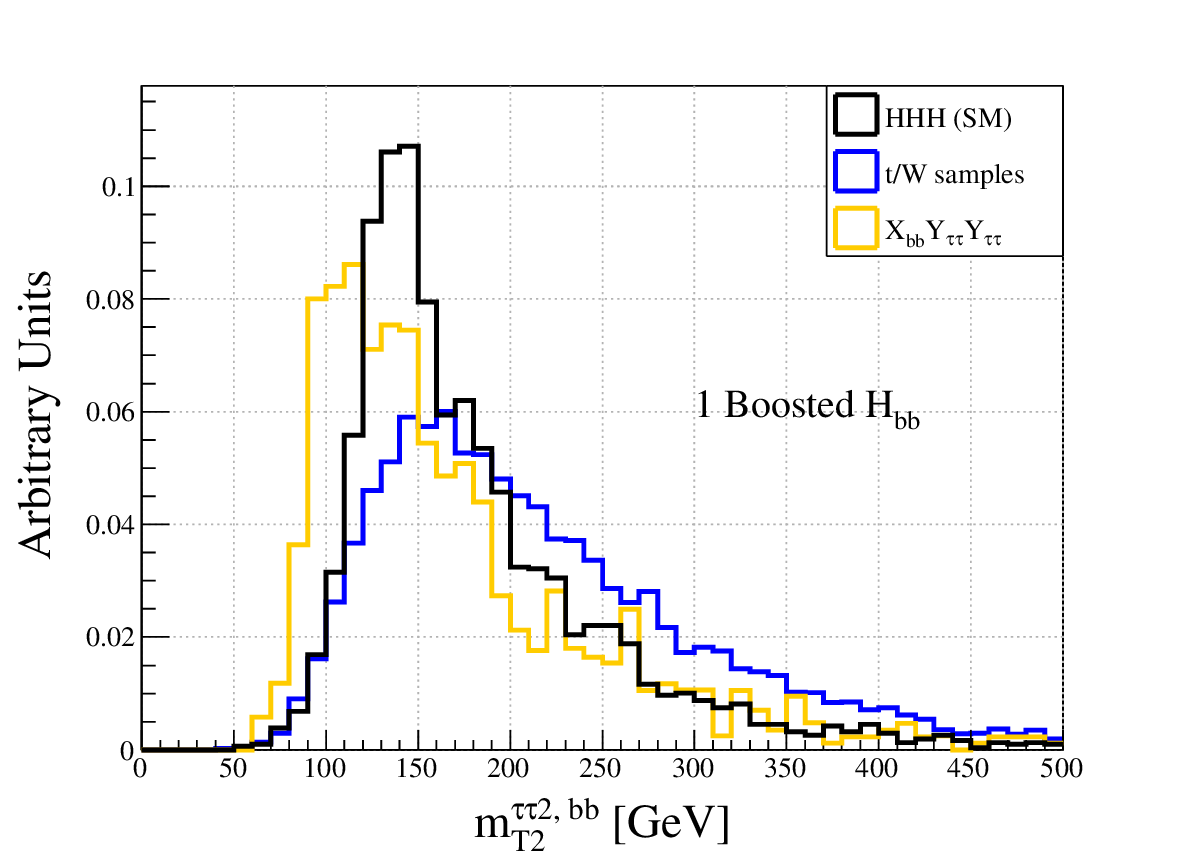}
    \includegraphics[width=.32\textwidth]{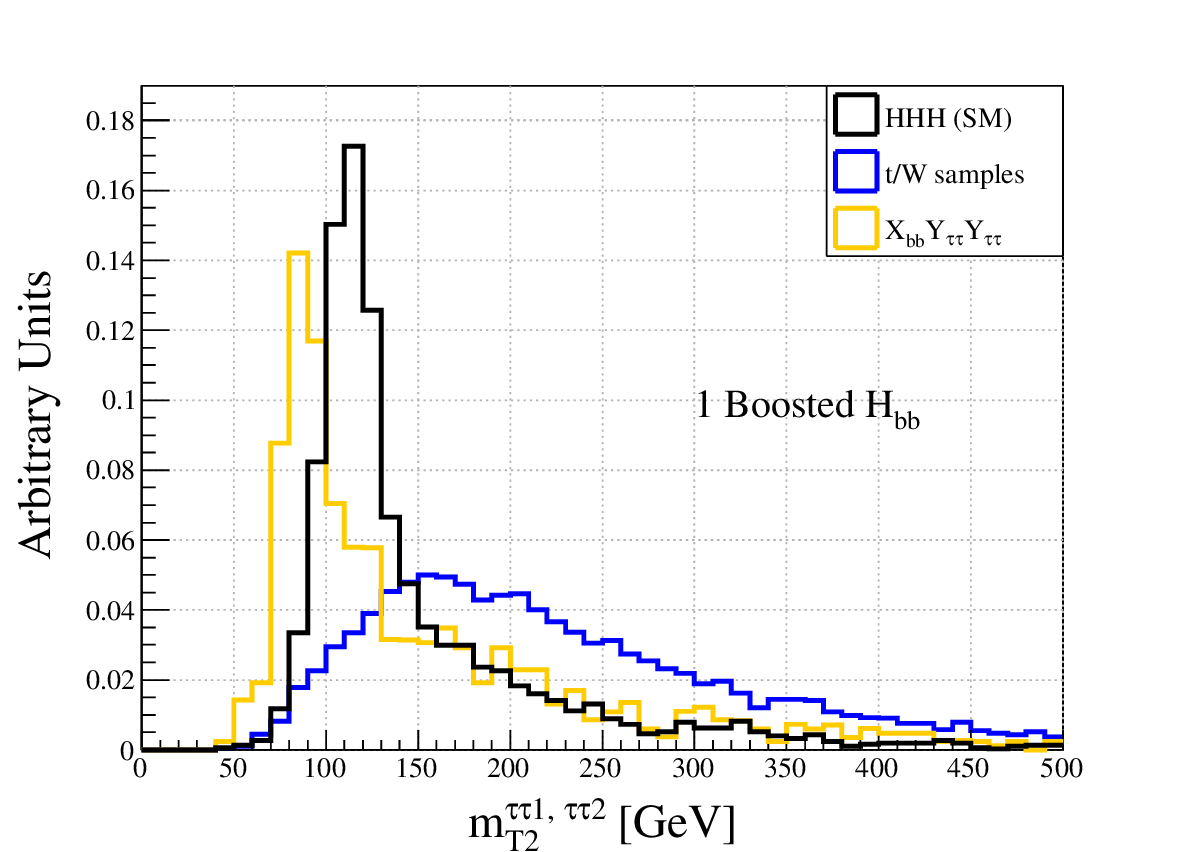}
    \caption{The $m_{b\bar{b}}$, $m_{\tau\tau}$ and $ m_{T2}^{h_i, h_j} $ distributions for 1 Boosted $H_{bb}$ category. Black lines represent the signal, while other colors represent background processes, including $t/W$ samples (Orange), $X_{b\bar{b}}Y_{\tau\tau}Y_{\tau\tau}$ samples (Blue).}
    \label{fig:boosted_cate1}
\end{figure}

\begin{figure}[tp]
    \centering
    \includegraphics[width=.32\textwidth]{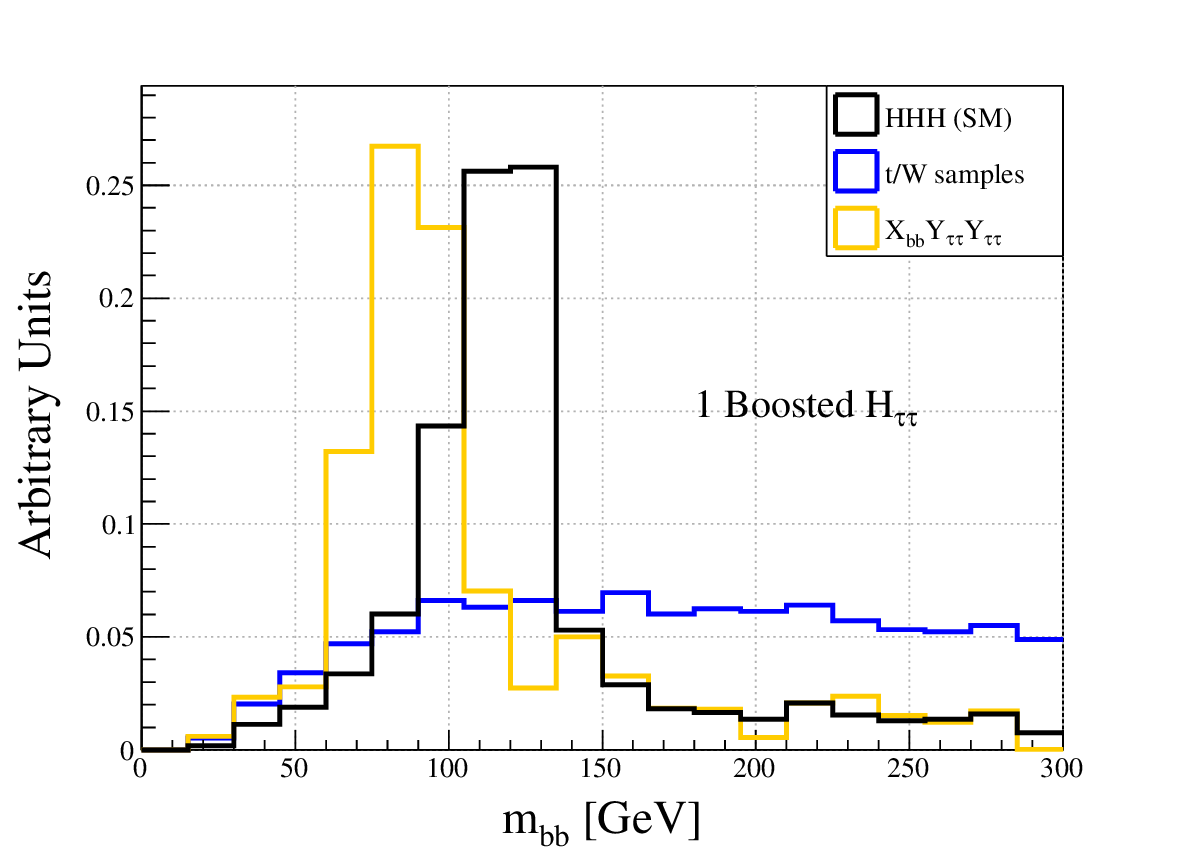}
    \includegraphics[width=.32\textwidth]{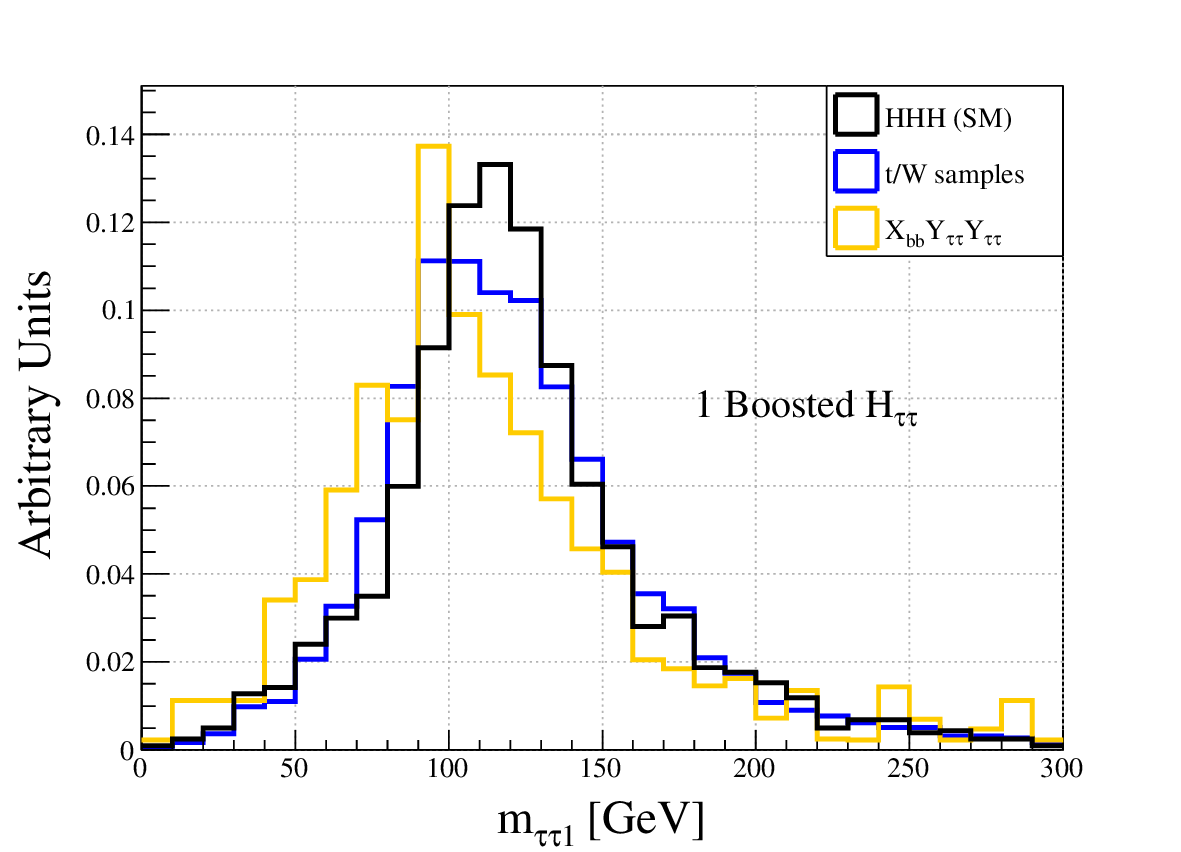}
    \includegraphics[width=.32\textwidth]{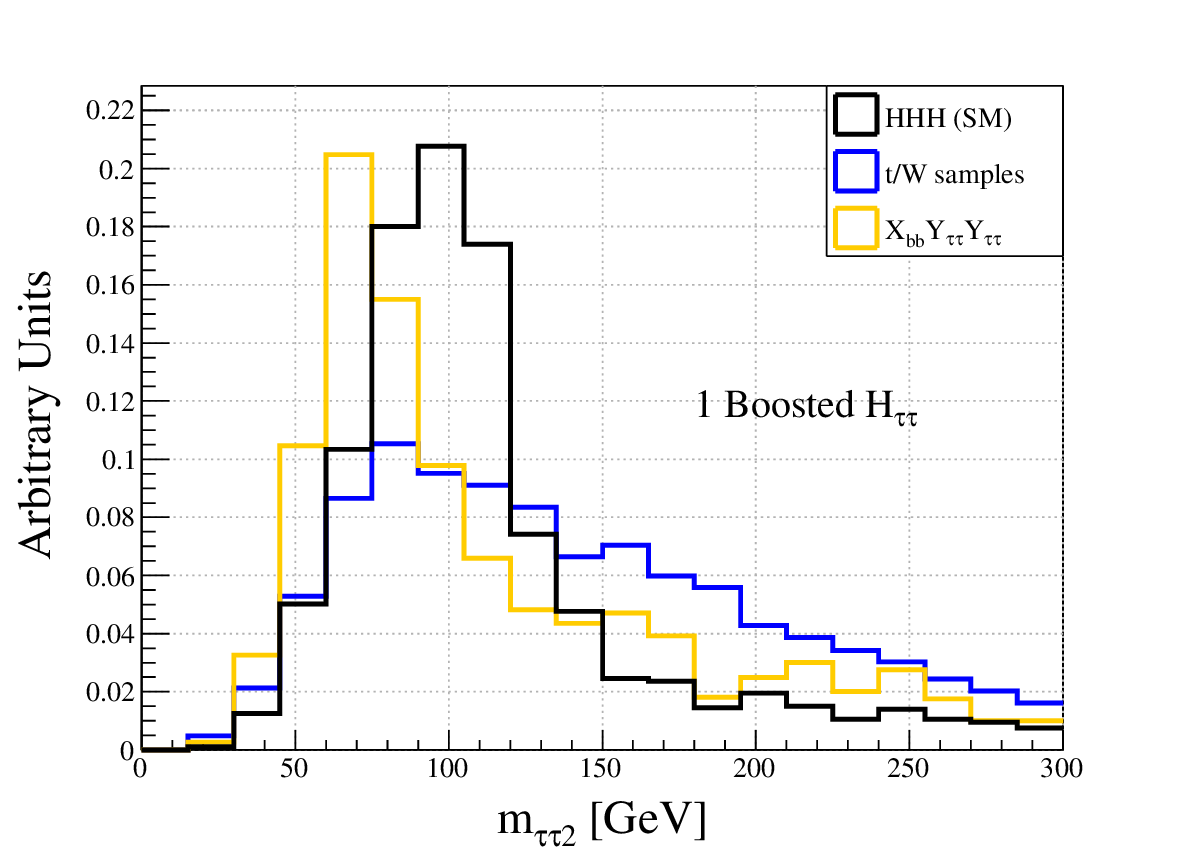}
    \qquad
    \includegraphics[width=.32\textwidth]{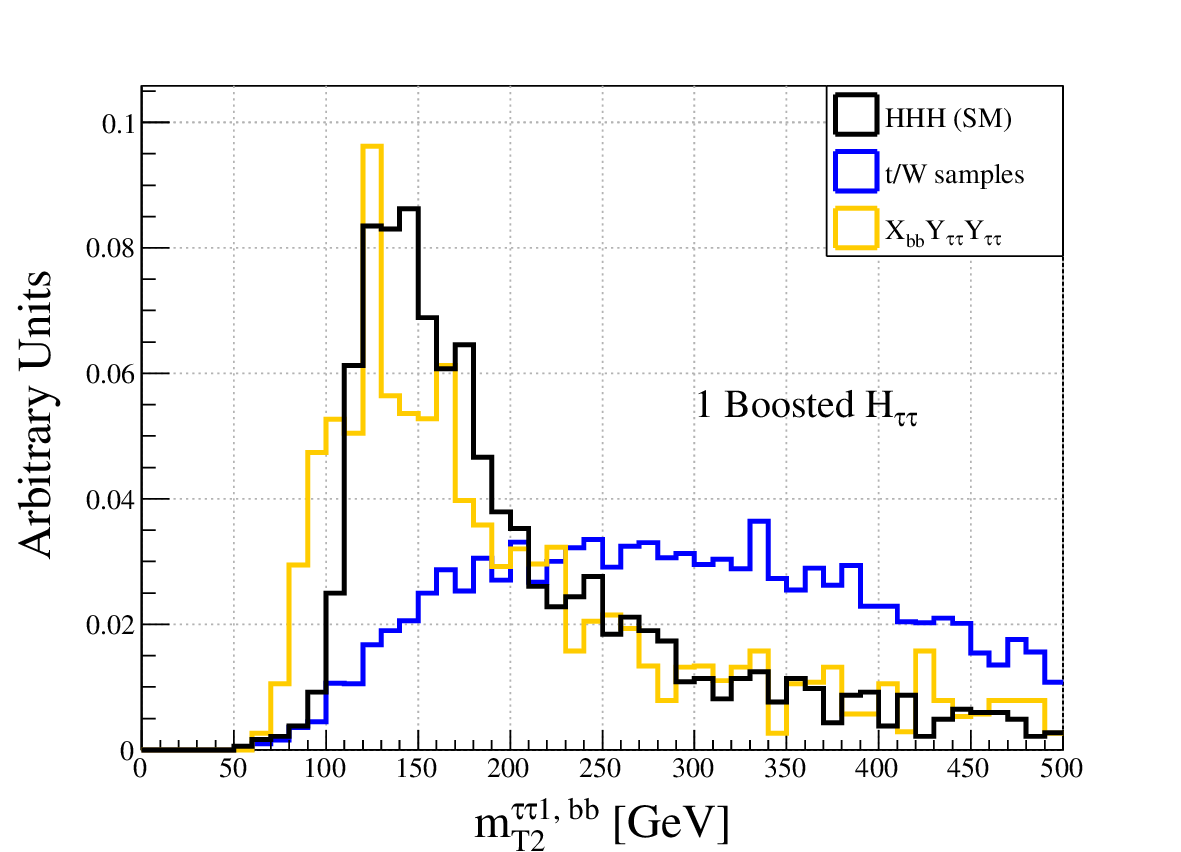}
    \includegraphics[width=.32\textwidth]{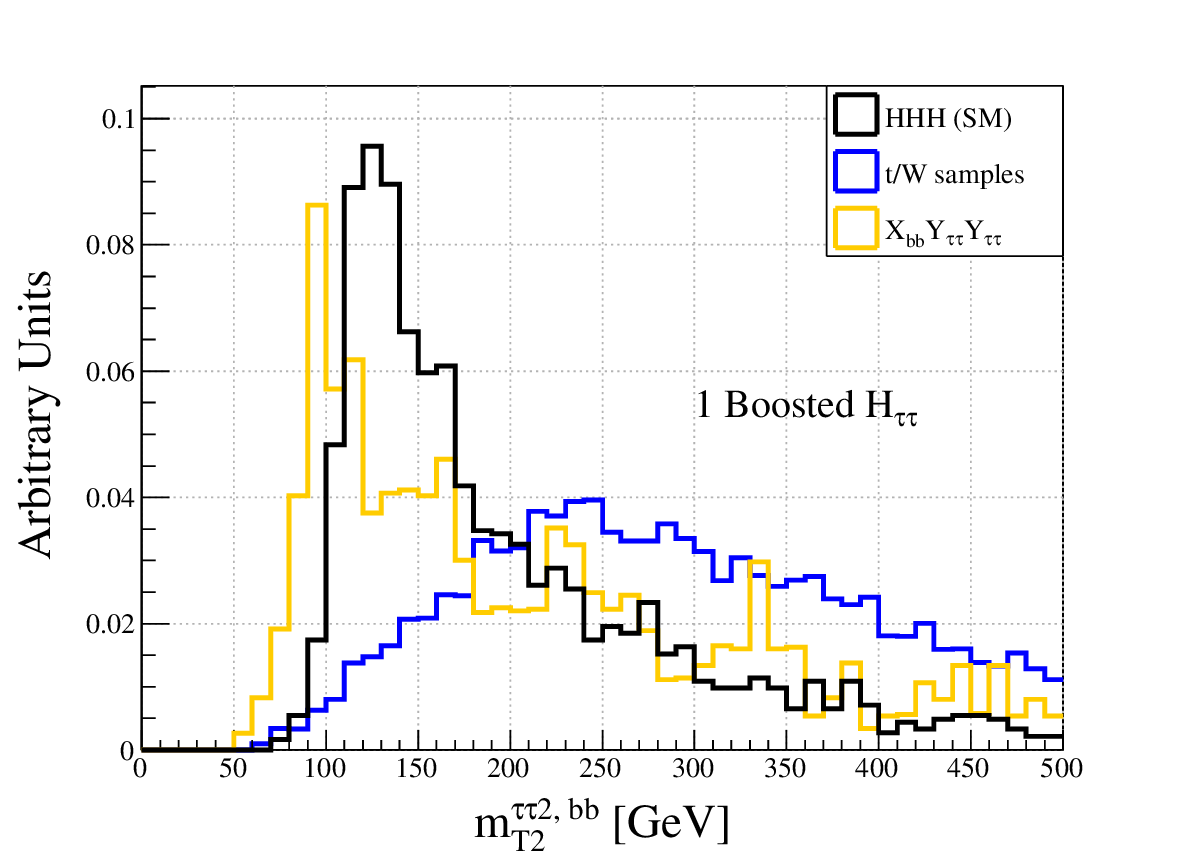}
    \includegraphics[width=.32\textwidth]{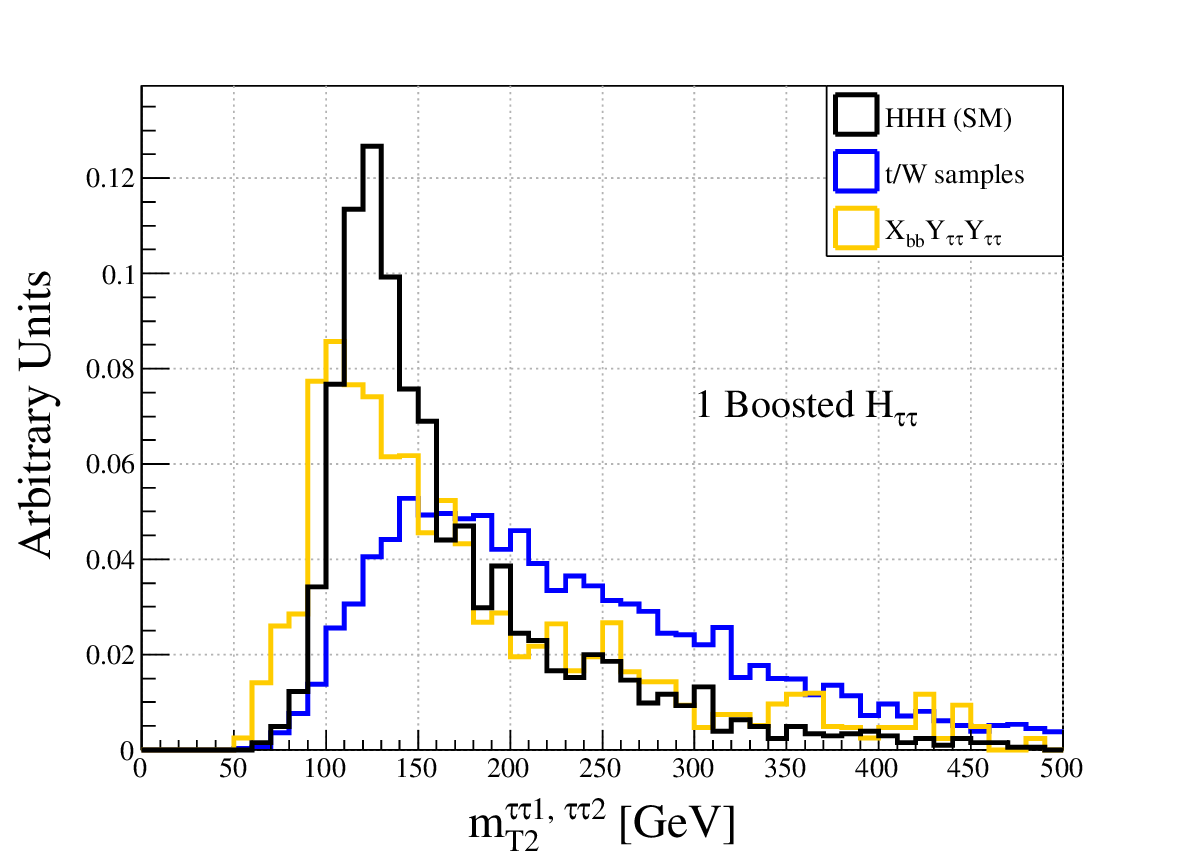}
    \caption{The $m_{b\bar{b}}$, $m_{\tau\tau}$ and $ m_{T2}^{h_i, h_j} $ distributions for 1 Boosted $H_{\tau\tau}^{1}$ category. Black lines represent the signal, while other colors represent background processes, including $t/W$ samples (Orange), $X_{b\bar{b}}Y_{\tau\tau}Y_{\tau\tau}$ samples (Blue).}
    \label{fig:boosted_cate2}
\end{figure}

\begin{figure}[tp]
    \centering
    \includegraphics[width=.32\textwidth]{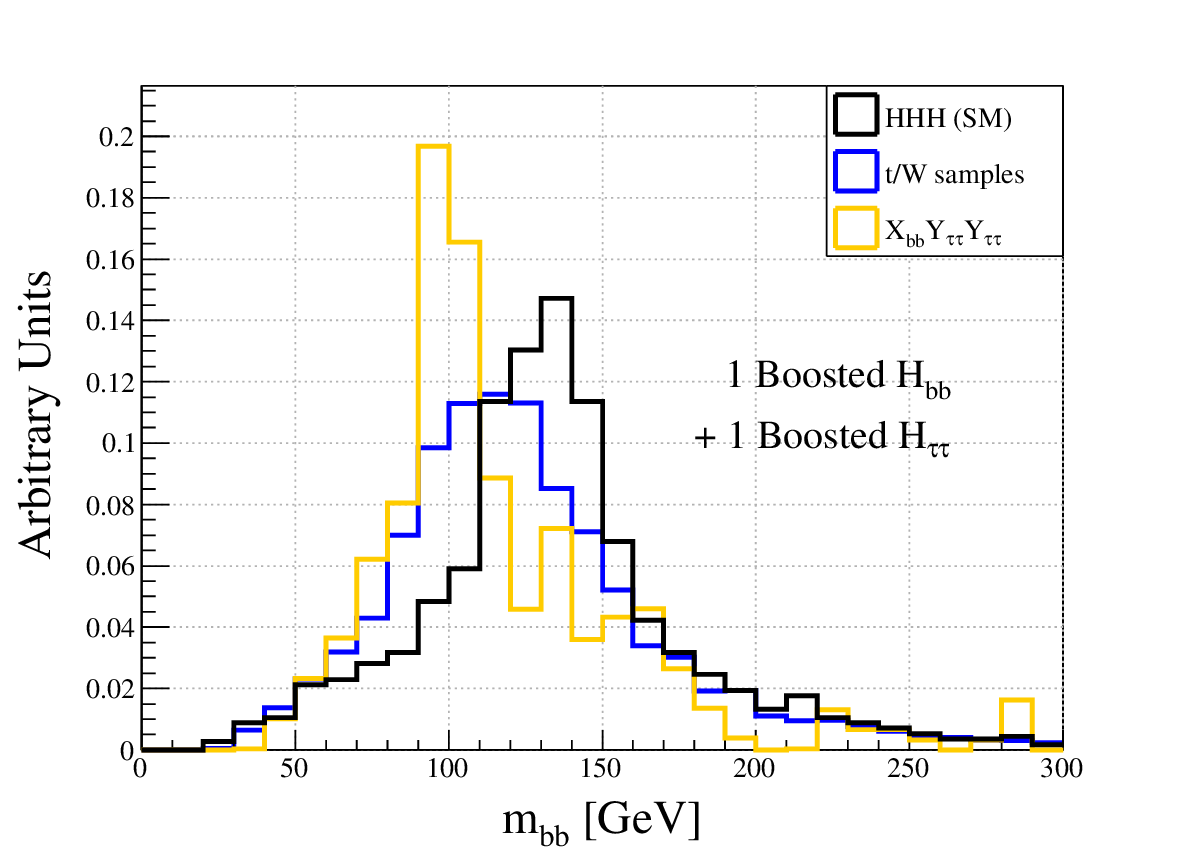}
    \includegraphics[width=.32\textwidth]{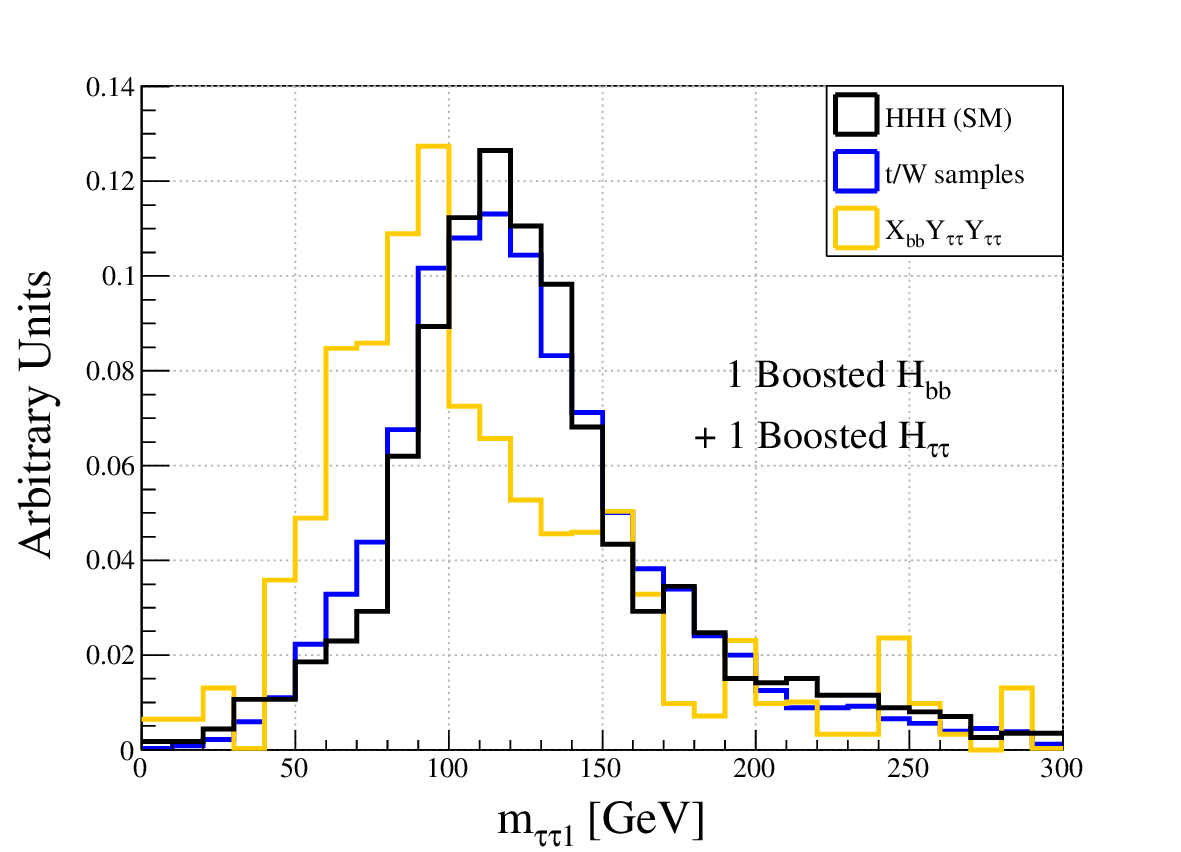}
    \includegraphics[width=.32\textwidth]{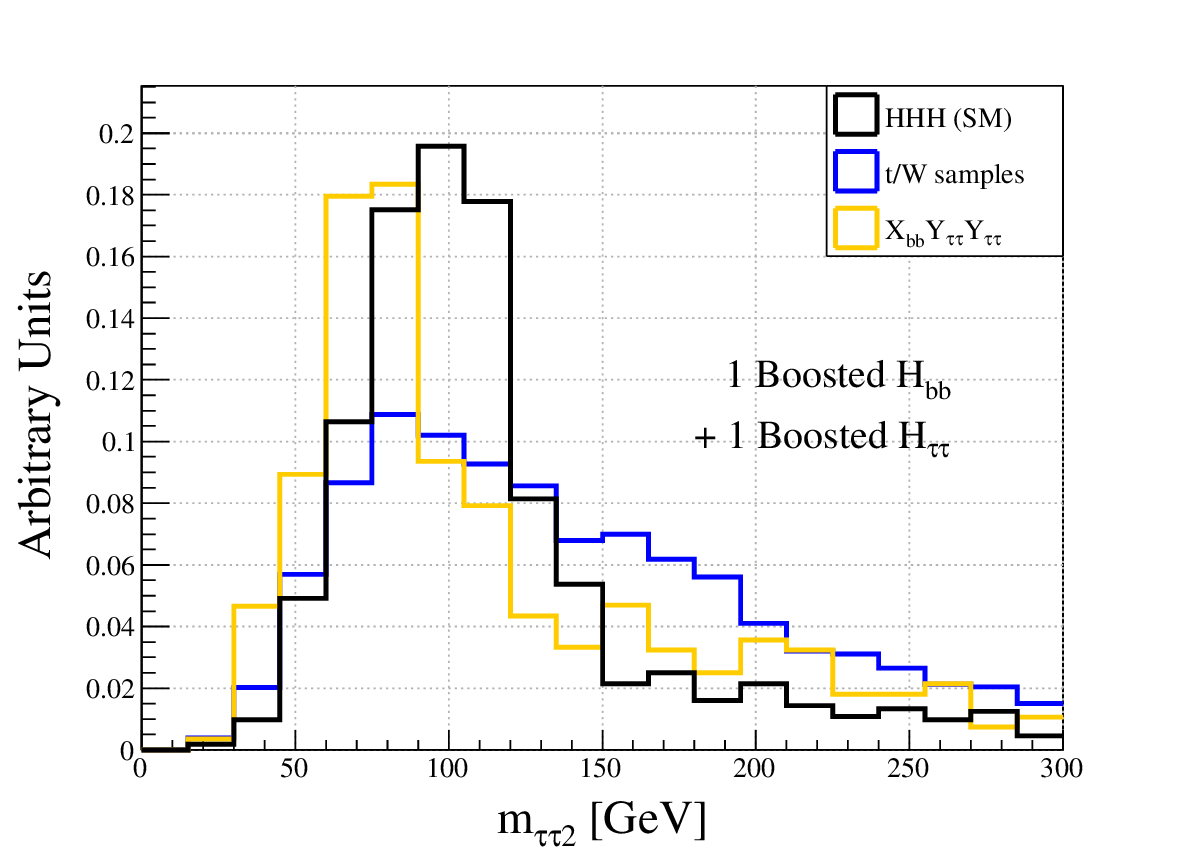}
    \qquad
    \includegraphics[width=.32\textwidth]{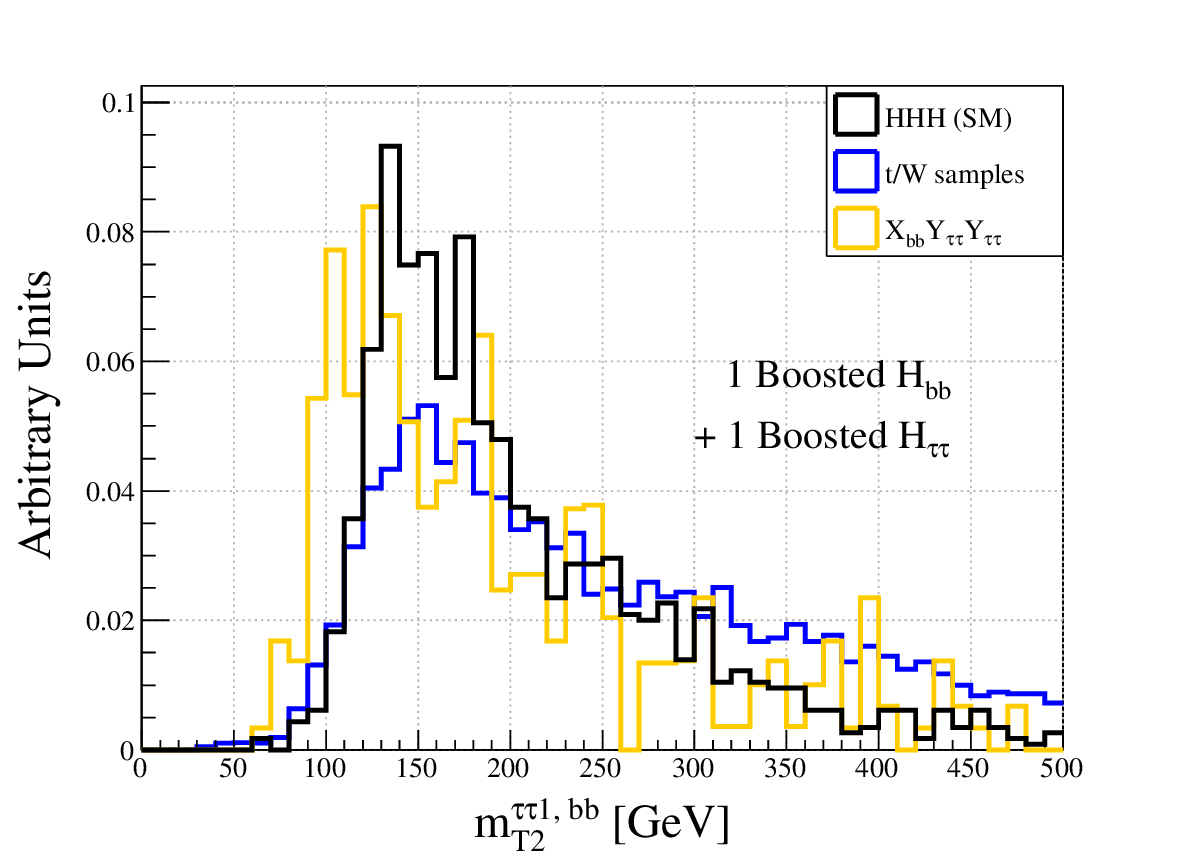}
    \includegraphics[width=.32\textwidth]{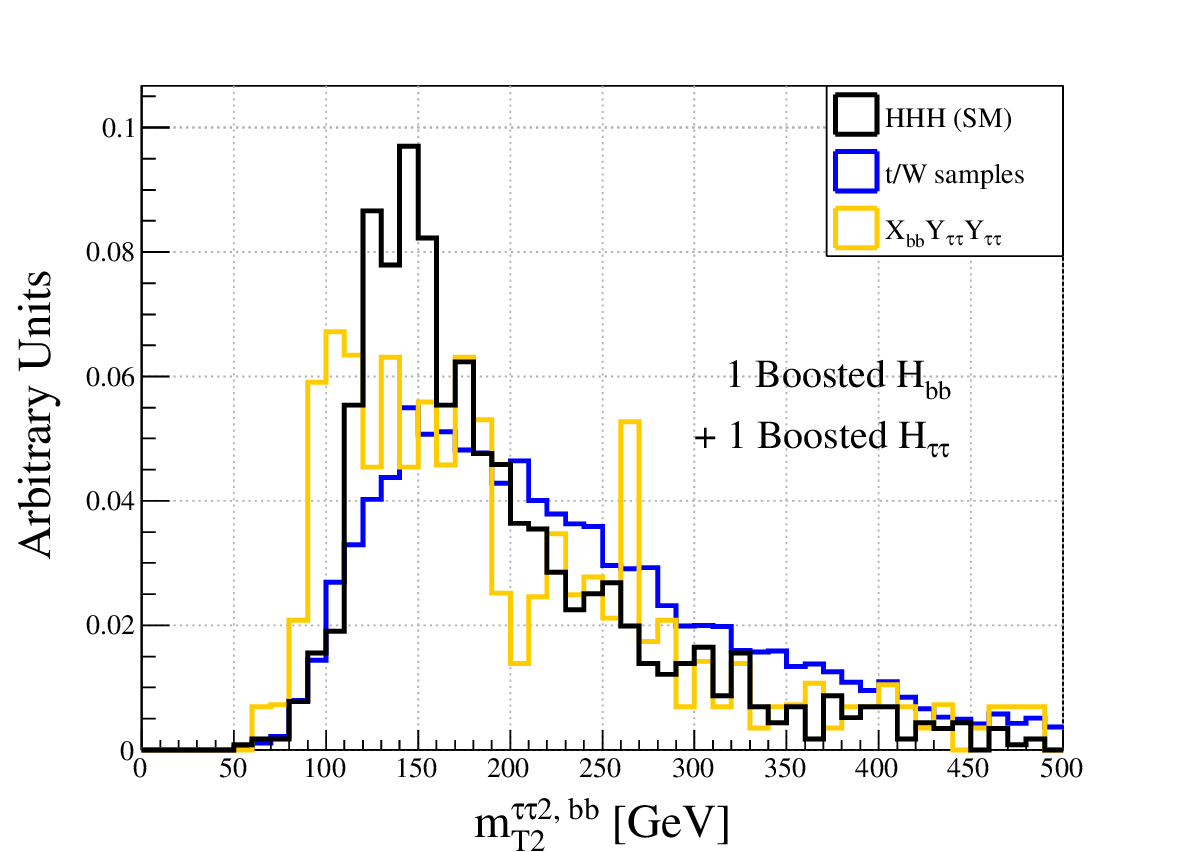}
    \includegraphics[width=.32\textwidth]{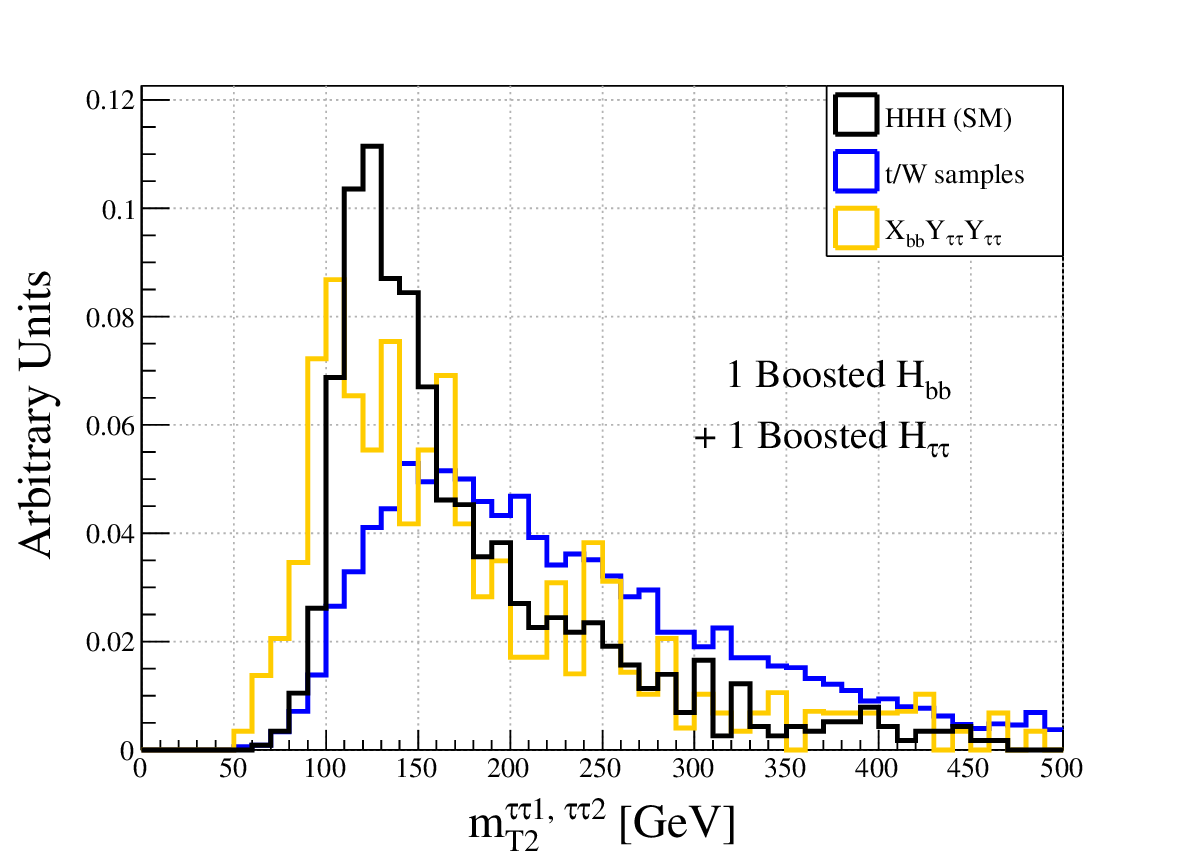}
    \caption{The $m_{b\bar{b}}$, $m_{\tau\tau}$ and $ m_{T2}^{h_i, h_j} $ distributions for 2 Boosted $H_{\tau\tau} H_{bb}$ category. Black lines represent the signal, while other colors represent background processes, including $t/W$ samples (Orange), $X_{b\bar{b}}Y_{\tau\tau}Y_{\tau\tau}$ samples (Blue).}
    \label{fig:boosted_cate3}
\end{figure}

\begin{figure}[tp]
    \centering
    \includegraphics[width=.32\textwidth]{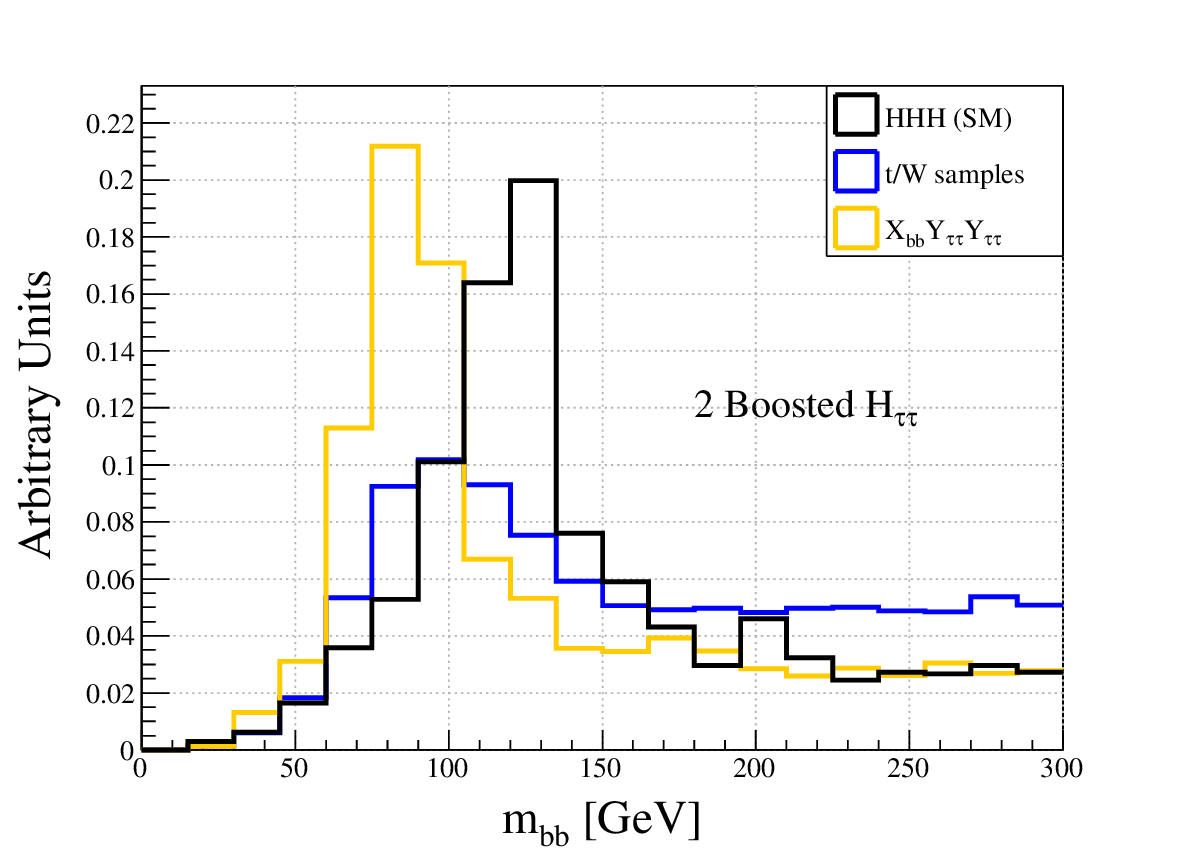}
    \includegraphics[width=.32\textwidth]{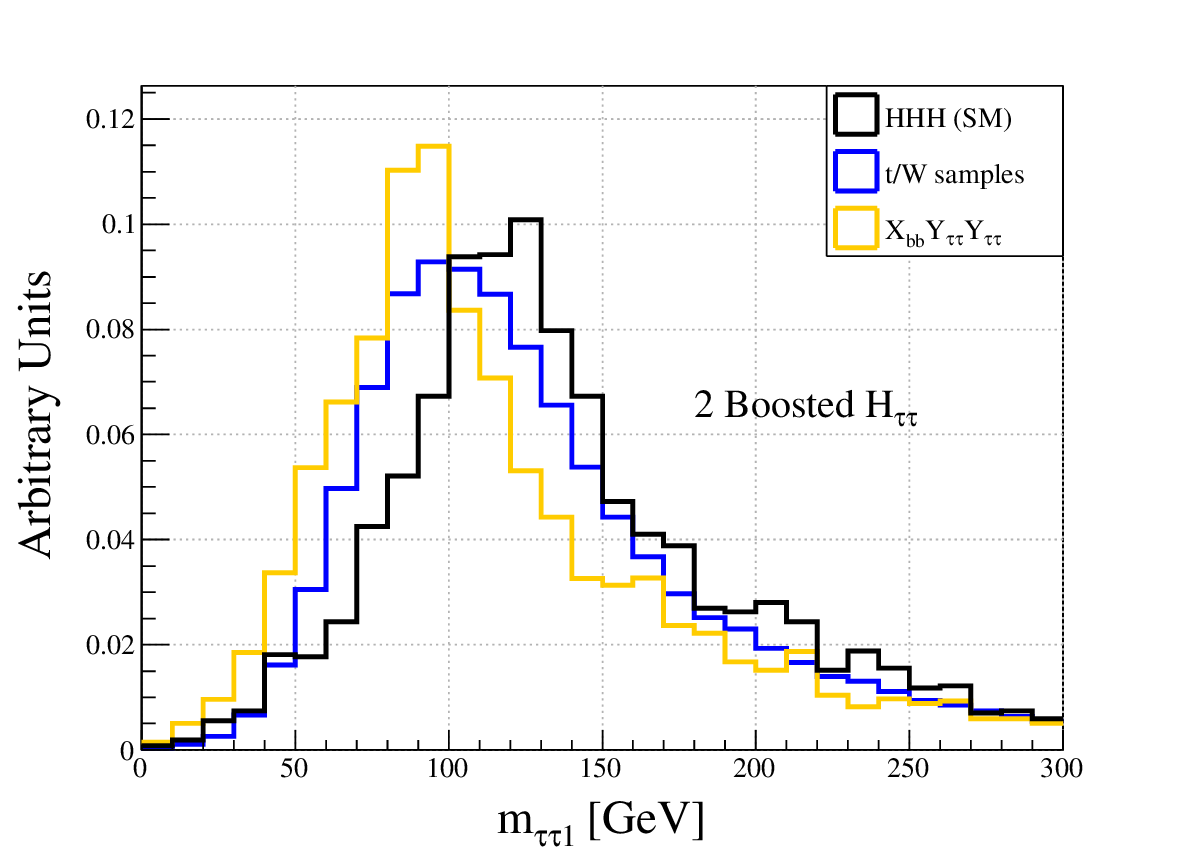}
    \includegraphics[width=.32\textwidth]{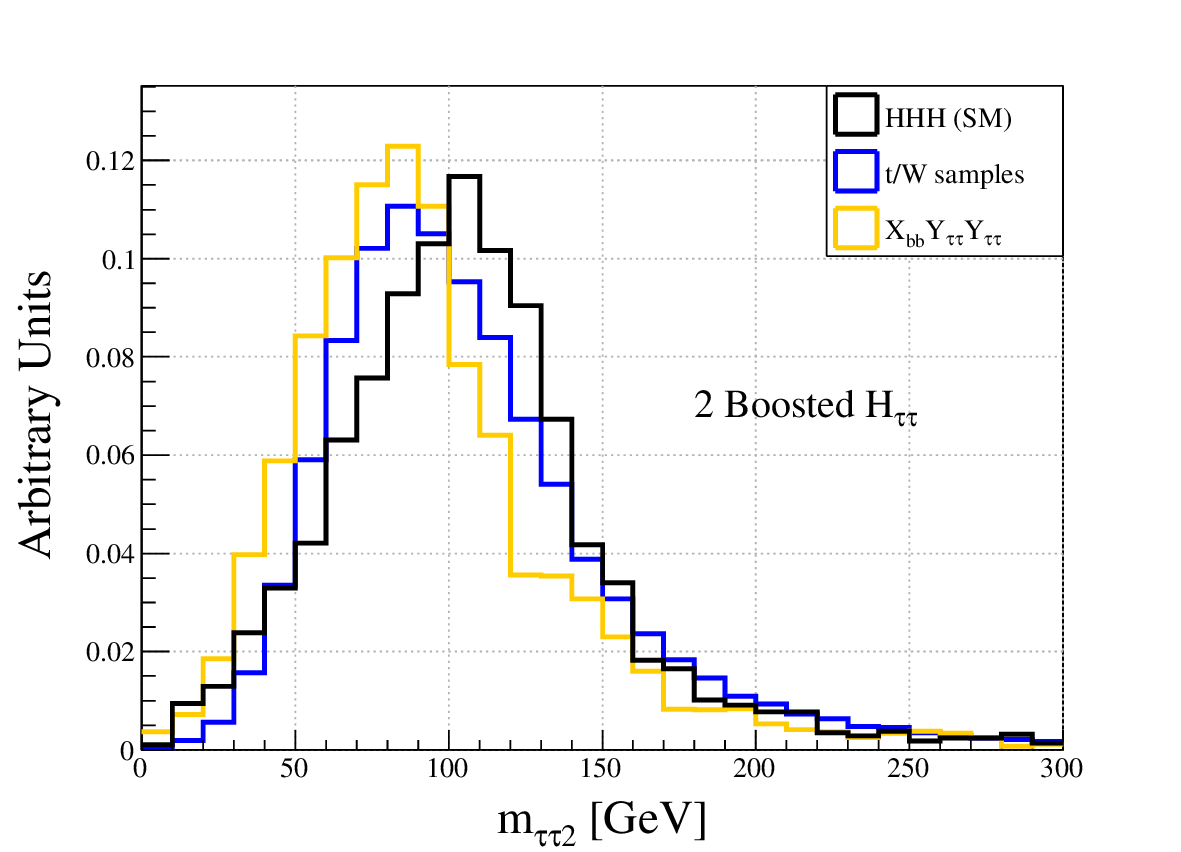}
    \qquad
    \includegraphics[width=.32\textwidth]{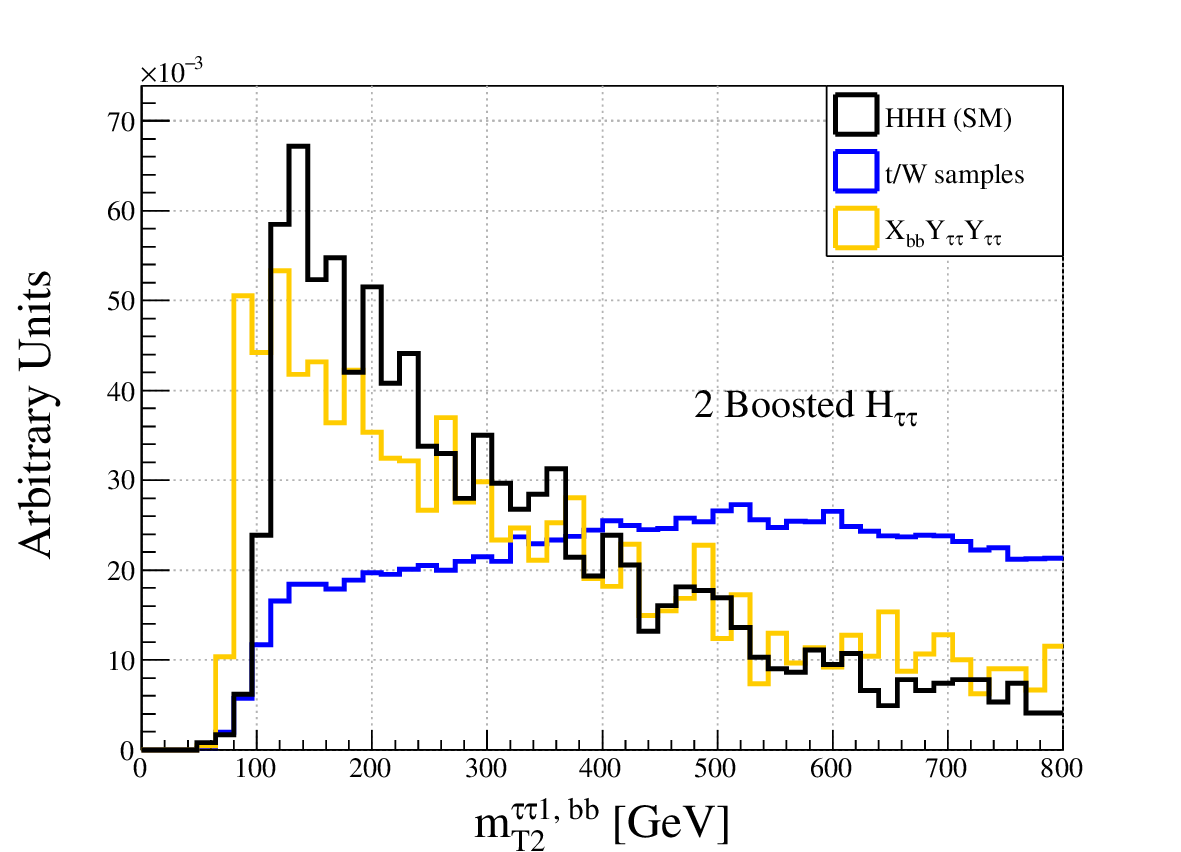}
    \includegraphics[width=.32\textwidth]{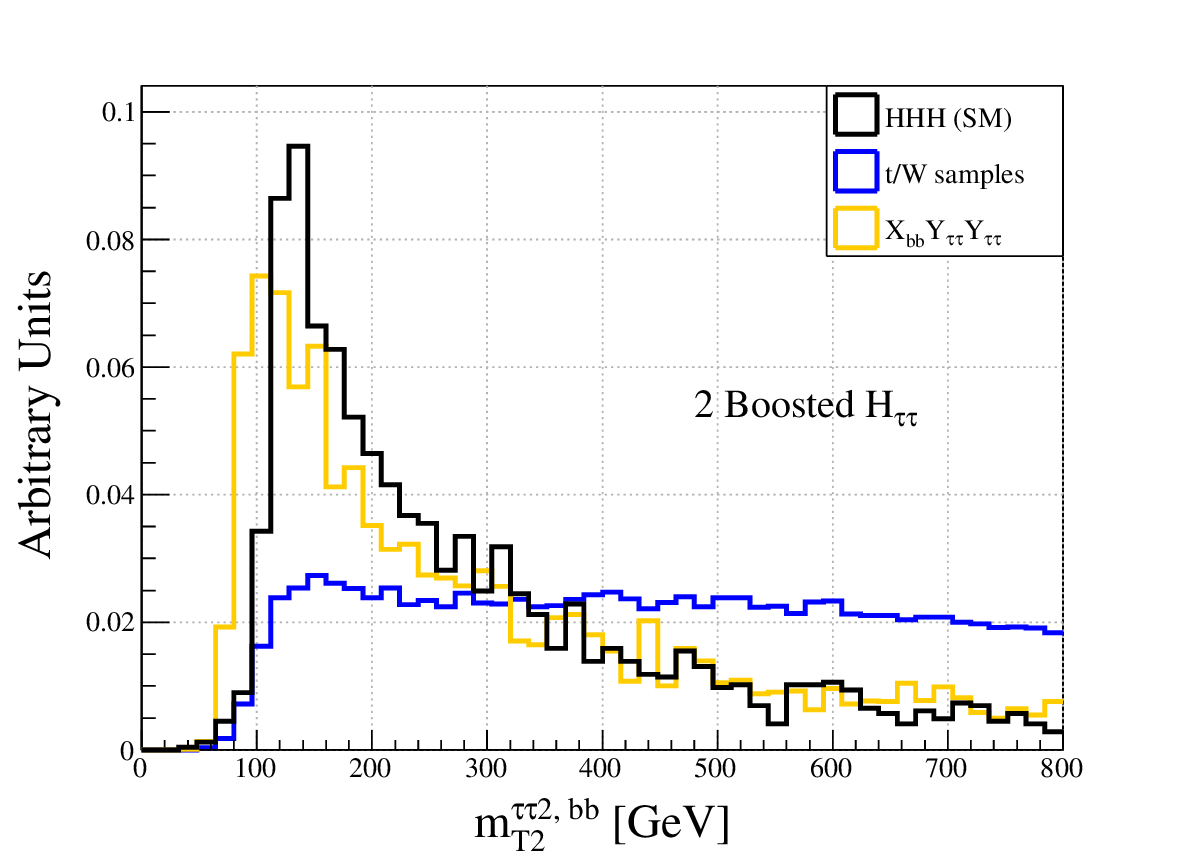}
    \includegraphics[width=.32\textwidth]{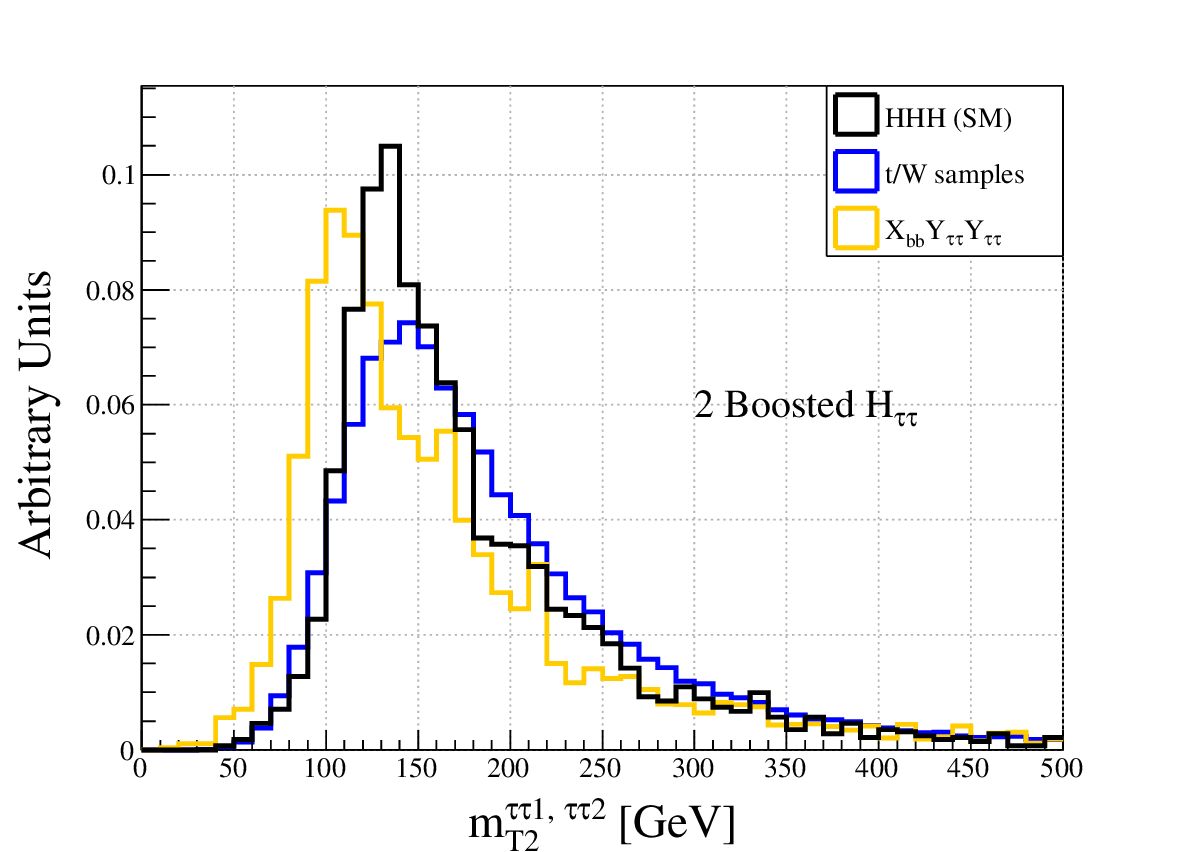}
    \caption{The $m_{b\bar{b}}$, $m_{\tau\tau}$ and $ m_{T2}^{h_i, h_j} $ distributions for 2 Boosted $H_{\tau\tau}^{1,2}$ category. Black lines represent the signal, while other colors represent background processes, including $t/W$ samples (Orange), $X_{b\bar{b}}Y_{\tau\tau}Y_{\tau\tau}$ samples (Blue).}
    \label{fig:boosted_cate4}
\end{figure}

\begin{figure}[tp]
    \centering
    \includegraphics[width=.32\textwidth]{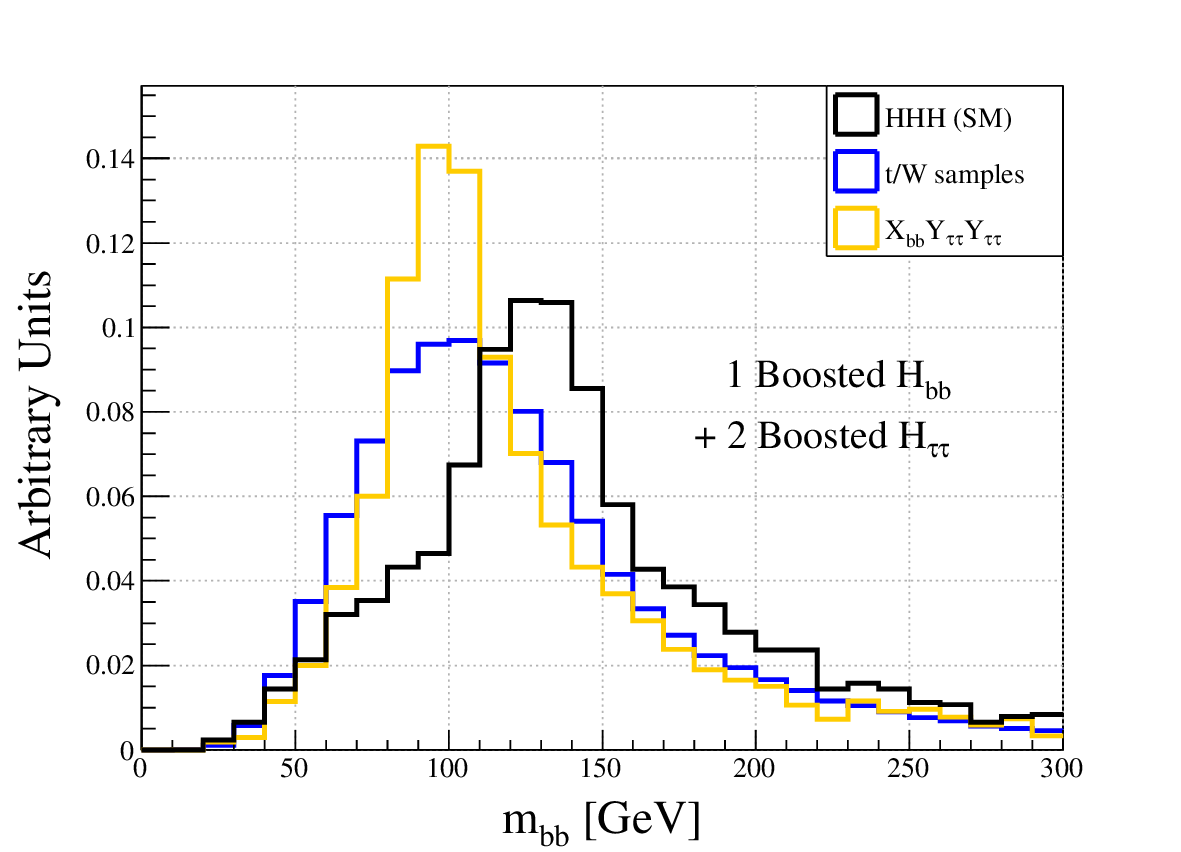}
    \includegraphics[width=.32\textwidth]{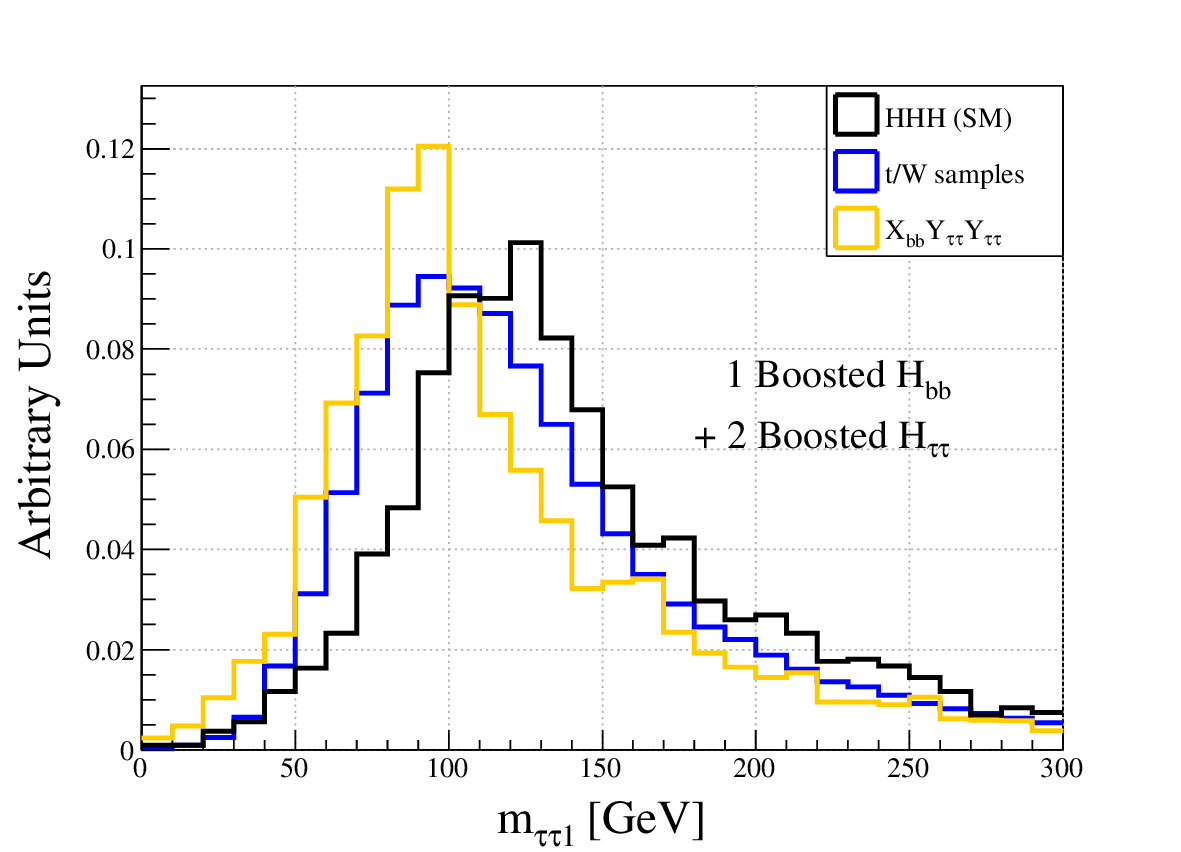}
    \includegraphics[width=.32\textwidth]{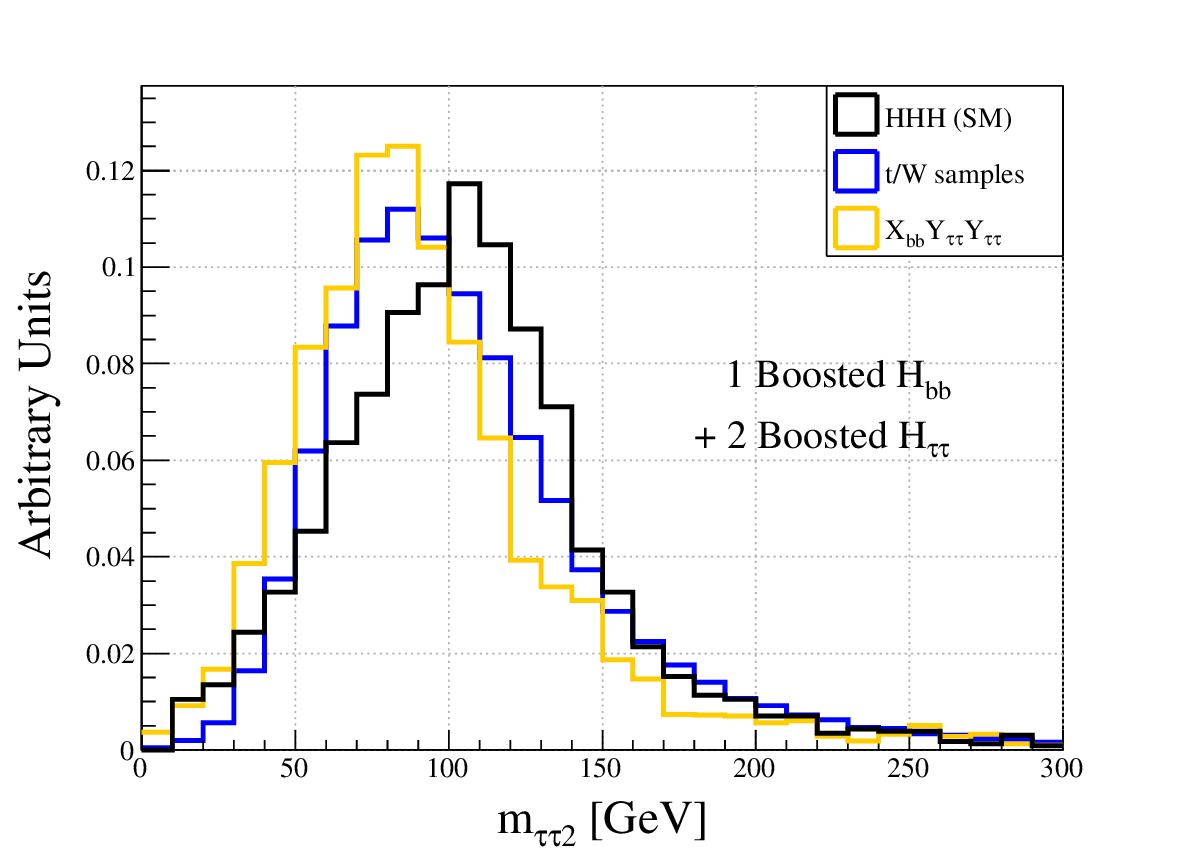}
    \qquad
    \includegraphics[width=.32\textwidth]{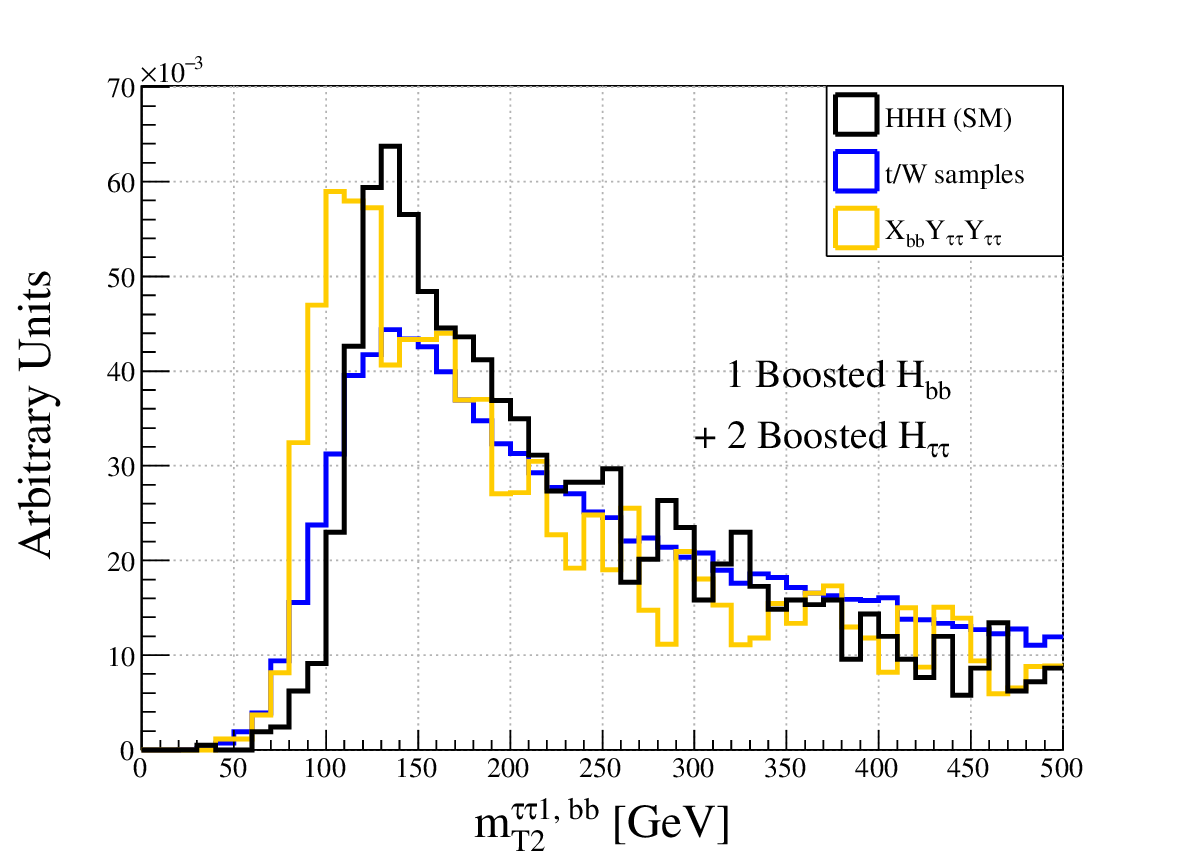}
    \includegraphics[width=.32\textwidth]{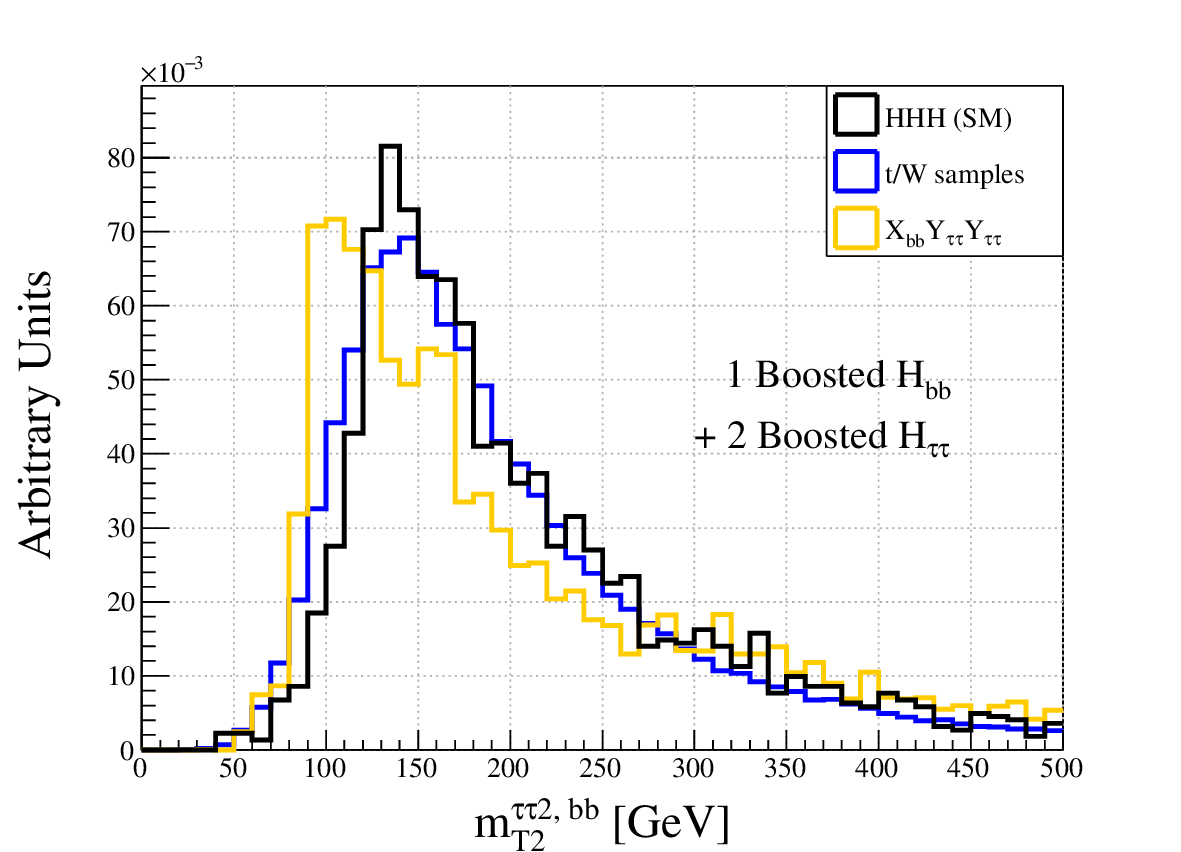}
    \includegraphics[width=.32\textwidth]{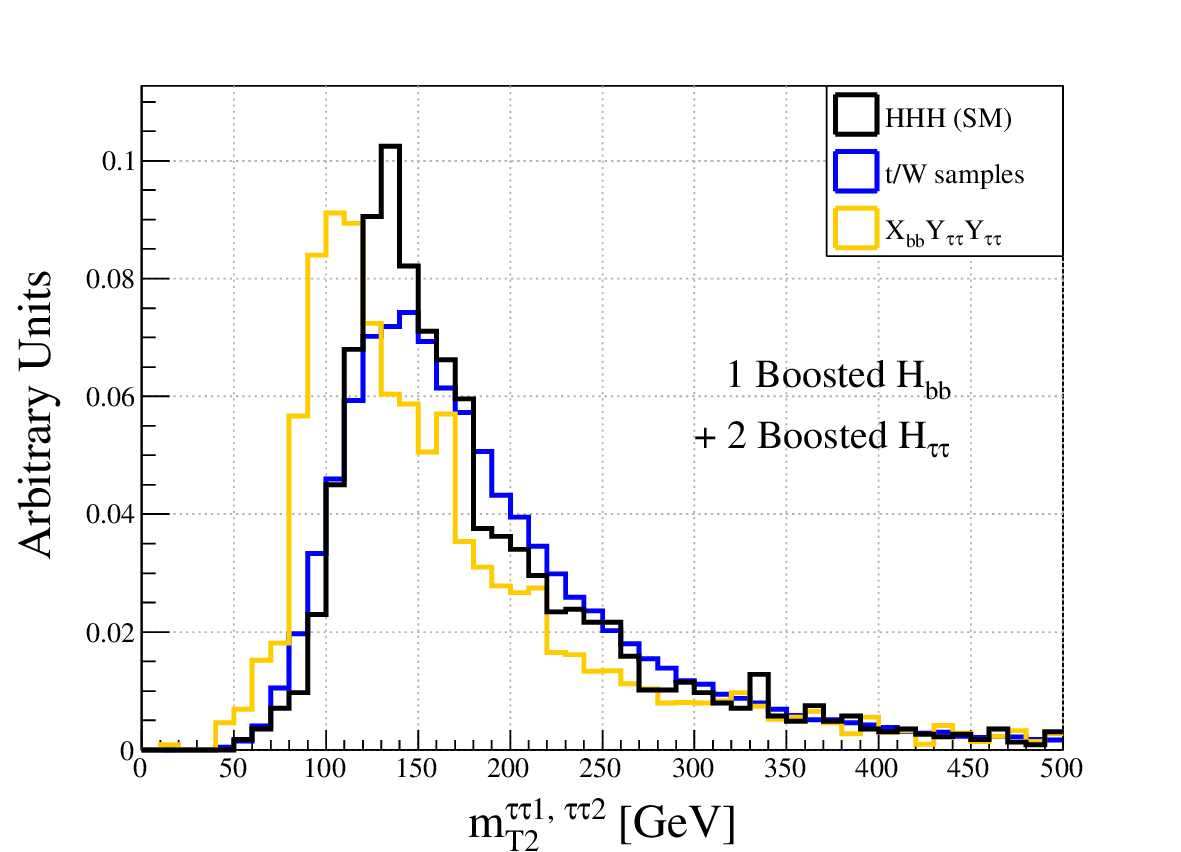}
    \caption{The $m_{b\bar{b}}$, $m_{\tau\tau}$ and $ m_{T2}^{h_i, h_j} $ distributions for 3 Boosted $H_{\tau\tau}^{1,2} H_{bb}$ category. Black lines represent the signal, while other colors represent background processes, including $t/W$ samples (Orange), $X_{b\bar{b}}Y_{\tau\tau}Y_{\tau\tau}$ samples (Blue).}
    \label{fig:boosted_cate5}
\end{figure}

Fig.~\ref{fig:boosted_cate1} to~\ref{fig:boosted_cate5} display the kinematic distributions of the boosted Higgs mass and $ m_{T2} $ for five categories based on the number of boosted Higgs. These distributions reveal that with a single boosted Higgs, certain regions in the mass and $ m_{T2} $ spectrum offer relative strong discriminative power. However, as the number of boosted Higgs bosons increases beyond one, separating the signal from the background effectively, especially for the t/W related background, becomes more challenging.

\subsection{Cut-based analysis}

Given the separation between signal and background observed in the last section, we conducted an optimization of event selection using the reconstructed Higgs mass and the $m_{T2}$ variable. This involves tuning the selection criteria to maximize their discriminative power, measured by the $Z_A$ significance, which is defined as: 
\begin{equation}
\label{eq:Za}
\begin{aligned}
    Z_A = \sqrt{2 \left[ (s + b) \ln \left(1 + \frac{s}{b}\right) - s \right]},
\end{aligned}
\end{equation}
where $s$ and $b$ represent the signal and background yields, respectively.
The optimization was performed in a bin-wise manner to take into account the shape dependence in both signal and background.

To ensure the robustness of the results and avoid statistical fluctuations in low-background regions, an additional constraint of number of background events is introduced during the optimization process. Only situation with at least one background event after selections were considered valid for optimization. This constraint prevents the selection of regions with artificially inflated significance due to very low background counts down to fractional, which could lead to unreliable results and overfitting.

Tab.~\ref{tab:resolved_selections} and Tab.~\ref{tab:boosted_selections} summarize the baseline selection criteria and the optimized mass windows and $m_{T2}$ results. Tab.~\ref{tab:resolved_selections} details the resolved selections for low and high $m_{\text{HHH}}$ categories. Tab.~\ref{tab:boosted_selections} presents the boosted selections, categorizing events based on the number of boosted Higgs bosons. 
In the resolved high $m_{\text{HHH}}$ category, the tightened mass windows are adopted given its relatively sharper distributions from more energetic events. In the boosted categories, the mass windows are broad to accommodate more signal events, resulting in increased overlap between the signal and $t/W$-related background events.

\begin{table}[tp]
    \centering
    \begin{tabular}{ccc}
        \toprule
        \text{Observable} & \text{Low $m_{\text{HHH}}$ category} & \text{High $m_{\text{HHH}}$ category}\\ 
        \midrule
            $p_{T}^{b, \tau}$ [GeV] & $> 20$  & $> 20$ \\ 
            $|\eta|$ & $< 2.5$ & $< 2.5$\\ 
            $m_{b\bar{b}}$ [GeV] & $\in [80, 135]$ & $\in [90, 135]$ \\ 
            $m_{\tau\tau1}$ [GeV] & $\in [80, 135]$ & $\in [90, 135]$ \\ 
            $m_{\tau\tau2}$  [GeV] & $\in [70, 145]$ & $\in [70, 200]$ \\ 
            $m_{T2}^{\tau\tau1,\tau\tau2}$ [GeV] & $< 130$ & $ < 130$ \\ 
            $m_{T2}^{\tau\tau1,b\bar{b}}$ [GeV] & $< 150$ & $< 300$ \\ 
            $m_{T2}^{\tau\tau2,b\bar{b}}$ [GeV] & $< 180$ & $< 300$ \\ 
        \bottomrule
    \end{tabular}
    \caption{Cut-based selections applied in the resolved categories.}
    \label{tab:resolved_selections}
\end{table}

\begin{table}[tp]
    \centering
    \resizebox{\textwidth}{!}{
    \begin{tabular}{cccccc}
        \toprule
        \text{Observable} & \text{1 Boosted $H_{bb}$} & \text{1 Boosted $H_{\tau\tau}^{1}$} & \text{2 Boosted $H_{\tau\tau} H_{bb}$} & \text{2 Boosted $H_{\tau\tau}^{1,2}$} & \text{3 Boosted $H_{\tau\tau}^{1,2} H_{bb}$} \\ 
        \midrule
            $p_{T}^{b, \tau}$ [GeV] & &&$> 20$  &&\\ 
            $p_{T}^{H_i}$ [GeV] & &&$> 300$ &&\\ 
            $|\eta|$ & &&$< 2.5$ &&\\ 
            $m_{b\bar{b}}$ [GeV] & $\in [110, 200]$ & $\in [110, 130]$ & $\in [110, 300]$ & $\in [100, 150]$  &$\in [110, 300]$ \\ 
            $m_{\tau\tau1}$ [GeV] & $\in [85, 135]$ & $\in [100, 300]$ & $\in [50, 300]$ & $\in [50, 300]$ &$\in [50, 300]$ \\ 
            $m_{\tau\tau2}$ [GeV] & $\in [60, 200]$ & $\in [70, 130]$ & $\in [50, 130]$ & $\in [50, 300]$ &$\in [50, 300]$ \\ 
            $m_{T2}^{\tau\tau1,\tau\tau2}$ [GeV] & $< 130 $ & $< 180$ & $< 300$ & $< 300$ &$< 300$ \\ 
            $m_{T2}^{\tau\tau1,b\bar{b}}$ [GeV] & $<300$ & $< 150$ & $< 300$  & $< 300$  &$< 300$  \\ 
            $m_{T2}^{\tau\tau2,b\bar{b}}$ [GeV] & $< 300$ & $< 300$  & $< 300$  & $< 300$  &$< 300$  \\
         \bottomrule
    \end{tabular}
    }
    \caption{Cut-based selections applied in the boosted categories.}
    \label{tab:boosted_selections}
\end{table}

\subsection{BDT analysis}
\label{sec:bdt}
In the Resolved category, the number of events is sufficient to enable the machine learning (ML) techniques to further improvement the sensitivity. In this analysis, we employed an XGBoost Boosted Decision Tree (BDT) algorithm\cite{DBLP:journals/corr/ChenG16}, which is an efficient and scalable gradient boosting method widely used for classification and regression tasks.

As described in Fig~\ref{fig:truth_HHH}, the resolved events are divided into two categories based on the invariant mass of the triple Higgs system: Low mass region: $m_{\text{HHH}} \leq 550$ GeV, High mass region: $m_{\text{HHH}} > 550$ GeV. Separate XGBoost BDT models were trained for each of these regions to optimize the sensitivity in their respective kinematic regimes.
The signal sample definition for the two regions is based on the self-coupling parameter $c_3$. For the Low mass category, the signal samples generated with $c_3 < 0$ or $c_3 > 3$ are used in the training given that they contribute more in the low invariant mass region of the triple Higgs system. For the High mass region, the signal sample generated with $0 \leq c_3 \leq 3$ are used in the training for their relatively harder spectrum of the invariant mass of the triple Higgs.
The boosted event categories, on the other hand, suffer from limited signal statistics, making it impractical to train a reliable ML model.

\begin{figure}[tp]
    \centering
    \begin{subfigure}{.45\textwidth}
        \centering
        \includegraphics[width=\textwidth]{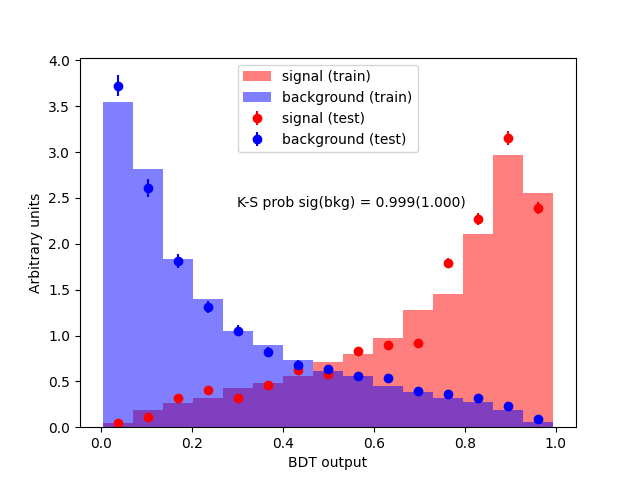}
        \caption{Low $m_{\text{HHH}}$ category}
    \end{subfigure}%
    \qquad
    \begin{subfigure}{.45\textwidth}
        \centering
        \includegraphics[width=\textwidth]{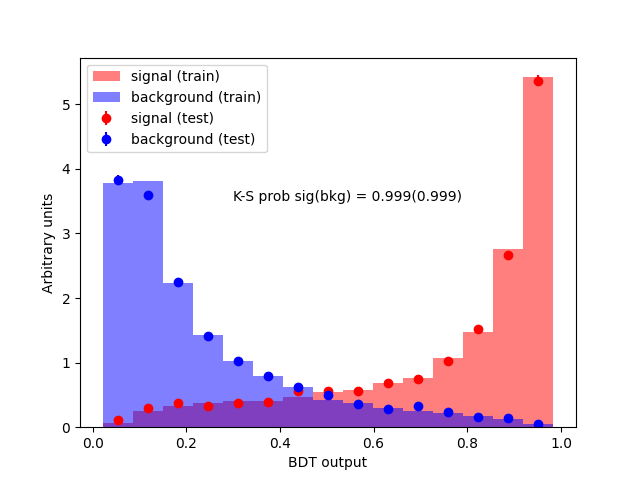}
        \caption{High $m_{\text{HHH}}$ category}
    \end{subfigure}
    \caption{The BDT score distributions.}
    \label{fig:bdt_hists}
\end{figure}

The MC samples were split into three subsets. Training set, 64\% of the total dataset, used to train the BDT model. Testing set, 16\% of the total dataset, used to validate the model during training and prevent overfitting. Application set, 20\% of the total dataset, used to evaluate the performance of the trained model on unseen data. The input features used for training are listed in Tab.~\ref{tab:xgb_vars}. The performance of the final BDT models was evaluated using the output distributions of the training and testing samples, as shown in Fig.~\ref{fig:bdt_hists}. To avoid overfitting, the BDT output distributions of the training and testing samples were compared using the Kolmogorov–Smirnov (KS) test, with the KS test scores exceeding 0.9 in both categories. Additionally, a scan of the hyper-parameters was performed to identify the optimal settings for each category. The final model hyper-parameters for Low $m_{\text{HHH}}$(High $m_{\text{HHH}}$) category include 2400 (500) trees, tree depth of 3 (5), learning rate of 0.01 (0.01). These parameters were chosen to balance model complexity and training efficiency while minimizing the risk of overfitting.  

The BDT scores are subsequently used in a likelihood-based fit to compute the final signal significance. Details of the fit and the resulting signal significance will be discussed in the following section. 

\begin{table}
    \centering
    \begin{tabular}{>{\centering\arraybackslash}p{0.5\linewidth}>{\centering\arraybackslash}p{0.5\linewidth}}
    \toprule
        Input variables & Description\\
    \midrule
        $p_T^{b_{1,2}}, p_T^{\tau_{1,2,3,4}}$, $p_T^{bb}, p_T^{\tau\tau1}, p_T^{\tau\tau2}, p_T^{4\tau}, p_T^{4\tau 2b}$& Transverse momentum ($p_T$) of the two b-jets, four $\tau$, di-b-jets, two $\tau$ pairs,  4$\tau$ and 4$\tau$2b system.\\
        $\eta_{b_{1,2}}, \eta_{\tau_{1,2,3,4}}$, $\eta_{bb}, \eta_{\tau\tau1}, \eta_{\tau\tau2}$& Pseudorapidity ($\eta$) of the two b-jets, four $\tau$, di-b-jets, two $\tau$ pairs.\\
        $m_{bb}, m_{\tau\tau1}, m_{\tau\tau2},  m_{4\tau}, m_{4\tau 2b}$ &Invariant mass of the di-b-jets, two $\tau$ pairs,  4$\tau$ and 4$\tau$2b system.\\
        $\Delta R_{bb}$, $\Delta R_{\tau\tau1}$, $\Delta R_{\tau\tau2}$, $\Delta R_{bb, \tau\tau1}$, $\Delta R_{bb, \tau\tau2}$, $\Delta R_{\tau\tau1, \tau\tau2}$&Angular distance ($\Delta R=\sqrt{\Delta\eta^2+\Delta\phi^2}$) between the constituents of the di-b-jets, two $\tau$ pairs. $\Delta R$ between the di-b-jets and $\tau$ pairs ($\tau\tau1,\tau\tau2$), $\Delta R$ between the two $\tau$ pairs.\\
        $\Delta \eta_{bb}$, $\Delta \eta_{\tau\tau1}$, $\Delta \eta_{\tau\tau2}$, $\Delta \eta_{bb, \tau\tau1}$, $\Delta \eta_{bb, \tau\tau2}$, $\Delta \eta_{\tau\tau1, \tau\tau2}$&Difference in pseudorapidity ($\Delta\eta$) for the  di-b-jets, two $\tau$ pairs. $\Delta\eta$ between the bb-system and each $\tau$ pairs , $\Delta \eta$ between two $\tau$ pairs.\\
        $\Delta \phi_{bb}$, $\Delta \phi_{\tau\tau1}$, $\Delta \phi_{\tau\tau2}$, $\Delta \phi_{bb, \tau\tau1}$, $\Delta \phi_{bb, \tau\tau2}$, $\Delta \phi_{\tau\tau1, \tau\tau2}$&Difference in azimuthal angle ($\Delta\phi$) for the  di-b-jets, two $\tau$ pairs. $\Delta\phi$ between the bb-system and $\tau$ pairs, $\Delta \phi$ between two $\tau$ pairs.\\
        $\frac{p_T^{\tau_{1,2}}}{m_{\tau\tau1}}$, $\frac{p_T^{\tau_{3,4}}}{m_{\tau\tau2}}$, $\frac{p_T^{b_{1,2}}}{m_{bb}}$, $\frac{p_T^{bb}}{m_{bb}}$&Ratios of the $p_T$ of single $\tau$/b-jet/di-b-jets  to the invariant mass of $\tau$ pairs /di-b-jets.\\
        $\frac{p_T^{\tau\tau_{1,2}}}{m_{4\tau}}$, $\frac{p_T^{\tau\tau_{1,2}}}{m_{4\tau2b}}$, $\frac{p_T^{4\tau}}{m_{4\tau2b}}$, $\frac{p_T^{2b}}{m_{4\tau2b}}$ &Ratios of the $p_T$ of $\tau$ pairs/$4\tau$ system/4$\tau$2b system to the invariant mass of $4\tau$ system/$4\tau2b$ system.\\
        $m_T^{\tau_{1,2,3,4}}, m_T^{\tau\tau_{1,2}}, m_T^{bb}, m_T^{total}$ &Transverse mass ($m_T$) of the single $\tau$, $\tau$ pairs, di-b-jets and total $4\tau2b$ system.\\
        $m_{T2}^{\tau\tau1}, m_{T2}^{\tau\tau_{2}}, m_{T2}^{bb},$ $m_{T2}^{\tau\tau1, bb}, m_{T2}^{\tau\tau_{2}, bb}, m_{T2}^{\tau\tau_{1}, \tau\tau_{2}}$&Stransverse mass ($m_{T2}$) of $\tau$ pairs, di-b-jets, $m_{T2}$ between the $\tau$ pairs and di-b-jets, and between the two $\tau$ pairs.\\
    \bottomrule
    \end{tabular}
    \caption{Summary of input variables for the XGBoost BDT training.}
    \label{tab:xgb_vars}
\end{table}

\section{Results}
\label{sec:results}
We present the results of both cut-based and BDT-based analyses for triple Higgs boson production (HHH), as well as its combination with vector boson associated di-Higgs production (VHH), at a 100 TeV proton-proton collider with an integrated luminosity of $30 \,\text{ab}^{-1}$. The cut-based analysis is performed across all resolved and boosted event categories, while the BDT-based analysis is applied only to the resolved categories due to limited statistics in the boosted regions.

To extract the final signal significance, we perform a statistical fit using the Pyhf framework~\cite{pyhf,pyhf_joss}. In the cut-based analysis, a binned likelihood fit is applied to the $m_{\text{HHH}}$ distribution, while in the BDT-based approach, the fit is performed on the BDT score distribution. This bin-wise strategy allows better exploitation of shape information, thereby improving the precision of significance estimation.

The expected event yields and significance from cut-based analysis are summarized in Tables~\ref{tab:results} and \ref{tab:results_2} for the SM benchmark ($c_3= 0$, $d_4= 0$) and a representative BSM scenario ($c_3= -2$, $d_4= -11$). These tables report the number of signal and background events remaining after each selection step — from baseline cuts to optimized mass window and $m_{T2}$ requirements — for both resolved and boosted categories.
The significance is computed using the $Z_A$ formula~\ref{eq:Za}, where $Z_A(\text{S}_1)$ refers to the HHH signal significance and $Z_A(\text{S}_2)$ includes the combined contribution from HHH and VHH productions. In the $\text{S}_1$ case, the SM VHH process is treated as part of the background. The cut-based analysis shows varying levels of sensitivity across different categories. Among all, the resolved category with high $m_{\text{HHH}}$ demonstrates the strongest sensitivity to the triple Higgs signal. This strong performance is largely attributed to the discriminating power of the $m_{T2}$ variable. In the boosted scenario, events with one boosted Higgs boson show moderate yet complementary sensitivity, contributing to the overall signal significance. 
Scenarios involving multiple boosted Higgs bosons are more challenging and contribute negligibly to the overall sensitivity, primarily due to limited statistics and reduced background separation power.



\begin{table}[tp]
\centering
\resizebox{\textwidth}{!}{
\begin{tabular}{lccccccl}
\toprule
\text{Category}& \text{Cut flow}& \text{HHH (SM)}&  $\text{HHH+VHH (SM)}$& $\text{t/W samples}$& $X_{b\bar{b}}Y_{\tau\tau}Y_{\tau\tau}$ & $Z_A(\text{S}_1)$& $Z_A(\text{S}_2)$\\
\midrule
\multirow{4}{*}{Resolved low $m_{\text{HHH}}$}& Baseline & 1.72&  2.86&6437.39& 5.86& 0.02&0.03\\
 & Mass window & 0.72&  1.13&741.15& 1.01& 0.03&0.04\\
 & $m_{T2}$ & $0.51$&  0.82&$228.40$& $0.80$& 0.04&0.06\\
\midrule
\multirow{4}{*}{Resolved high $m_{\text{HHH}}$}& Baseline & 3.53&  5.14&13560.52& 8.37& 0.03&0.04\\
 & Mass window & 0.99&  1.23&222.69& 0.40& 0.08&0.09\\
 & $m_{T2}$ & $0.72$&   0.90&$37.74$& $0.24$& 0.22&0.23\\
\midrule
\multirow{4}{*}{1 Boosted $H_{b\bar{b}}$} & Baseline & 1.48&  2.13&3740.43& 3.23&  0.03&0.03\\
 & Mass window & 0.54&  0.67&438.26& 0.30&  0.03&0.03\\
 & $m_{T2}$ & 0.37&  0.46&70.45& 0.17& 0.04&0.05\\
\midrule
\multirow{4}{*}{1 Boosted $H_{\tau\tau}$} & Baseline & 0.95&  1.43&2954.61& 0.04&  0.03&0.03\\
 & Mass window & 0.16&  0.17&33.23& 0.02&  0.03&0.03\\
 & $m_{T2}$ & 0.06&  0.06&2.26& $<0.01$& 0.04&0.04\\
\midrule
\multirow{4}{*}{2 Boosted $H_{b\bar{b}}H_{\tau\tau}$} & Baseline & 0.53&  0.80&1823.57& 0.02&  0.02&0.02\\
 & Mass window & 0.28&   0.37&483.37& 0.02&  0.02&0.02\\
 & $m_{T2}$ & 0.25&  0.32&264.86& 0.18& 0.02&0.02\\
\midrule
\multirow{4}{*}{2 Boosted $H_{\tau\tau}^{1, 2}$} & Baseline & 1.36&  3.78&30653.17& 10.22&  $<0.01$&0.02\\
 & Mass window & 0.36&  0.69&1680.29& 1.00&  $<0.01$&0.02\\
 & $m_{T2}$ & 0.27&  0.49&1021.44& 0.68& $<0.01$&0.02\\
\midrule
\multirow{4}{*}{3 Boosted $H_{b\bar{b}}H_{\tau\tau}^{1, 2}$ } & Baseline & 1.12&  3.38&33788.98& 8.22&  $<0.01$&0.02\\
 & Mass window &  0.68&  1.85&15968.47& 3.02&  $<0.01$&0.01\\
 & $m_{T2}$ & 0.41&  0.89&6110.52& 0.04& $<0.01$&0.01\\
\bottomrule
\end{tabular}
}
\caption{Event yields and significance for different categories after sequential cuts. The yields are shown for two signal scenarios (HHH and HHH+VHH in SM) and main backgrounds (t/W samples and $X_{b\bar{b}}Y_{\tau\tau}Y_{\tau\tau}$ samples) at a 100 TeV collider with 30 $\text{ab}^{-1}$ integrated luminosity. $S_1$ and $S_2$ represent the signal of HHH and HHH+VHH, respectively.}
\label{tab:results}
\end{table}

\begin{table}[tp]
\centering
\resizebox{\textwidth}{!}{
\begin{tabular}{lccccccl}
\toprule
\text{Category}& \text{Cut flow}& \text{HHH (BSM)}&  \text{HHH+VHH (BSM)}& \text{t/W samples} & $X_{b\bar{b}}Y_{\tau\tau}Y_{\tau\tau}$ & $Z_A(\text{S}_1)$& $Z_A(\text{S}_2)$\\
\midrule
\multirow{4}{*}{Resolved low $m_{\text{HHH}}$}& Baseline & 11.02&  11.36&6437.39& 5.86& 0.14&0.14\\
 & Mass window & 5.03&  5.14&741.15& 1.01& 0.19&0.19\\
 & $m_{T2}$ & 3.56&  3.65&$228.40$& $0.80$& 0.25&0.26\\
\midrule
\multirow{4}{*}{Resolved high $m_{\text{HHH}}$}& Baseline & 16.06&  16.95&13560.52& 8.37& 0.14&0.15\\
 & Mass window & 4.31&  4.44&222.69& 0.40& 0.30&0.31\\
 & $m_{T2}$ & 3.03&   3.13&$37.74$& $0.24$& 0.70&0.75\\
\midrule
\multirow{4}{*}{1 Boosted $H_{b\bar{b}}$} & Baseline & 6.48&  6.90&3740.43& 3.23&  0.10&0.11\\
 & Mass window & 2.36&  2.44&438.26& 0.30&  0.11&0.12\\
 & $m_{T2}$ & 1.60&  1.66&70.45& 0.17& 0.19&0.20\\
\midrule
\multirow{4}{*}{1 Boosted $H_{\tau\tau}$} & Baseline & 4.10&  4.43&2954.61& 0.04&  0.08&0.08\\
 & Mass window & 0.63&  0.65&33.23& 0.02&  0.10&0.11\\
 & $m_{T2}$ & 0.23&  0.23&2.26& $<0.01$& 0.15&0.15\\
\midrule
\multirow{4}{*}{2 Boosted $H_{b\bar{b}}H_{\tau\tau}$} & Baseline & 2.44&  2.64&1823.57& 0.02&  0.06&0.06\\
 & Mass window & 1.28&  1.36&483.37& 0.02&  0.06&0.06\\
 & $m_{T2}$ & 1.13&  1.20&264.86& 0.18& 0.07&0.07\\
\midrule
\multirow{4}{*}{2 Boosted $H_{\tau\tau}^{1, 2}$} & Baseline & 5.66&   8.00&30653.17& 10.22&  0.03&0.05\\
 & Mass window & 1.55&  1.88&1680.29& 1.00&  0.04&0.05\\
 & $m_{T2}$ & 1.15&  1.37&1021.44& 0.68& 0.04&0.04\\
\midrule
\multirow{4}{*}{3 Boosted $H_{b\bar{b}}H_{\tau\tau}^{1, 2}$ } & Baseline & 4.72&  6.91&33788.98& 8.22&  0.03&0.04\\
 & Mass window & 2.75&  3.88&15968.47& 3.02&  0.02&0.03\\
 & $m_{T2}$ & 1.73&  2.20&6110.52& 0.04& 0.02&0.03\\
\bottomrule
\end{tabular}
}
\caption{Event yields and significance for different categories after sequential cuts. The yields are shown for two signal scenarios (HHH and HHH+VHH in a BSM case $c_3=-2, d_4=-11$) and main backgrounds (t/W samples and $X_{b\bar{b}}Y_{\tau\tau}Y_{\tau\tau}$ samples) at a 100 TeV collider with 30 $\text{ab}^{-1}$ integrated luminosity. $S_1$ and $S_2$ represent the signal of HHH and HHH+VHH, respectively.}
\label{tab:results_2}
\end{table}

\begin{figure}[tp]
    \centering
    \begin{subfigure}{.45\textwidth}
        \centering
        \includegraphics[width=\textwidth]{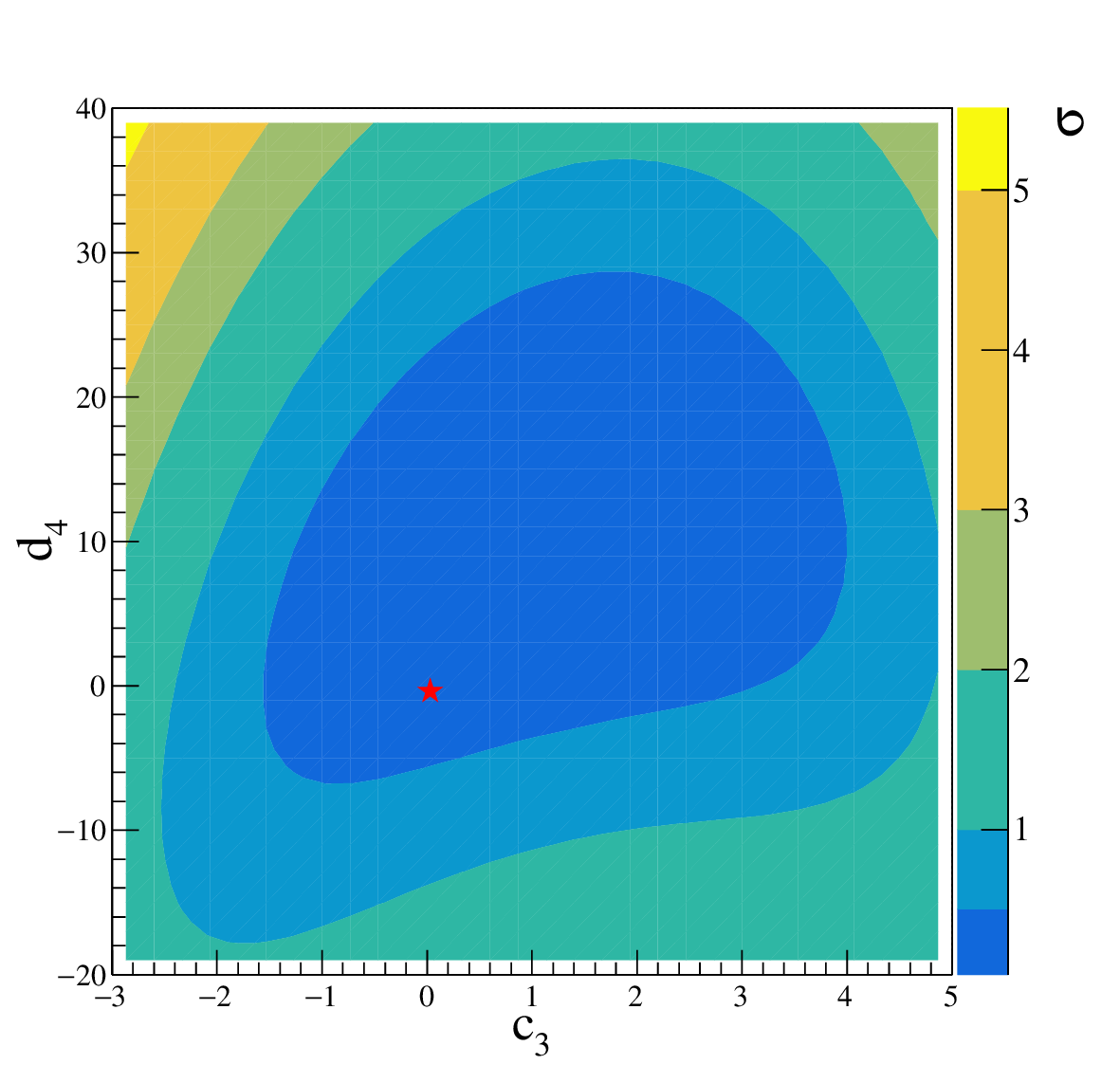}
        \caption{}
    \end{subfigure}%
    \qquad
    \begin{subfigure}{.45\textwidth}
        \centering
        \includegraphics[width=\textwidth]{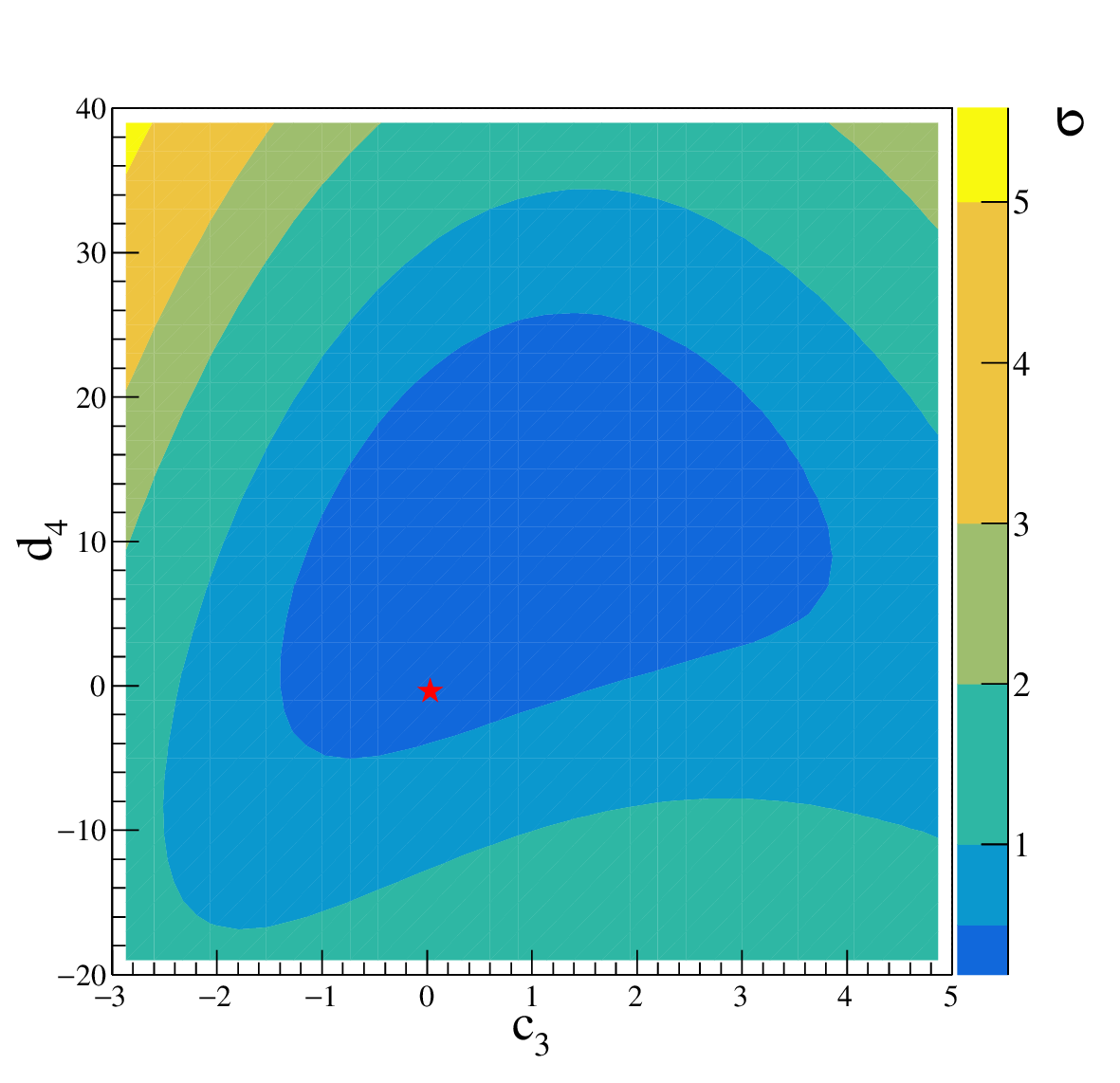}
        \caption{}
    \end{subfigure}
    \caption{Significance contour on the $(c_3, d_4)$ plane expected for a luminosity of 30~$\text{ab}^{-1}$ at a 100 TeV proton-proton collider, based on cut-based optimization, combining all resolved and boosted categories. (a) represents the HHH-only signal; (b) represents the combined HHH+VHH signal. The red star indicates the SM point.}
    \label{fig:significance_contour}
\end{figure}

Fig.~\ref{fig:significance_contour} presents the cut-based significance contour plots on the $(c_3, d_4)$ plane for two scenarios: (a) HHH-only signal and (b) combined HHH+VHH signal. 
These significance values are derived by combining all resolved and boosted categories. The results reveal the sensitivity variation as a function of the trilinear and quartic Higgs self-couplings. The signal significance in this scenario would reach around 2$\sigma$ in the corner where $c_3 < -0.5$ and $d_4 > 10$ in the scanned range.



\begin{figure}[tp]
    \centering
    \begin{subfigure}{.45\textwidth}
        \centering
        \includegraphics[width=\textwidth]{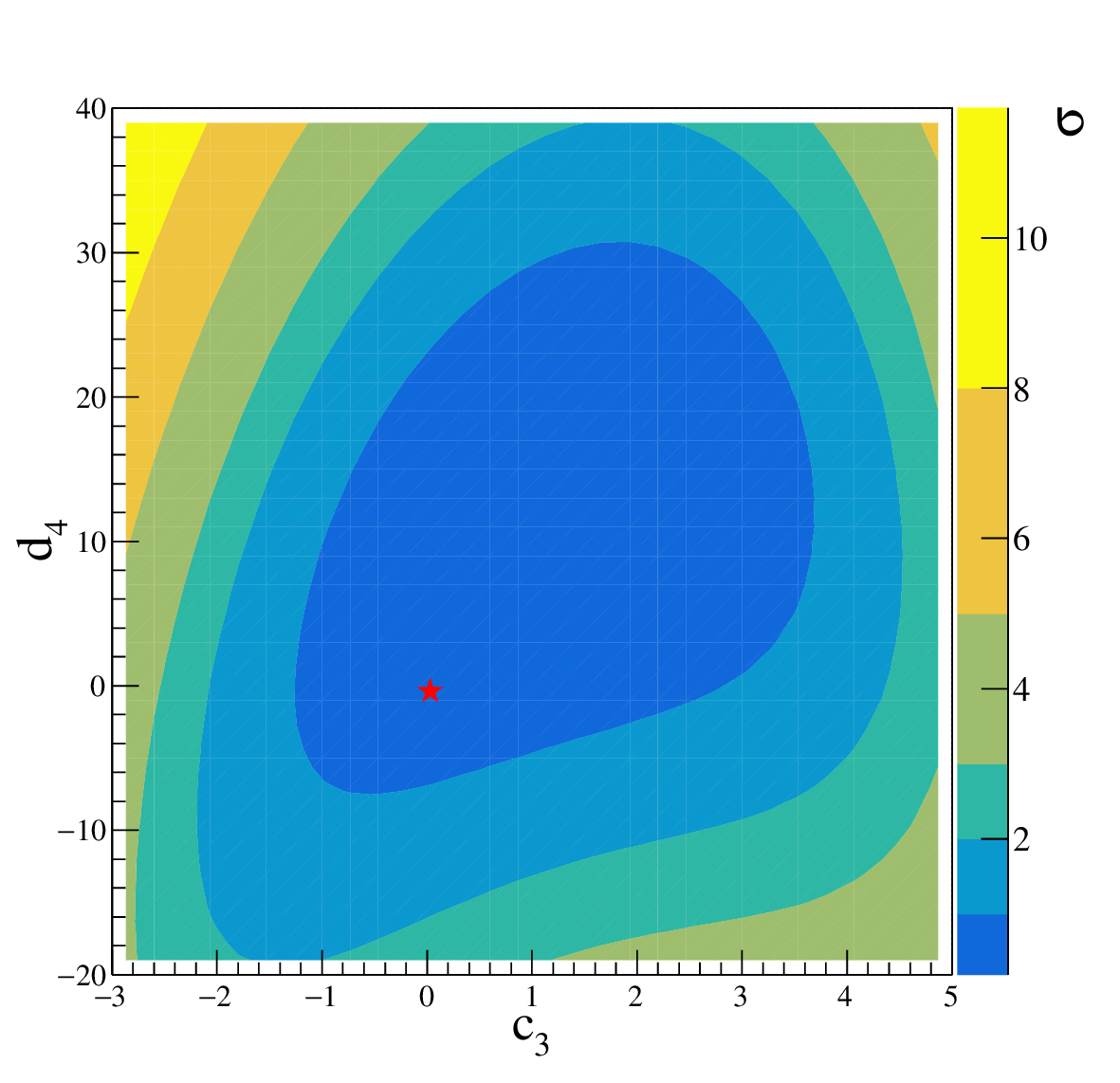}
        \caption{}
    \end{subfigure}%
    \qquad
    \begin{subfigure}{.45\textwidth}
        \centering
        \includegraphics[width=\textwidth]{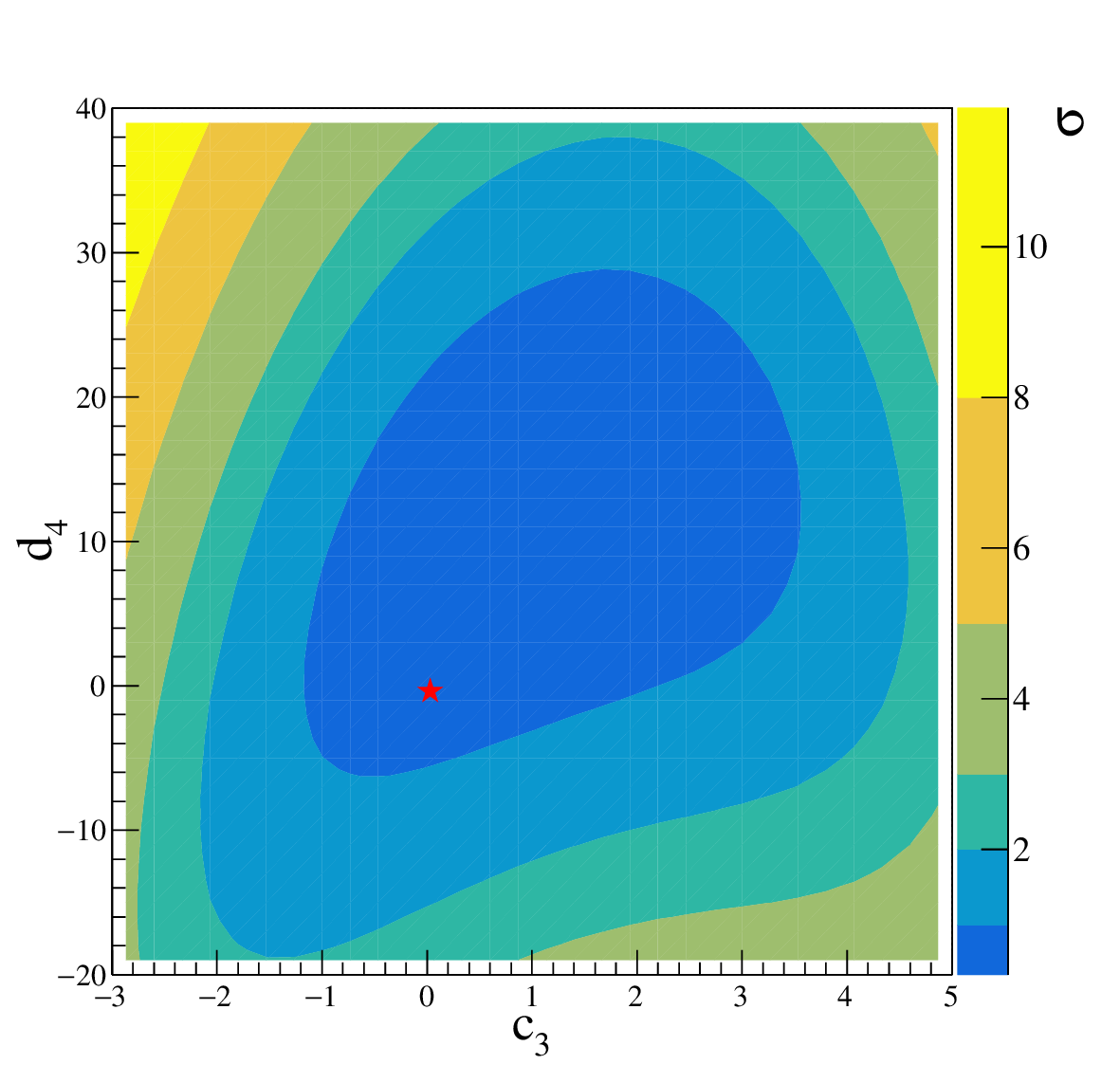}
        \caption{}
    \end{subfigure}
    \caption{Significance contour on the $(c_3, d_4)$ plane expected for a luminosity of 30~$\text{ab}^{-1}$ at a 100 TeV proton-proton collider, based on XGBoost BDT training with resolved events. (a) represents the HHH-only signal; (b) represents the combined HHH+VHH signal. The red star indicates the SM point.}
    \label{fig:xgb_significance_contour}
\end{figure}

In addition to the results based on cut-based optimization, this analysis incorporates BDT training for the resolved event categories. 
In the BDT-based approach, the signal significance is extracted with a binned fit to the BDT score distributions, 
allowing the multivariate classifier to exploit correlations among multiple kinematic variables and thereby enhance sensitivity.
Tables~\ref{tab:cmp_HHH} and \ref{tab:cmp_VHH} summarize the expected significances for various benchmark points in the $(c_3, d_4)$ plane, comparing the performance of cut-based and BDT-based methods. The BDT-based method consistently outperforms the cut-based approach, achieving improvements in significance ranging from approximately 60\% up to around 160\%, depending on the benchmark. This highlights the power of multivariate analysis in enhancing sensitivity to triple Higgs production.
Figure~\ref{fig:xgb_significance_contour} presents the BDT-based significance contours for the HHH-only signal (left) and for the combined HHH+VHH signal (right). Compared to the cut-based results, the BDT contours exhibit a visibly improved sensitivity across the entire parameter space. The signal significance with BDT method would reach around 2$\sigma$ for almost entire $d_4$ phase space when $c_3 < -2$. A significance of 5$\sigma$ can be reached in the corner where $c_3 < -1$ and $d_4 > 10$.

\begin{table}
    \centering
    \begin{tabular}{lccc}
        \toprule
         HHH signal&  $\sigma$ (cut-based)&  $\sigma$ (BDT)& Improvement\\
        \midrule
         SM $c_3=0, d_4=0$&  0.239&  0.385& 61.44\%\\
         BSM $c_3=4, d_4=9$&  0.478&  1.237& 158.82\%\\
         BSM $c_3=-2, d_4=-11$&  0.806 &  1.825 & 126.42\%\\
         BSM $c_3=3, d_4=-21$&  2.140 &  4.095 & 91.34\%\\
         BSM $c_3=0, d_4=-21$&  1.714&  3.139 & 83.21\%\\
         BSM $c_3=-2, d_4=19$& 1.548 & 3.622&134.1\%\\
         BSM $c_3=-3, d_4=9$&  2.171 &  5.504 & 153.4\%\\
         BSM $c_3=-3, d_4=14$&  2.578 &  6.327 & 145.4\%\\
        \bottomrule
    \end{tabular}
    \caption{Comparison of signal significance between the cut-based and BDT-based analyses for SM and several BSM scenarios of HHH signal.}
    \label{tab:cmp_HHH}
\end{table}

\begin{table}
    \centering
    \begin{tabular}{lccc}
        \toprule
         HHH+VHH signal&  $\sigma$ (cut-based)&  $\sigma$ (BDT)& Improvement\\
        \midrule
         SM $c_3=0, d_4=0$&  0.309 &  0.498 & 61.17\%
\\
         BSM $c_3=4, d_4=9$&  0.506 &  1.282 & 153.36\%
\\
         BSM $c_3=-2, d_4=-11$&  0.853 &  1.875 & 119.81\%
\\
         BSM $c_3=3, d_4=-21$&  2.187 &  4.220 & 92.96\%
\\
         BSM $c_3=0, d_4=-21$&  1.791 &  3.254 & 81.69\%
\\
         BSM $c_3=-2, d_4=19$& 1.604& 3.684&129.68\%
\\
         BSM $c_3=-3, d_4=9$&  2.168&  5.569& 156.87\%
\\
         BSM $c_3=-3, d_4=14$&  2.585&  6.399& 147.54\%\\
        \bottomrule
    \end{tabular}
    \caption{Comparison of signal significance between the cut-based and BDT-based analyses for SM and several BSM scenarios of HHH+VHH signal.}
    \label{tab:cmp_VHH}
\end{table}

The cut-based analysis is primarily optimized for the HHH production process, with no dedicated optimization for the VHH signal. As a result, the VHH signal yields remains low across all categories. A similar limitation applies to the BDT-based analysis, where the VHH process is not included in the training as part of the signal. When VHH signal is included in the $(c_3, d_4)$ parameter space, a relatively modest improvement in the overall significance is observed, typically in the range of 5\% to 50\%. This limited enhancement is mainly attributed to the low VHH signal efficiency after selections, which restricts its impact on the final results. Nonetheless, the residual VHH events provide a modest but non-negligible contribution to the overall sensitivity, particularly for the trilinear coupling parameter $c_3$.


\section{Conclusion}
\label{sec:conclusion}

In this study, we presented the first study of triple Higgs production in the $4\tau 2b$ decay channel at a future 100~TeV proton-proton collider, incorporating both resolved and boosted reconstruction techniques. 
The coupling-dependent partitioning of the $m_{\text{HHH}}$ phase space proved particularly effective, enabling optimized sensitivity across different regions of the Higgs self-coupling parameter space.

The $4\tau 2b$ channel offers complementary sensitivity to previously studied channels such as $6b$ and $4b 2\tau$. While this channel benefits from relatively lower SM backgrounds compared to b-enriched final states, the advantage is partially mitigated by lower signal yield inherent to $\tau$-lepton final states.

In this study, we employed two complementary approaches: a traditional cut-based optimization and a more advanced BDT-based analysis. The BDT training, applied to the resolved event categories, demonstrated significant improvements in sensitivity by leveraging the multivariate information in the event topology. By combining variables such as invariant masses, transverse mass variables, the BDT was able to exploit subtle correlations that are challenge to capture with simple rectangular cuts. The BDT-based analysis improved the signal significance by approximately 60\%–160\% compared to the cut-based approach. This highlights the power of machine learning techniques in enhancing the sensitivity of complex searches such as triple Higgs production.


In summary, this work establishes the $4\tau 2b$ channel as a promising avenue for probing the Higgs self-couplings at a future 100 TeV collider. By combining traditional reconstruction techniques with modern machine learning approaches like XGBoost BDT training, we demonstrate the ability to achieve good sensitivity in this challenging final state, especially in $c_3 \lesssim -1$ and $d_4 \gtrsim 10$ of the scanned range where 5 $\sigma$ in significance could be reached. With further advancements in analysis strategies, including the adoption of deep learning, the $4\tau 2b$ channel is expected to play a viable role in constraining the Higgs potential and exploring the nature of electroweak symmetry breaking.


\acknowledgments

The work is supported in part by the National Science Foundation of China under Grants No. 12175006, No.~12188102, No.~12061141002 and by the Ministry of Science and Technology of the People's Republic of China under Grants No.~2023YFA1605800.


\bibliographystyle{JHEP}
\bibliography{main}






\end{document}